\documentclass[preprint2,twocolumn,times,tighten]{aastex62}
\usepackage{pifont}
\usepackage{color}
\usepackage{enumitem}
\usepackage{hyperref}
\usepackage{verbatim}
\usepackage{amsmath,amstext,mathrsfs}
\usepackage[all]{hypcap} 
\usepackage{afterpage}    
\usepackage{marginnote}
\usepackage{makecell}
\usepackage{subfigure}

\def\X#1{ \raisebox{.9pt}{\textcircled{\raisebox{-.9pt}{#1}}}%
}
\graphicspath{{./}{figures/}}

\shorttitle{Identifying gravitational collapse with Velocity Gradient Technique}
\shortauthors{Hu, Lazarian \& Yuen}

\begin{document}
	
	\title{Velocity Gradient in the Presence of Self-Gravity: Identifying Gravity-induced Inflow and Determining Collapsing Stage }
	
	\email{yue.hu@wisc.edu, alazarian@facstaff.wisc.edu, kyuen@astro.wisc.edu}
	
	\author[0000-0002-8455-0805]{Yue Hu}
	\affiliation{Department of Physics, University of Wisconsin-Madison, Madison, WI 53706, USA}
	\affiliation{Department of Astronomy, University of Wisconsin-Madison, Madison, WI 53706, USA}
	\author{A. Lazarian}
	\affiliation{Department of Astronomy, University of Wisconsin-Madison, Madison, WI 53706, USA}
	\affiliation{Korea Astronomy and Space Science Institute, Daejeon 34055, Republic of Korea}
	\author{Ka Ho Yuen}
	\affiliation{Department of Astronomy, University of Wisconsin-Madison, Madison, WI 53706, USA}
	
	\begin{abstract}
		Understanding how star formation is regulated requires studying the energy balance between turbulence, magnetic fields, stellar feedback, and gravity within molecular clouds. However, identifying the transition region where the gravity takes over remains elusive. Recent studies of the Velocity Gradient Technique (VGT), which is an advanced tool for magnetic field studies, reveal that the gradients of spectroscopic observables change their directions by 90$^\circ$ with respect to the magnetic fields in the regions of gravitational collapse. In this study, we perform 3D MHD numerical simulations. We observe that star formation successfully proceeds in strongly magnetized and fully ionized media. We confirm that the self-gravity induces the change of gradients' orientation and high gradients' amplitude. We explore two ways of identifying collapsing self-gravitating regions through the double-peak feature in the histogram of gradients' orientation and the curvature of gradients. We show that velocity gradients' morphology and amplitude can be synthetically used to trace the convergent inflows. By comparing with the column density Probability Density Functions (N-PDFs) method, we show that VGT is a powerful new tool for studying the gas dynamics and tracing magnetic field in star-forming regions. By analogy with VGT, we extend the Intensity Gradient Technique (IGT) to locate the gravitational collapsing region and shocks. We demonstrate the synergy of VGT and IGT can determine the collapsing stages in a star-forming region. 
	\end{abstract}
	
	\section{Introduction} 
	\label{sec:intro}
	The molecular clouds in Milky Way are permeated by ubiquitous turbulent magnetic fields \citep{1981MNRAS.194..809L,2004ARA&A..42..211E,2007prpl.conf...63B,MO07,CL09}. Turbulence, magnetic fields, and gravity significantly impact the critical properties of the star formation processes and stellar initial mass distribution in molecular clouds \citep{1966ApJ...146..480J,1965ApJ...142..584P,1979cmft.book.....P,1998ARA&A..36..189K,1998ApJ...498..541K,2011Natur.479..499L,2013ApJ...768..159H,Aetal15,2014ApJ...783...91C,2016ApJ...832..199X, 2016ApJ...824..113X}. To understand the complex interplay of gravity, turbulence, and magnetic fields, it is essential to identify and study the transition regions where gravity takes over and collapse occurs \citep{1992pavi.book.....S,1994ApJ...429..781S,1977ApJ...214..488S,2005ApJ...630..250K,2011ApJ...743L..29H,2011ApJ...730...40P,2012ApJ...761..156F,2013ApJ...763...51F,2020arXiv200107721X,2020MNRAS.491.4310T}. Nevertheless, observational studies of the self-gravitating transition region in molecular clouds have still not yet been fully developed.
	
	Most analytic star formation theories rely on the column density Probability Density Functions (N-PDFs) of supersonic, magnetized, and isothermal turbulence to get an insight of self-gravity \citep{2000ApJ...535..869K,2017ApJ...840...48P,2011ApJ...727L..20K,2018ApJ...863..118B,1995ApJ...441..702V,2008ApJ...680.1083R,2012ApJ...750...13C,2018ApJ...863..118B}. It was believed that in the presence of self-gravitating gas, the N-PDF evolves to a combination of log-normal format for low-density gas and a power-law tail for high-density gas \citep{2011MNRAS.416.1436B,2011ApJ...727L..21P,1995ApJ...441..702V,2008ApJ...680.1083R,2012ApJ...750...13C,2018ApJ...863..118B}. The transition from log-normal to power-law N-PDF reveals the density threshold, above which the gas is likely to become self-gravitating. However, in addition to self-gravity, the N-PDFs are also shaped into a power-law distribution due to the insufficient optical depth in self-absorbing media \citep{pdf} or the effects of line-of-sight contamination on the column density structure \citep{2015A&A...575A..79S,2015A&A...578A..29S,2019MNRAS.484.3604L}, as well as  stellar feedback via ionization \citep{2014A&A...564A.106T}. The applicability of N-PDFs is therefore constrained.
	
	The Velocity Gradient Technique (VGT) is initially developed as an advanced tool for magnetic field studies \citep{2017ApJ...835...41G,YL17a,LY18a,PCA}. It uses the fact that within magnetohydrodynamic (MHD) turbulence, the turbulent eddies are anisotropic  \citep{GS95} and their anisotropy reveals the local direction of the magnetic field that percolates the eddy \citep{LV99,2000ApJ...539..273C}. As a result, the velocity gradients of the eddies are perpendicular to the magnetic fields.
	\begin{figure*}[t]
		\centering
		\includegraphics[width=.89\linewidth,height=0.73\linewidth]{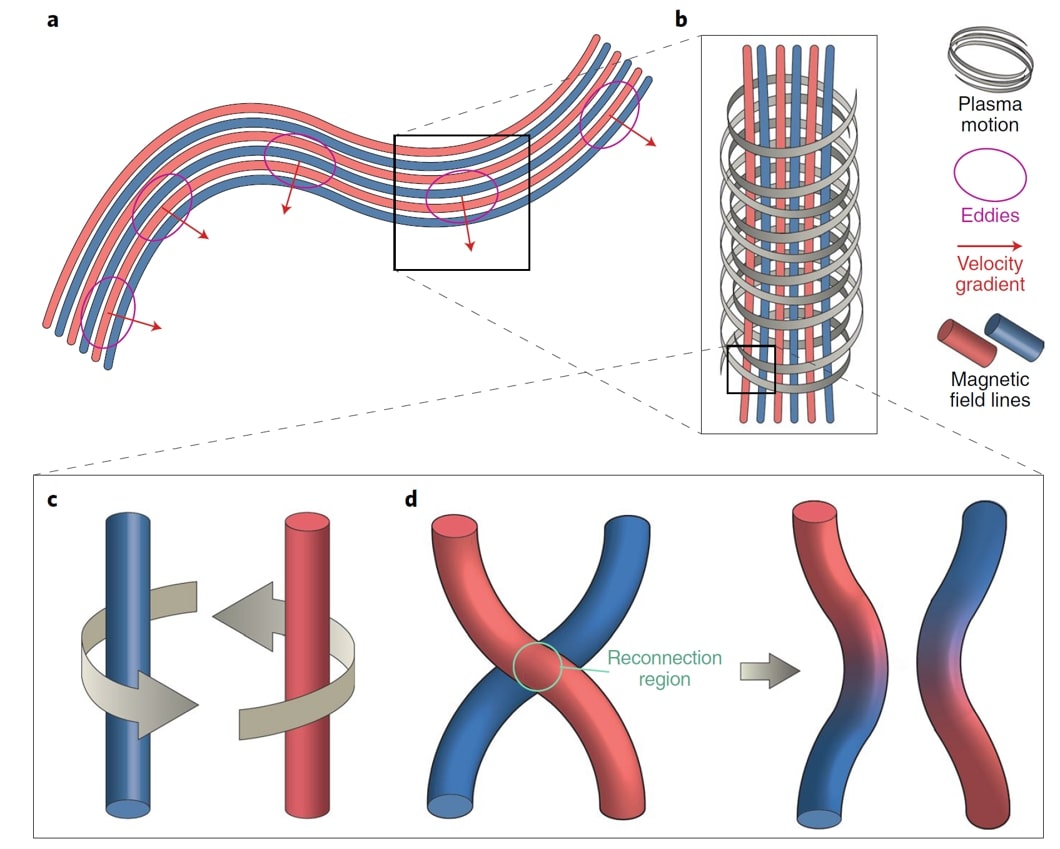}
		\caption{\label{fig:cartoon} A cartoon extracted from \citet{2019NatAs...3..692P}, explaining the principles behind the velocity gradient technique. \textbf{Panel a:} The VGT technique considers Alfv\'{e}nic fluid motions, wherein a magnetic field and the conducting fluid in which it is embedded move together. \textbf{Panel b:} Turbulent eddies are elongated along magnetic field lines. \textbf{Panel c:} Embedded magnetic field lines are preferentially moved perpendicular to their local direction by small-scale turbulent eddy motions. \textbf{Panel d:} Turbulent reconnection is an essential part of the dynamics of turbulent eddies, which enables the mixing of magnetic field lines perpendicular to the magnetic field direction. The mixing motions within the eddies (shown as magenta circles in Panel a) induce changes of the fluid velocities perpendicular to the magnetic field lines. Therefore, the local gas velocity gradients (red arrows in Panel a) are directed perpendicular to the local directions of the magnetic field. As a result, it is possible to predict the direction of the magnetic field by measuring the direction of gas flow velocity gradients. }
	\end{figure*}
	
	The ability of VGT to trace magnetic fields in turbulent media has been extensively tested in numerical simulations \citep{YL17a,2018ApJ...865...46L,PCA,2018ApJ...865...54Y,2019ApJ...873...16H} and in observations in, for example, in molecular clouds \citep{survey,velac} and neutral hydrogen gases \citep{2019ApJ...874...25G,PGCA}. \citet{LY18a} \& \citet{survey} also reveals that a new effect takes place in the presence of self-gravity, which results in a new method for measuring the transition from strongly magnetized to gravitational collapsing gas. The matter infall induces a change of the direction of velocity gradients concerning the magnetic field. In other words, towards regions where star formation is taking place, the gradients of velocity observables induced by the infall motions parallel to the magnetic field gradually begin dominating over the velocity gradients arising from turbulence. The observational signature of this change in the cloud dynamics is that the VGT orientation flips by up to 90$^\circ$ to align parallel with the direction of the magnetic fields \citep{YL17b,LY18a, survey}.

	By identifying the change of the velocity gradients, one can, therefore, identify the regions in which dynamics are dominated by self-gravity. If magnetic field direction is already known from polarimetry, this identification is trivial. However, the VGT was introduced as a technique for studying the magnetic field independently of high-cost and frequently impossible polarimetric studies. Therefore in this paper, we discuss the pure applicability of VGT in identifying collapsing cloud regions and, very importantly, studying magnetic fields in self-gravitating molecular clouds. 
	
	One of the most important tools in identifying the signature of self-gravity in molecular clouds would be the use of N-PDFs. Our study of N-PDFs of of synthetic CO emissions in \citet{pdf} has revealed that the shape of the N-PDFs will be skewed in the presence of self-absorption. This is a severe limitation of the N-PDFs. Moreover, even in the absence of the effects of radiation transfer, as we discuss in the following sections, VGT is a more sensitive tool compared to the N-PDFs, as VGT is sensitive to the convergent infall flows induced by the gravitational collapse. In addition, VGT can be used synthetically with intensity gradients that also change their orientation with respect to the magnetic field in the presence of self-gravity \citep{YL17b, IGs}. However, the change of density field is an accumulating process, while the velocity field is significantly changed only when the gravitational energy dominates over the kinematic energy of turbulence. This helps to identify the stage of gravitational collapsing for molecular clouds using the relative orientation of intensity gradients and velocity gradients. In addition, self-gravity induces high gradient amplitudes, but this is not the case for shocks. This high amplitude feature paves the way to distinguish shocks and self-gravitating materials.
	
	For this study, we use several computational tools that are developed for the gradient studies. For instance, to identify the boundary of the collapsing region, we develop the double-peak algorithm and also use the curvature algorithm presented in \citet{2020arXiv200201926Y}. At the same time, we use the Principal Component Analysis (PCA) to extract the most important channel map data from spectroscopic position-position-velocity cubes \citep{PCA}. To find gradient directions, apart from the classical fixed size sub-block averaging \citep{YL17a}, we experiment with the adaptive sub-block averaging (see \S~\ref{subsec:asb}). These methods improve the ways that the gradients are calculated.

	In what follows, we theoretically illustrate the properties of MHD turbulence in the presence and absence of self-gravity, in \S~\ref{sec:theory}. In \S~\ref{sec:method}, we describe the full algorithm in calculating velocity gradients with the filtering process, i.e., the PCA. In \S~\ref{Sec.data} we give details about the numerical simulation used in this work. In \S~\ref{sec:results}, we study the responses of N-PDFs and velocity gradients with respect to self-gravity, as well as identify gravitational collapsing regions. In \S~\ref{sec:IGT}, we extend our study of velocity gradients to intensity gradients and show how to determine collapsing stages and shocks. In \S~\ref{sec.dis}, we discuss the application and further development of VGT. In \S~\ref{sec:con}, we give our conclusions.
	
	
	\section{Theoretical Consideration}
	
	\label{sec:theory}
	\subsection{VGT and MHD turbulence theory}
	
	The VGT utilizes the properties of MHD turbulence, and below we explain those essential for understanding this technique (see Fig.~\ref{fig:cartoon}). A more detailed discussion of MHD turbulence can be found in a recent book by \citet{BL19}. 
	
	The theory of MHD turbulence has been given a boost by the prophetic study by \citet{GS95}, denoted as GS95 later. In particular, GS95 predicted the turbulent eddies are anisotropic and showed that the degree of turbulence anisotropy increases as the scale of turbulent motions decreases (see Fig.~\ref{fig:cartoon}b). The subsequent study of turbulent reconnection in \citet{LV99} demonstrated that turbulent reconnection of the magnetic field, which takes place over just one eddy turnover time, is an intrinsic part of the MHD turbulent cascade. The reconnection enables the mixing of magnetic field lines perpendicular to the magnetic field direction (see Fig.~\ref{fig:cartoon}c). The mixing motions within the eddies further induce changes of the fluid velocities perpendicular to the magnetic field lines (see Fig.~\ref{fig:cartoon}d). Therefore, the local gas velocity gradients are directed perpendicular to the local directions of the magnetic field \footnote{The derivations in \cite{GS95} for the anisotropy are done using the mean-field reference frame. The GS95 scaling is not observable in this mean-field frame.}. This phenomenon has been confirmed by the numerical studies in \citet{2000ApJ...539..273C} and \citet{2001ApJ...554.1175M}. This notion of a local system of reference is critical for understanding the VGT. Indeed, for the detailed tracing magnetic field using the velocity field, it is essential that the velocities are oriented in respect to the direction of the magnetic field in the sampled volumes, rather than the direction of the mean magnetic field.
	
	In the process of observations, the contributions from eddies of different scales are summed up. The smallest eddies trace the fine magnetic structure, while the larger eddies sample magnetic field on a coarser grid scale. For the VGT, it is essential that the gradients arising from the smallest eddies dominate. This follows from the scaling of MHD turbulence.
	
	According to \citet{LV99}, magnetic fields give minimal resistance to the motions of eddies with scale $l_{\bot}$ perpendicular to the local direction of the magnetic field, i.e., the magnetic field passing through the eddy at scale $l$. Thus, the eddies obey the hydrodynamic Kolmogorov law $v_{l,\bot}\sim l_\perp^{\frac{1}{3}}$, where $v_{l,\bot}$ is the turbulence's injection velocity perpendicular to the local direction of the magnetic field. The relation between the scale $l_\|$ along the magnetic field and $l_\perp$ perpendicular the magnetic field is expressed as \citep{LV99}:
	\begin{equation}
	\label{eq.1}
	l_\|\simeq L_{inj}(\frac{l_{\perp}}{L_{inj}})^{\frac{2}{3}}M_A^{-\frac{4}{3}}
	\end{equation}
	where $L_{inj}$ is the injection scale of turbulence and $M_A = v_l/v_A$ is the Alfv\'{e}nic Mach number, i.e., the ratio bewteen the injection velocity $v_{l}$ of turbulence and Alfv\'{e}nic velocity $v_A$. By equating the period of Alfv\'{e}nic wave and turbulent eddy's turnover time, the scaling of turbulent velocity of eddies is derived as \citep{LV99} :
	\begin{equation}
	\frac{l_{\|}}{v_A}\simeq\frac{l_{\bot}}{v_{l,\bot}}
	\end{equation}
	
	\begin{equation}
	\label{eq:KS}
	v_{l,\bot}\simeq v_l(\frac{l_{\perp}}{L_{inj}})^{\frac{1}{3}}M_A^{\frac{1}{3}}
	\end{equation}
	when $M_A = 1$ the relation returns to the GS95 version. Explicitly, since the anisotropic relation indicates $l_\bot \ll l_\parallel$, the velocity gradient scales as:
	\begin{equation}
	\label{eq:4}
	\nabla v_l\simeq\frac{v_{l,\bot}}{l_\bot}\simeq \frac{v_l}{L_{inj}}(\frac{l_{\perp}}{L_{inj}})^{-\frac{2}{3}}M_A^{\frac{1}{3}}
	\end{equation}
	Eq.~\ref{eq:4} is a crucial element for VGT technique, as it testifies that (i) the gradients of velocity amplitude scale as $v_{l,\bot}/l_{\bot}\propto l_{\bot}^{-2/3}$, i.e., the smallest resolved scales are most important in calculating the gradients; (ii) the measured velocity gradients are perpendicular to the magnetic field at the smallest resolved scales, i.e., they well trace the magnetic field in the turbulent volume. 
	\begin{figure*}[t]
		\centering
		\includegraphics[width=.89\linewidth,height=0.55\linewidth]{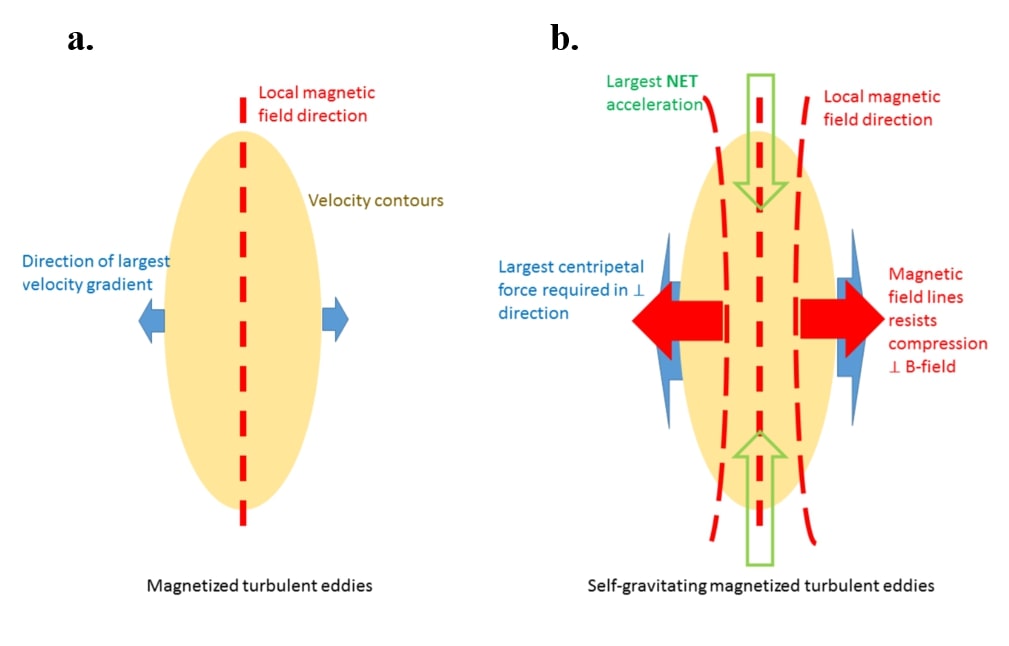}
		\caption{\label{fig:gravity} Illustrations on how self-gravity changes the maximum gradient direction, extracted from \cite{YL17b}. \textbf{Panel a:} Eddies are elongated parallel to the local magnetic field direction. When gravity is absent (left), the maximum change of the velocity amplitudes (i.e., velocity gradient) is in the direction perpendicular to the local magnetic field. \textbf{Panel b:} the gravitational pull produces the most significant acceleration of the plasma in the direction parallel to the magnetic field, and the velocity gradients are parallel to the magnetic field. }
	\end{figure*}
	
	\subsection{Velocity gradient for sub- and super-Alfv\'{e}nic turbulence}
	The magnetic field plays a crucial role in the anisotropic scaling relation, as shown in Eq.~\ref{eq.1}. It is therefore important o discuss the turbulence scalings in different regimes, i.e. sub- and super-Alfv\'{e}nic turbulence. We briefly describe the regimes below. A more extensive discussion can be found in the review by \citet{2013SSRv..178..163B}.
	
	For super-Alfv\'{e}nic turbulence, the anisotropic relation appears when the scale of turbulence gets smaller than the transitional scale $l_A$, which is expressed as \citep{2006ApJ...645L..25L}:
	\begin{equation}
	\label{eq.4}
	l_A=L_{inj}M_A^{-3}, M_A>1
	\end{equation}
	At scales smaller than $l_A$, the turbulence transfers to the trans-Alfv\'{e}nic turbulence described by GS95 relation and the anisotropy of the turbulent eddies can be observed, i.e., Eq.~\ref{eq:KS} still holds on. When turbulence's scale is larger than $l_A$, the turbulence cascade is essentially a hydrodynamic Kolmogorov cascade showing isotropic property instead of anisotropic  \citep{2006ApJ...645L..25L}. 
	
	In the strong magnetic field case $M_A<1$, the corresponding turbulence is always highly anisotropic. However, turbulence in the range from the injection scale $L_{inj}$ to transitional scale $l_{tr}$ becomes wave-like rather than eddy-like:
	\begin{equation}
	\label{eq.5}
	l_{tr}=L_{inj}M_A^{2}, M_A<1
	\end{equation}
	which is termed as weak Alfv\'{e}nic turbulence \citep{BL19}. This type of turbulence means that the velocities change to \citep{LV99,2000JPlPh..63..447G}:
	\begin{equation}
	v_{l,\bot}\simeq v_l(\frac{l_{\perp}}{L_{inj}})^{\frac{1}{2}}
	\end{equation}
	and the velocity gradient becomes:
	\begin{equation}
	\label{eq:3}
	\nabla v_l\simeq\frac{v_{l,\bot}}{l_\bot}\simeq  \frac{v_l}{L_{inj}}(\frac{l_{\perp}}{L_{inj}})^{-\frac{1}{2}}
	\end{equation} 
	At the scale smaller than $l_{tr}$, the turbulence transforms into the strong Alfv\'{e}nic regime, which obeys the GS95 anisotropic relation. In both regimes, the velocity gradients are perpendicular to the magnetic field. Note that turbulence can be also injected locally. Then the scale of turbulent motions $L_{inj}$ gets different, as well as $l_A$. To trace the magnetic field direction using gradients, it's important that a telescope can resolve the turbulence at a scale smaller than $l_A$ for super-Alfv\'{e}nic turbulence. In the case that resolution is not sufficient, the contribution from large scales should be removed using spatial filtering \citep{YL17b}.

	\subsection{Velocity gradient in the presence of self-gravity}
	In the vicinity of gravitational collapse, the self-gravity radically modifies the nature of the turbulent flow. In the case of strong self-gravity, velocity gradients are expected to change their orientation from perpendicular to magnetic fields to align with magnetic fields \citep{YL17b, survey}. As shown in Fig.~\ref{fig:gravity}, assuming the gravitational center is located at the center of a turbulent eddy, when gravity is sub-dominant to magnetic and turbulent energy, the magnetized turbulent eddies are elongated in the direction parallel to the magnetic field surrounding the eddies. As a result, the maximum change of the velocity amplitudes (i.e., velocity gradient), is in the direction perpendicular to the local magnetic field, and by rotating the velocity gradient by 90$^\circ$ we can trace the
	magnetic field. In regions where gravitational collapse has begun, the dynamics are different. If the magnetic field is strong enough to provide support, we expect that in the direction perpendicular to the magnetic field, any gravitational pull inducing the acceleration is counteracted by a magnetic force. Hence, the gravitational pull produces the most significant acceleration of the plasma in the direction parallel to the magnetic field, and the velocity gradients are parallel to the magnetic field. Alternatively, in a case where the magnetic field support is weak compared to gravity, the infall motions of the plasma will alter the magnetic field geometry so that it tends to align parallel to the direction of gravitational collapse.  Nevertheless, the acceleration induced by the collapse is still along the magnetic field direction. As a result, the velocity gradients are also parallel to the magnetic field. In this paper, we focus on exploring the ability of VGT to trace magnetic field in the magnetically dominated environments with sub-Alfv\'{e}nic turbulence. The turbulence in molecular clouds can be super-Alfv\'{e}nic \citep{1999ApJ...526..279P}. For numerically studying the applicability of the VGT in super-Alfv\'{e}nic setting, we require higher resolution resolving scales less than the transitional scale $l_A$. The numerical study of the VGT for super-Alfv\'{e}nic molecular clouds will be done elsewhere.
	
	A similar argument also holds for intensity gradients, i.e., intensity gradients are perpendicular to the magnetic field due to the anisotropic relation of turbulent eddies. However, intensity gradients exhibit different properties from velocity gradients. A prior study in \cite{Soler2013} reported that the intensity gradient is rotating with respect to magnetic fields when arriving density threshold $n_T \sim 50\langle n\rangle$ (where $\langle n\rangle$ denotes the average column density) in the case of M$_s$=10 super-sonic media, and $n_T \sim 500\langle n\rangle$ in the case of moderately magnetized media ($\beta$ = 1.0) for super-Alfv\'{e}nic and self-gravitating MHD simulation. These findings are at the foundations of the Histograms of Relative Orientation (HRO) technique proposed in \cite{Soler2013} and elaborated further in \citet{2017A&A...607A...2S}.
	
	The difficulty of employing HRO compared to our way of identifying the gravitational collapse are stems from two issues. First of all, HRO does not identify the change of the intensity gradients on its own. Instead, it requires polarization measurements to provide the magnetic field direction. Second, the change of the intensity gradient direction with respect to the magnetic field is also happening in the presence of shocks. Indeed, it was demonstrated that the intensity gradient tends to be parallel to its local magnetic fields when getting close to the dense shock front in the absence of gravity \citep{YL17b,2019ApJ...878..157X,IGs}. Therefore, the intensity gradient is sensitive to both self-gravity and shock. The change of intensity gradient induced by shocks does not specify a density threshold, which calls the value of $n_T$ into questions. It, therefore, encourages a detailed study of intensity gradients, i.e., distinguishing shocks and self-gravitating regions. In contrast, VGT provides, as we discuss further the {\it polarimetry free} way of determining self-gravitating regions. In addition, using IGT, as we discuss later, there is a reliable way of distinguishing the shocks and self-gravitating regions 
	
	We note, that in this paper we also use the intensity gradients. However, our IGT is {\it different} from the HRO \citep{IGs}. The IGT is an offshoot of the VGT technique and it uses the VGT procedures to trace magnetic field directions. In this paper we show the synergy of using the VGT together with the IGT.
	
	In this paper, we address the issue of how velocity gradients and IGs trace magnetic fields in the regions of self-dominant gravity for the case of sub- and trans-Alfv\'{e}nic clouds, and further discuss how the stage of collapse is correlated to the alignment of density gradients and velocity gradients.
	
	\subsection{Distinguishing shocks from gravity in spectroscopic velocity channels}
	\label{subsec:shockvschannel}
	
	Our previous works show that both regions with shock and self-gravity will make the gradients of either intensity or centroid gradients to be parallel to the magnetic field \citep{YL17b,IGs}. In this subsection, we formulate qualitatively how shocks are different from self-gravitating regions in velocity channel maps and how can we make use of this simple yet important principle to extract self-gravitating regions in observations.
	
	
	In obtaining the velocity-gradient predicted magnetic field directions, we have to perform a weighted sum along the line of sight together with the magnetic field angle we predicted by VGT on channels. Shock and self-gravitating regions act differently under this "Stokes" sum (see \S~\ref{sec:method}). MHD shocks are mostly formed with its velocity dispersion being rather small. Therefore when we are observing a shock region in PPV space, it occupies only a few velocity channels. Comparatively, the self-gravitating region accelerates fluid particles in its vicinity. That means when we are observing the self-gravitating region in PPV space, we can find this strongly self-gravitating region in almost all velocity channels. 
	
	This simple phenomenological effect has a deep consequence in searching for self-gravitating regions using velocity gradients. The shock density enhancement, unlike its self-gravity counterpart, is bounded by the sonic Mach number of the environment, but that for self-gravity is unbounded. In the case of shocks, there will be a limited number of channels that are affected by shocks, subjected to the sonic Mach number. In this case, we will find most of the shocks in a limited amount of channels together with non-shock materials that have their gradients perpendicular to magnetic fields. As a result, for the channels with shocks, the alignment measure decreases, while for the channels without shocks the alignment measure stays the same. Since our prediction of the magnetic field is a weighted sum along the line of sight, the effective alignment measure is only slightly decreased and is a function of $M_s$.
	
	Self-gravity acts differently in velocity channels. Since we can find the self-gravitating region in all velocity channels depending on its stage of collapse, as a result, the velocity gradients of every channel are rotated according to the strength of acceleration, causing the alignment of each channel to be significantly low. That would mean that in the weighted sum there are no channels that have high enough alignment measure to counterbalance the gravity-induced low alignment measure channels, causing the resultant Stokes map to have a very low alignment measure. That means only the change of orientation caused by gravity is retained under Stokes's addition. Using this simple fact we can, therefore, determine the self-gravitating regions by spotting the locations where there are changes of orientation of Stokes-weighted velocity gradients.

	\subsection{Velocity fluctuations in thin velocity channel}
	The information about turbulent velocities is contained in Position-Position-Velocity (PPV) data which is available through spectroscopic observations. The statistics of the intensity fluctuations in PPV and their relations to the underlying statistics of turbulent velocity and density are presented in \cite{LP00}. There it was shown the velocity effects are most prominent in thin channel maps, i.e. in thin velocity slices of PPV cubes.
	
	Within thin channel maps, the field of velocity produces intensity fluctuations due to the effect of velocity caustics. These fluctuations in particular regimes identified in \cite{LP00} can dominate the effects of intensity fluctuations arising from density clumping. In fact, if the energy spectrum of magnetic fluctuations is steep, e.g. Kolmogorov spectrum, the velocity fluctuations determine the statistics of the intensities in thin channels \cite{LP00}.  It can be demonstrated that this assumption holds not only in the single-phase self-absorption media but also in two-phase \ion{H}{1} media \citep{LP04,KLP16,KLP17a,KLP17b}, provided that the cold \ion{H}{1} clumps are moved together with the warm media. Therefore, by varying the thickness of velocity channels, one can extract the statistical density fluctuation and velocity fluctuation in PPV cubes \citep{L09, 2019arXiv190403173Y}. The criterion for distinguishing the thin channel and the thick channel is given as :
	\begin{equation}
	\label{eq1}
	\begin{aligned}
	\Delta v<\sqrt{\delta v^2}, \thickspace\mbox{thin channel}\\
	\Delta v\ge\sqrt{\delta v^2},\thickspace\mbox{thick channel}\\
	\end{aligned}
	\end{equation}
	where $\Delta v$ is the velocity channel width, $\delta v$ is the velocity dispersion calculated from velocity centroid \citep{LP00}. Therefore, we extract the velocity gradient from all thin channels that satisfy with the criteria as listed in Eq.\eqref{eq1}.
	
	However, in PPV cubes, only the velocity component parallel to LOS is achievable. If the flow induced by the gravitational collapse is exactly perpendicular to LOS and the flow does not allow the transfer of energy between different components, then there may have problems with detecting the velocity gradient’s change.  However, in reality, it is not possible to accelerate the turbulent flow regularly in one direction without changing all the components of velocity. Different from the convergent flow induced the self-gravity which spreads in the space, shock only induces the jump of velocity in the direction perpendicular to shocks' front and the velocity has limited range. Therefore, shocks' contribution to velocity field is insignificant in PPV cubes as we discussed in \S \ref{subsec:shockvschannel}.
	
	\citet{2019ApJ...874..171C} recently questioned the validity of velocity caustics effect in \ion{H}{1} emission lines. They propose that the \ion{H}{1} intensity features in thin channels are real density structures of the cold neutral medium instead of velocity caustics. The arguments of the authors are based on special properties of two-phase \ion{H}{1} gas and they are not applicable to the study of velocity gradients in isothermal conditions of molecular clouds that we deal in this paper. For a number of molecular clouds, we have obtained good correspondence of the magnetic field structure obtained from VGT and the one from Planck polarization data \citep{survey} and from BLASTPOL polarization \citep{velac}.

	A detailed discussion of the nature of the intensity fluctuations in channel maps for galactic \ion{H}{1} studies is beyond the scope of the present paper. We just mention that the corresponding answer has been presented in \citet{2019arXiv190403173Y}. The fact is that the formation of velocity caustics predicted in \citet{LP00} theory is the effect that has been studied extensively both theoretically and numerically for the last two decades \citep{2001ApJ...555..130L,KLP17b,2018MNRAS.479.1722C,2006ApJ...653L.125P,2003MNRAS.342..325E,CL09}.
	
	The predictions of the velocity caustic theory were successfully used to find the velocity spectrum by different groups in both two-phases \ion{H}{1} media and one-phase media of CO isotopes, H$_\alpha$, \ion{N}{2}, \ion{S}{2} and other species \citep{2001ApJ...551L..53S,2006ApJS..165..512K,2006ApJ...653L.125P,KLP17b,2001ApJ...561..264D,2006ApJS..165..512K}. Therefore, it is an established fact that the velocity effects are important for the formation of intensity structure of the velocity channel maps. The discussion initiated by \citet{2019ApJ...874..171C} should be viewed in terms of the relative contributions of density and velocity for the production of the filaments observed in the channel maps. Realistic simulations of the two-phase media turbulent will clarify this issue. We also note that applying the VGT to \ion{H}{1} data we obtained predictions of foreground polarization that is in good correspondence with Planck measurements \citep{PGCA}.    
	
	\subsection{Probability density function}
	The Probability Density Functions (PDF) is one alternative way to identify gravitational collapsing regions utilizing the information of the density field. Indeed, several studies revealed that the density distribution is a log-normal ($P_{N}$) for supersonic magnetized isothermal turbulence \citep{1995ApJ...441..702V,2008ApJ...680.1083R,2012ApJ...750...13C,2018ApJ...863..118B}:
	\begin{equation}
	P_{N}(s)=\frac{1}{\sqrt{2\pi\sigma_s^2}}e^{-\frac{(s-s_0)^2}{2\sigma_s^2}}
	\end{equation}
	where s$=ln(\rho/\rho_0)$ is the logarithmic density and $\sigma_s$ is the standard deviation of the log-normal, while $\rho_0$ and $s_0$ denote the mean density and mean logarithmic density.
	
	In the case of self-gravitating MHD turbulence, the gas density probability distribution function (PDF) evolves to a combination of log-normal ($P_{N}$) PDF  at low densities and a power-law ($P_{L}$) PDF  at high densities:
	\begin{equation}
	P_{L}(s)\propto e^{-\alpha s},s>S_t
	\end{equation}
	where S$_t = ln(\rho_t/\rho_0 )=(\alpha-\frac{1}{2})\sigma_s^2$ is the logarithm of the normalized transitional density between the log-normal and power-law forms of the density PDF \citep{1994ApJ...423..681V,2005MNRAS.356..737S,2011MNRAS.416.1436B,2019MNRAS.482.5233K}. The transition density $S_t$ value depends on the slope $\alpha$ of the power law and the width of the log-normal. As $\alpha$ shallows (i.e. becomes less steep, which is expected for strong self-gravitating turbulence), the transition density (S$_t$) between the PDF log-normal and power law moves towards lower density \citep{2018MNRAS.477.4951L}. Nevertheless, there still exist limitations of the N-PDFs in distinguishing regions of gravitational collapse in self-absorbing media, which is discussed in \citet{pdf}.
	\begin{figure*}
		\centering
		\includegraphics[width=1.00\linewidth,height=0.76\linewidth]{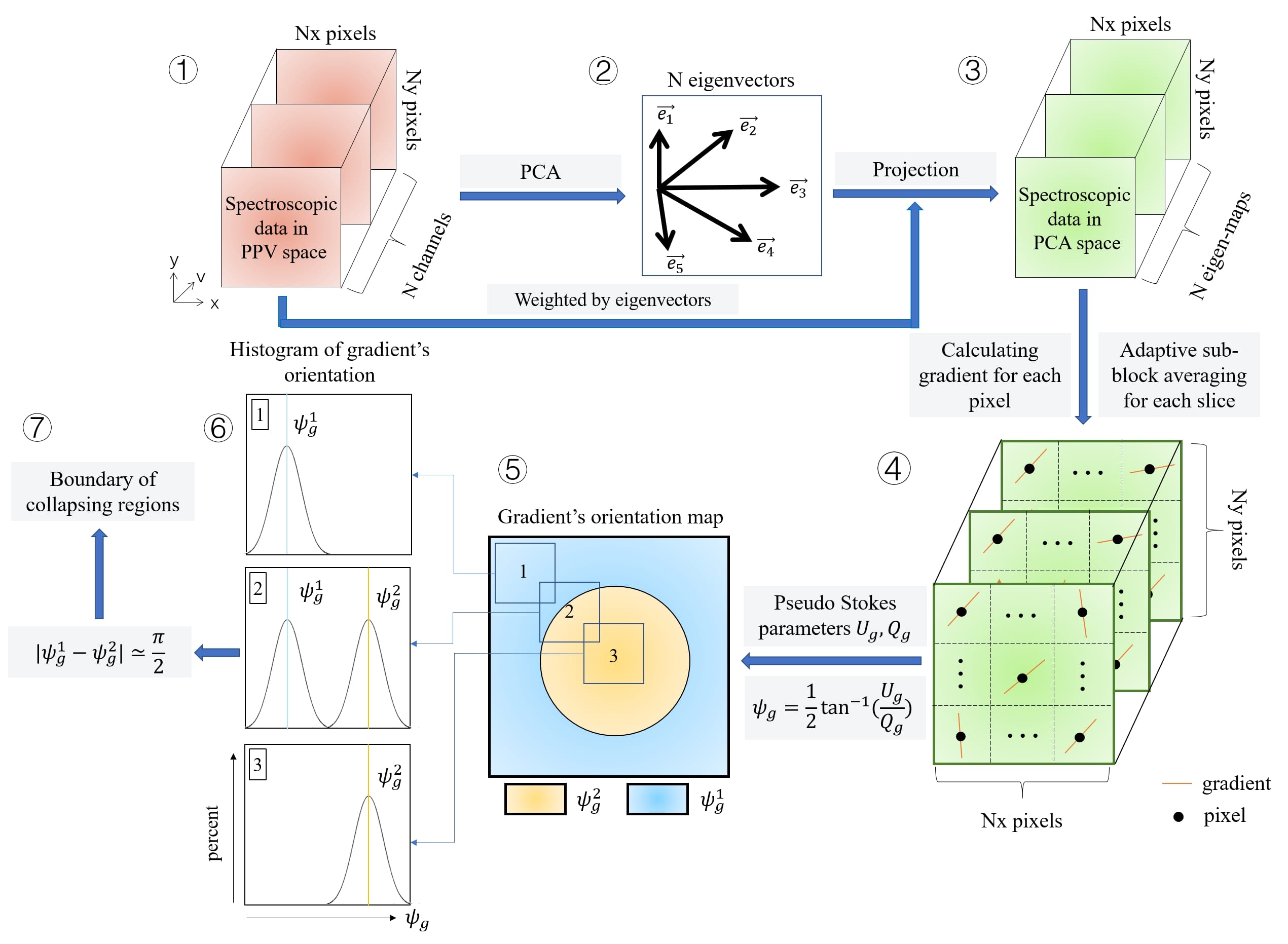}
		\caption{\label{fig:vgt} Diagram of the VGT procedure to identify the gravitational collapsing regions (see \S~\ref{sec:method}). Step 1 to step 3 are preprocessing of the spectroscopic data with dimension $N_x\times N_y\times N$ through PCA. Assuming that a PPV cube $\rho(x,y,v)$ is a probability density function of three random variables $x,y,v$, we can construct a new orthogonal coordinate system formed by the eigenvectors ($\Vec{e_1}$,$\Vec{e_2}$ ... $\Vec{e_N}$) of PPV's covariance matrix (see Eq.~\ref{eq:13} and Eq.~\ref{eq:14}). By weighting the original PPV cube with the eigenvectors we project the data into PCA's eigenspace. In the new space, the significance of important eddies is separated and signified. Step 4 and step 5 are constructing the pixelized 2D gradients map (see Eq.~\ref{Eq.16} and Eq.~\ref{eq.17}). With the preprocessed data cube, we calculate the per-pixel gradients in each slice through the convolution with Sobel kernels. The per-pixels gradient slice then is processed by the per-pixel adaptive sub-block averaging method (see \S~\ref{subsec:asb}), which takes the Gaussian fitting peak value of the gradient distribution in a selected sub-block to statistically define the mean magnetic field in the corresponding sub-region. This step outputs a $N_x\times N_y\times N$ gradient cube. The final 2D gradient's orientation map results from the summation of the gradient cube in a way similar to Stokes parameters (see Eq.~\ref{eq.17}). Step 6 and step 7 are identifying the boundary of gravitational collapsing regions through the double-peak algorithm (see \S~\ref{subsec:dp}). Based on the theoretical considerations, the gradients change their orientation by $\pi/2$ in gravitational collapsing regions. Therefore, we can see in the diffuse region 1, the gradient's histogram is approximately a single Gaussian distribution with the peak value locates at $\psi_g^1$. In gravitational collapsing region 3, the peak value of the histogram becomes $\psi_g^2$. However, in the transitional region 2, the histogram appears bi-modality showing two peak values $\psi_g^1$ and $\psi_g^2$. The corresponding pixel is labeled as the boundary of collapsing regions once the difference between $\psi_g^1$ and $\psi_g^2$ is approximate $\pi/2$. Step 6 and step 7 can be achieved by the curvature algorithm (\S~\ref{subsec:cur}).}
	\end{figure*}
	\section{Methodology}
	\label{sec:method}
	\subsection{Velocity Gradient Technique}
	\label{subsec:vgt}
	The VGT initially employed either Velocity Centroid map \textbf{C(x,y)} \citep{2017ApJ...835...41G,YL17a} or Velocity Channel map \textbf{Ch(x,y)} \citep{LY18a} to extract velocity information in PPV cubes:
	\begin{equation}
	\begin{aligned}
	C(x,y)&=\frac{\int v\cdot\rho(x,y,v)dv}{\int \rho(x,y,v)dv}\\
	Ch(x,y)&=\int_{v_0-\Delta v/2}^{v_0+\Delta v/2}\rho(x,y,v) dv
	\end{aligned}
	\label{eq2}
	\end{equation}
	where $\rho$ is gas density, $v$ is the velocity component along the line of sight, $\Delta v$ is the velocity channel width satisfied with Eq.~\ref{eq1}, $v_0$ is the velocity corresponding to the peak position in PPV's velocity profile. Note the validity of these two methods have been tested in both numerical simulations \citep{YL17a, YL17b, 2019ApJ...873...16H,LY18a} and observational data \citep{velac,survey}. 
	
	To improve the performance of VGT, \citet{PCA} used the Principal Component Analysis (PCA) as a tool to extract the most crucial $n$ velocity components in a PPV cube. The implementation of PCA constructs a new orthogonal basis, which consists of $n$ eigenvectors and $n$ eigenvalues. The contribution from a principal component corresponding to small eigenvalues is also insignificant. By omitting small eigenvalues and their corresponding principal components in the PPV cube,  we can, therefore, remove the noise. Also, it is possible to enhance the contribution from crucial components by projecting the original dataset into the new orthogonal basis formed by the eigenvectors \citep{PCA,PGCA}. The PCA algorithm is implemented as follows. 
	
	As illustrated in Fig.~\ref{fig:vgt} \X1 \X2, assuming that a PPV cube $\rho(x,y,v)$ is a probability density function of three random variables $x,y,v$, we can obtain its covariance matrix and the eigenvalue equation for this covariance matrix from \citep{2002ApJ...566..276B,2002ApJ...566..289B,PCA,PGCA}:
	\begin{equation}
	\label{eq:13}
	\begin{aligned}
	S(v_i,v_j) \propto &\int dxdy \rho(x,y,v_i)\rho(x,y,v_j)\\
	&-\int dxdy \rho(x,y,v_i)\int dxdy\rho(x,y,v_j)
	\end{aligned}
	\end{equation}
	\begin{equation}
	\label{eq:14}
	\textbf{S}\cdot \textbf{u}=\lambda\textbf{u}
	\end{equation}
	where \textbf{S} is the co-variance matrix with matrix element $S(v_i,v_j)$, with $i,j=1,2,...,n_v$. $n_v$ is the number of channel in PPV cubes and $\lambda$ is the eigenvalues associated with the eigenvector $\textbf{u}$ \citep{PCA,PGCA}. By weighting channel $\rho(x,y,v_j)$ with the corresponding eigenvector element $u_{ij}$, we project the PPV cube into a new orthogonal basis, in which the corresponding eigen-map $I_i(x,y)$ is:
	\begin{equation}
	I_i(x,y)=\sum_j^{n_v}u_{ij}\cdot \rho(x,y,v_j)
	\end{equation}
	
	Repeating the procedure for each eigenvector $\textbf{u}_i$ resultants in a set of eigen-maps $I_i(x,y)$, with $i=1,2,...,n_v$ (see Fig.~\ref{fig:vgt}\X3). From an individual eigen-map $I_i(x,y)$, the gradient orientation at individual pixel $(x,y)$ is calculated by convolving the image with 3 $\times$ 3 Sobel kernels:
	$$
	G_x=\begin{pmatrix} 
	-1 & 0 & +1 \\
	-2 & 0 & +2 \\
	-1 & 0 & +1
	\end{pmatrix},
	G_y=\begin{pmatrix} 
	-1 & -2 & -1 \\
	0 & 0 & 0 \\
	+1 & +2 & +1
	\end{pmatrix}
	$$
	\begin{equation}
	\label{Eq.16}
	\begin{aligned}
	\bigtriangledown_x I_i(x,y)=G_x * I_i(x,y)  \\  
	\bigtriangledown_y I_i(x,y)=G_y * I_i(x,y)  \\
	\psi_{gi}(x,y)=tan^{-1}(\frac{\bigtriangledown_y I_i(x,y)}{\bigtriangledown_x I_i(x,y)})
	\end{aligned}
	\end{equation}
	where $\bigtriangledown_x I_i(x,y)$ and $\bigtriangledown_y I_i(x,y)$ are the x and y components of gradients respectively. $\psi_{gi}$ is the pixelized gradient map for each channel $I_i(x,y)$(see Fig.~\ref{fig:vgt}\X4).
	
	Note that in the picture of turbulence, a single gradient in each pixel of $\psi_{gi}$ contains little statistical information and does not indicate the magnetic field direction directly. However, the distributions of the gradient orientation appear as an accurate Gaussian profile with an appropriate size of sub-regions. \citet{YL17a}, therefore, proposed the sub-block averaging method, i.e., taking the Gaussian fitting peak value of the gradient distribution in a selected sub-block, to statistically define the mean magnetic field in the corresponding sub-region. Different from \citet{YL17a}, we continuously implement the sub-block averaging method. Each pixel of $\psi_{gi}$ is taken as the center of a sub-block and applied the recipe of sub-block averaging. The size of each sub-block is determined by the fitting errors within the 95\% confidence level. We vary the sub-block size and check its corresponding fitting errors. When the fitting error gets its minimum value, the corresponding sub-block size is the optimal selection. The resolution of the resulting gradient map is unchanged. We refer to this procedure as adaptive sub-block averaging (see Fig.~\ref{fig:vgt}\X4).
	
	We denote the gradients' orientation after the adaptive sub-block averaging implemented as $\psi_{gi}^s(x,y)$. In analogy to the Stokes parameters of polarization, the pseudo Q$_g$ and U$_g$ of gradient-induced magnetic fields are defined as:
	\begin{equation}
	\label{eq.17}
	\begin{aligned}
	&Q_g(x,y)=\sum_{i=1}^n I_i(x,y)\cos(2\psi_{gi}^s(x,y))\\
	&U_g(x,y)=\sum_{i=1}^n I_i(x,y)\sin(2\psi_{gi}^s(x,y))\\
	&\psi_g=\frac{1}{2}tan^{-1}(\frac{U_g}{Q_g})
	\end{aligned}
	\end{equation}
	The pseudo polarization angle $\psi_g$ is then defined correspondingly, which gives a probe of plane-of-the-sky magnetic field orientation after rotating 90$^\circ$. The rotation is automatically performed without specification (see Fig.~\ref{fig:vgt}\X5). Note that in constructing the $Q_g(x, y)$ and $U_g(x, y)$, one can project the data onto the subset of the dominant principal components but not onto all of them, i.e., $n < n_v$, especially when the smallest eigenvalues of the covariance matrix are dominated by noise \citep{PGCA}. The noise can be removed by eliminating the the smallest eigenvalues. Also, the pseudo-Stokes parameters is advantageous in correcting abnormal gradients vectors induced by either noise or fitting error. \citet{LY18a} showed that magnetic field lines should be continuous, but a abnormal gradient vector cuts off the streamline comparing with the neighboring vectors. To smooth the outlying abnormal gradients, \citet{LY18a} proposed the moving window method. In terms of the moving window method, a sub-block is selected to move along the orientation of the predicted magnetic field. When there is an abnormal gradient vector, a rotation is applied to the abnormal vector so that a smooth field line is formed. The moving window method is mathematically equivalent to the convolution of a Gaussian kernel with the cosine and sine angular components of the vectors. Nevertheless, \citet{IGs} points out the non-linear cosine or sine transformation distorts the Gaussian properties of the vectors. The Gaussian convolution may further induce abnormal gradient vectors. In the contrary, the pseudo-parameters recover the Gaussian distribution because of the weighting term $I_i(x,y)$ (see Fig.~10 in \citet{IGs}). Therefore, instead of smoothing directly the cosine and sine components, we propose the Gaussian convolution can be applied to $Q_g(x,y)$ and $U_g(x,y)$. In this work, as no noise is introduced to the numerical simulation, we do not distinguish the noise subset selecting $n = n_v$ and do not employ the smoothing procedure.

	The relative alignment between magnetic fields orientation and rotated pseudo polarization angle ($\psi_g+\pi/2$) is quantified by the \textbf{Alignment Measure} \footnote{The Alignment measure given by Eq. (\ref{AM_measure}) was introduced in analogy with the grain alignment research (see \citet{2007JQSRT.106..225L}) in \citet{2017ApJ...835...41G} and then was used in all the subsequent publications by our group and was borrowed by other groups, e.g. for doing the Histogram of Oriented Gradients \citep{2019A&A...622A.166S} and the Rolling Hough Transform \citep{2019ApJ...887..136C}.}(AM): 
	\begin{align}
	\label{AM_measure}
	AM=2(\langle cos^{2} \theta_{r}\rangle-\frac{1}{2})
	\end{align}
	where $\theta_r$ is the angular difference in individual pixels, while $\langle ...\rangle$ denotes the average within a region of interests. In the case of a perfect alignment of the magnetic field and VGT, we get AM = 1, i.e., the global rotated $\psi_g$ is parallel to the POS magnetic field, while AM=-1 indicates global rotated $\psi_g$ is perpendicular to the POS magnetic field. The standard error of the mean gives the uncertainty $\sigma_{AM}$; that is, the standard deviation divided by the square root of the sample size.
	
	\subsection{Double-peak features in the histogram of velocity gradients orientation}
	\label{subsec:dp}
\begin{figure*}
		\centering
		\includegraphics[width=.99\linewidth,height=0.55\linewidth]{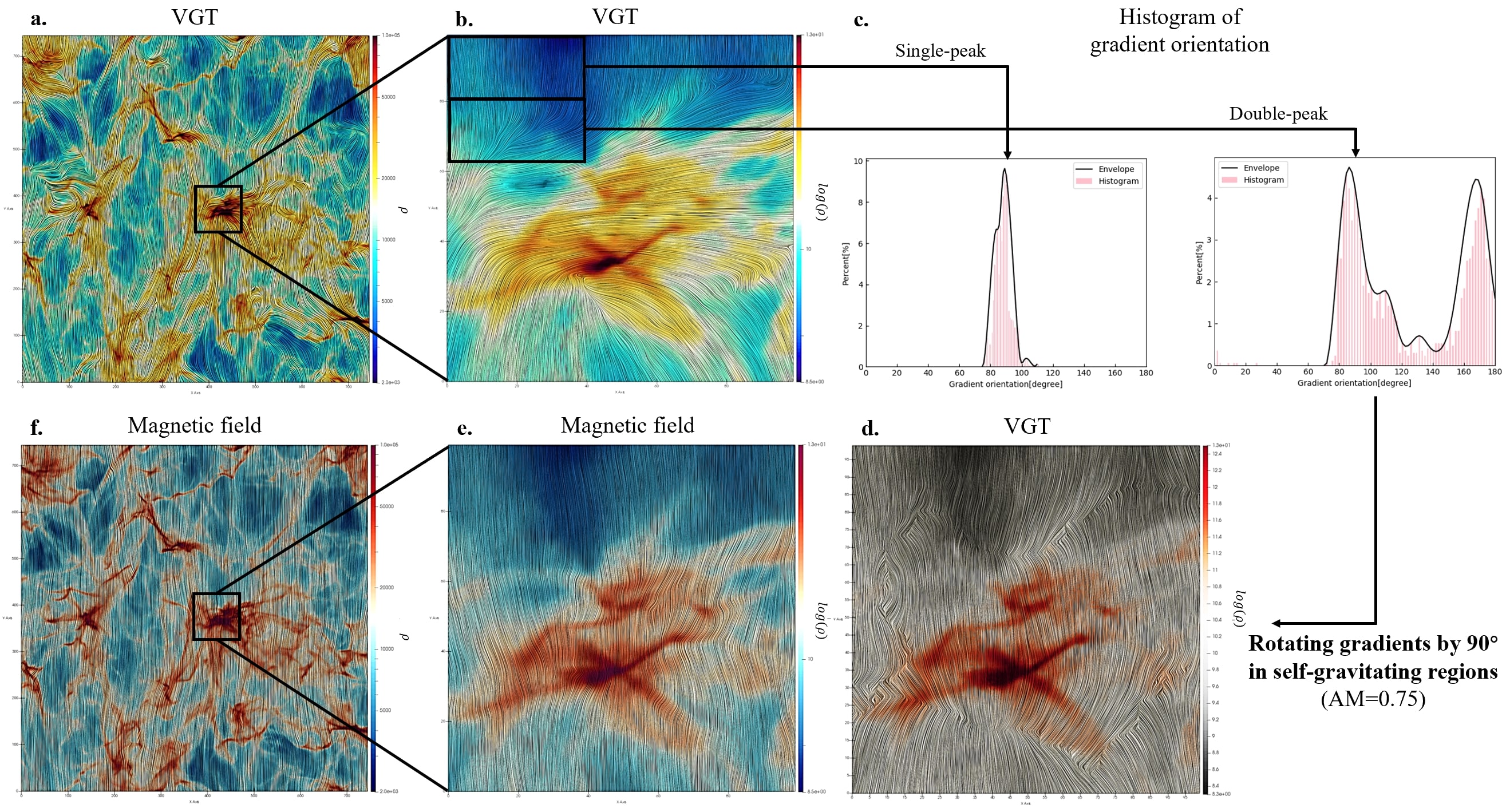}
		\caption{\label{fig:illu} An example of how velocity gradients change orientations at the gravitational collapsing region, using simulation A1 with $t_r\simeq0.8$ Myr (see Tab.~\ref{tab:sim}). \textbf{Panel a and b:} a global and a zoom-in magnetic field morphology inferred from VGT respectively. The magnetic field is superimposed on the projected intensity map and visualized using the Line Integral Convolution (LIC). The LIC is plotted using Paraview's default setting with steps 40 and step size 0.25. \textbf{Panel f and e:} a global and a zoom-in magnetic field morphology inferred from synthetic polarization respectively. \textbf{Panel c:} the histograms of velocity gradients orientation in the diffuse region which shows a single-peak Gaussian profile (left) and in the boundary of the gravitational collapsing region which shows a double-peak Gaussian profile (right). The same region has been shown to be gravitationally collapsing in Fig.~\ref{fig:pdf_vgt} through the N-PDFs. \textbf{Panel d:} the zoom-in magnetic field morphology inferred from VGT with a 90$^\circ$ re-rotation implemented for the gravitational collapsing region. The corresponding AM is 0.75. }
	\end{figure*}

	To identify the boundary of collapsing regions from gradients' orientation, we developed the "double-peak algorithm" the essence of which we describe below. Based on our theoretical consideration, velocity gradients change their orientation by $\pi/2$ in gravitational collapsing regions. This change can be extracted from the histogram of velocity gradients orientation: (i) The single peak of the histogram locates at $\theta$ in diffuse regions, (ii) The single peak of the histogram becomes $\theta+\pi/2$ in gravitational collapsing regions. (iii) In the transitional region, i.e., the boundary of collapsing regions, the histogram is therefore expected to show two peak values $\theta$ and $\theta+\pi/2$, denoted as a double-peak feature. As shown in Fig.~\ref{fig:illu}, the histogram of velocity gradients orientation in the diffuse region shows a single-peak Gaussian profile, while in the boundary of the gravitational collapsing region it shows a double-peak Gaussian profile. This distinct feature, therefore, gives information about gravitational collapsing regions. For magnetic field tracing, we perform a 90$^\circ$ re-rotation to the gradients in the collapsing region. The resultant magnetic fields agree with the one inferred from synthetic polarization, i.e., AM = 0.75. 
	
	To implement the double-peak algorithm, we define every single pixel of $\psi_g$ as a center of sub-block and draw the histogram of velocity gradients orientation within this sub-block (see Fig.~\ref{fig:vgt}\X6). Note the size of the sub-block can be different from the one for the sub-block averaging method (see \S~\ref{subsec:vgt}) and we denote it as the 2$^{nd}$ block. To suppress noise and the effect from insufficient bins' number, we plot the envelope for the histogram, which is a smooth curve outlining its extremes. The term whose histogram weight is less than the mean weight value of the envelope is masked. After masking, we work out the peak value of each consecutive profile. Once there exist more than one peak values, and the maximum difference of these peaks values is within the range $90^{\circ}\pm\sigma_\theta$ where $\sigma_\theta$ is the total standard deviation of each consecutive profile, the center of this sub-block is labeled as the boundary of a gravitational collapsing region (see Fig.~\ref{fig:vgt}\X7). 
	
	\subsection{The curvature of velocity gradients}
	\label{subsec:cur}
	In addition to the double-peak algorithm, the curvature of velocity gradients can also be used to identify the boundary of collapsing regions. For a gravitational collapsing region, velocity gradients rapidly change their direction by 90$^\circ$. In this case, the curvature of velocity gradients reaches its maximum value on the boundary of the collapsing region, but the ambient region usually has small curvature. By sorting out the curvature, one can, therefore, find the boundary of self-gravitating regions. Note that to have the best performance, the calculation of the velocity gradient should satisfy Eq.~\ref{eq1}. Otherwise, the contribution from density may be significant.
	
	To calculate the curvature, we follow the receipt used in \citet{2020arXiv200201926Y} by considering the magnetic field lines as the velocity field of an imaginary particle.  We firstly interpolate the pixelized gradient field $\psi_g$. Then a 2$^{nd}$ order Runge-Kutta (RK2) vector integrator is implemented to produce a sufficient stream-path that allows us to compute the curvature directly. 
	
	The method of RK2 vector integrator can be as simple as, given a vector field $\psi_g(x,y)$ and step size $dt$, where $(x,y)\in R$. For example, we can produce the next two coordinates $(x',y')$ and $(x'', y'')$ along the stream-path of gradients as:
	\begin{align}
	(x',y') &= (x,y) \pm \psi_g(x,y)*dt/2\\
	(x'',y'') &= (x,y) \pm \psi_g(x',y')*dt/2
	\end{align}
	\begin{table}
		\centering
		\label{tab:sim}
		\begin{tabular}{| c | c | c | c | c | c |}
			\hline
			Model & M$_s$ & M$_A$ & $\beta=2(\frac{M_A}{M_s})^2$ & Resolution & $t_r$ [Myr]\\ \hline \hline
			& 5.64 & 0.31 & 0.006 & $792^3$ & 0\\  
			& 6.03 & 0.31 & 0.005 & $792^3$ & 0.2\\ 
			A1 & 6.23 & 0.31 & 0.005 & $792^3$ & 0.4\\ 
			& 6.41 & 0.31 & 0.005 & $792^3$ & 0.6\\ 
			& 6.52 & 0.31 & 0.004 & $792^3$ & 0.8\\ \hline
			& 0.20 & 0.02 & 0.020 & $480^3$ & 0 \\
			& 0.20 & 0.02 & 0.020 & $480^3$ & 0.2\\ 
			& 0.21 & 0.02 & 0.018 & $480^3$ & 0.4\\ 
			& 0.25 & 0.02 & 0.013 & $480^3$ & 0.6\\ 
			& 0.29 & 0.03 & 0.021 & $480^3$ & 0.8\\ 
			& 0.35 & 0.03 & 0.015 & $480^3$ & 1.0\\ 
			& 0.42 & 0.04 & 0.018 & $480^3$ & 1.2\\ 
			& 0.52 & 0.04 & 0.012 & $480^3$ & 1.4\\ 
			A2 & 0.63 & 0.05 & 0.013 & $480^3$ & 1.6\\ 
			& 0.77 & 0.06 & 0.012 & $480^3$ & 1.8\\ 
			& 0.94 & 0.08 & 0.014 & $480^3$ & 2.0\\ 
			& 1.16 & 0.10 & 0.015 & $480^3$ & 2.2\\ 
			& 1.43 & 0.12 & 0.014 & $480^3$ & 2.4\\ 
			& 1.77 & 0.14 & 0.013 & $480^3$ & 2.6\\ 
			& 2.20 & 0.17 & 0.012 & $480^3$ & 2.8\\
			& 2.71 & 0.21 & 0.012 & $480^3$ & 3.0\\ 
			& 3.31 & 0.24 & 0.010 & $480^3$ & 3.2\\ 
			& 3.97 & 0.27 & 0.010 & $480^3$ & 3.4\\ \hline
		\end{tabular}
		\caption{Description of our MHD simulations. The self-gravitating module is switched on at the beginning, i.e., running time $t_r$ is 0. We a snapshot of the simulation every 0.2 Myr. M$_s$ and M$_A$ are the instantaneous values at each snapshot.}
	\end{table}
	where the plus sign means a forward integrator and minors sign is a backward integrator. We perform the integrator forward and backward for four steps respectively to establish a streamline $L(t) = \{(x_i,y_i,t)\}_{i \in \mathbf{N}}$ for each pixel, where $t$ is the parametrized variable defining the streamline. Then we can compute the unsigned curvature $\kappa(t)$ through the definition \citep{C1952}:
	\begin{equation}
	\begin{aligned}
	{\bf T}(t) &= \frac{\dot{L}(t)}{|\dot{L}(t)|}\\
	\kappa(t) &= |\dot{\bf T}(t)|\\
	\end{aligned}
	\end{equation}
	Since the computation of ${\bf T}(t)$ requires a normalization, we then will first obtain the normalized T.
	The derivative of ${\bf T}(t)$ is computed using the one dimensional five-point stencil:
	\begin{equation}
	\begin{small}
	\begin{aligned}
	\dot{T}_x(t) = \frac{-T_x(t+2dt)+8T_x(t+dt)-8T_x(t-dt)+T_x(t-2dt)}{12dt}\\
	\dot{T}_y(t) = \frac{-T_y(t+2dt)+8T_y(t+dt)-8T_y(t-dt)+T_y(t-2dt)}{12dt}
	\end{aligned}
	\end{small}
	\end{equation}
	
	By repeating this recipe for every pixel in the map, one can get a pixelized 2D curvature map of the gradient field $\psi_g(x,y)$. When there exists a 90$^\circ$ change of gradients, $\kappa(t)$ gets its maximum value. 
	
	
	\section{MHD simulation data}
	\label{Sec.data}
	
	We perform 3D MHD simulations through ZEUS-MP/3D code \citep{2006ApJS..165..188H}, which solves the ideal MHD equations in a periodic box. We use single fluid, operator-split, solenoidal turbulence injections, and staggered grid MHD Eulerian assumption. To emulate a part of an interstellar cloud, we use the barotropic equation of state, i.e., these clouds are isothermal with temperature T = 10.0 K, sound speed $c_s$ = 187 m/s and cloud size L = 10 pc. The sound crossing time $t_v = L/c_s$ is $\sim 52.0$ Myr, which is fixed owing to the isothermal equation of state. 
	
	MHD turbulence is characterized by sonic Mach number M$_{s}=v_{l}/c_{s}$ and Alfv\'{e}nic Mach numbers M$_{A}=v_{l}/v_{A}$, where $v_{l}$ is the injection velocity and $v_{A}$ is the Alfv\'{e}nic velocity. The turbulence is highly magnetized when the magnetic pressure of plasma is larger than the thermal pressure, i.e., M$_{A}<1$. The compressibility of turbulence is characterized by $\beta=2(\frac{M_A}{M_s})^2$. We refer to the simulations in Tab.~\ref{tab:sim} by their model name. For instance, our figures will have the model name indicating which data cube was used to plot the figure. The total mass $M_{tot}$ in the simulated cubes A1 is $M_{tot} \sim 18430.785$ M$_\odot$, as well as the magnetic Jean mass $M_{JB}\sim 286.95$ M$_\odot$, average magnetic field strength $B\sim30.68\mu G$, mass-to-flux ratio $\Phi\sim1.11$, and volume density $\rho\sim318.32$ cm$^{-3}$. As for A2, we have $M_{tot} \sim32765.74$ $M_\odot$, $M_{JB}\sim 96.68$ $M_\odot$, $B\sim 31.33\mu G$, $\Phi\sim1.92$, and  volume density $\rho\sim565.90$ cm$^{-3}$. Based on the initial setting parameters, the intrinsic free fall time $t_{ff}$ is $\sim1.88Myr$ and $1.40 Myr$ for simulation A1 and A2 respectively.
	
	The self-gravitating module, which employs a periodic Fast Fourier Transform Poisson solver, is switched on after turbulence inside the cube gets saturated  and the simulation has run for at least two sound crossing time. The scale of the simulation is with 10 pc represented by 792 and 480 pixels for A1 and A2 respectively. The thermal Jeans length, therefore, occupies 94 pixels (1.18 pc for A1) or 45 pixels (0.89 pc for A2). We keep driving both turbulence and self-gravity until the simulation violates the Truelove criterion \citep{1997ApJ...489L.179T}, which requires the self-gravitating core should at least occupy 4 pixels. The density, therefore, can be enhanced by a factor of $\sim$ 552 (A1) or $\sim$ 127 (A2) times, i.e., we can have maximum volume density $\rho_{max}\sim$ 175712.63 cm$^{-3}$ for A1, while $\rho_{max}\sim$ 71869.30 cm$^{-3}$ for A2, which informs us when to stop the simulation before having numerical artifacts due to self-gravitating collapse.
	
	\section{Results}
	\label{sec:results}
	\subsection{Evolution of the probability density functions}
	\begin{figure}
		\subfigure[]{
			\centering
			\includegraphics[width=.95\linewidth,height=0.77\linewidth]{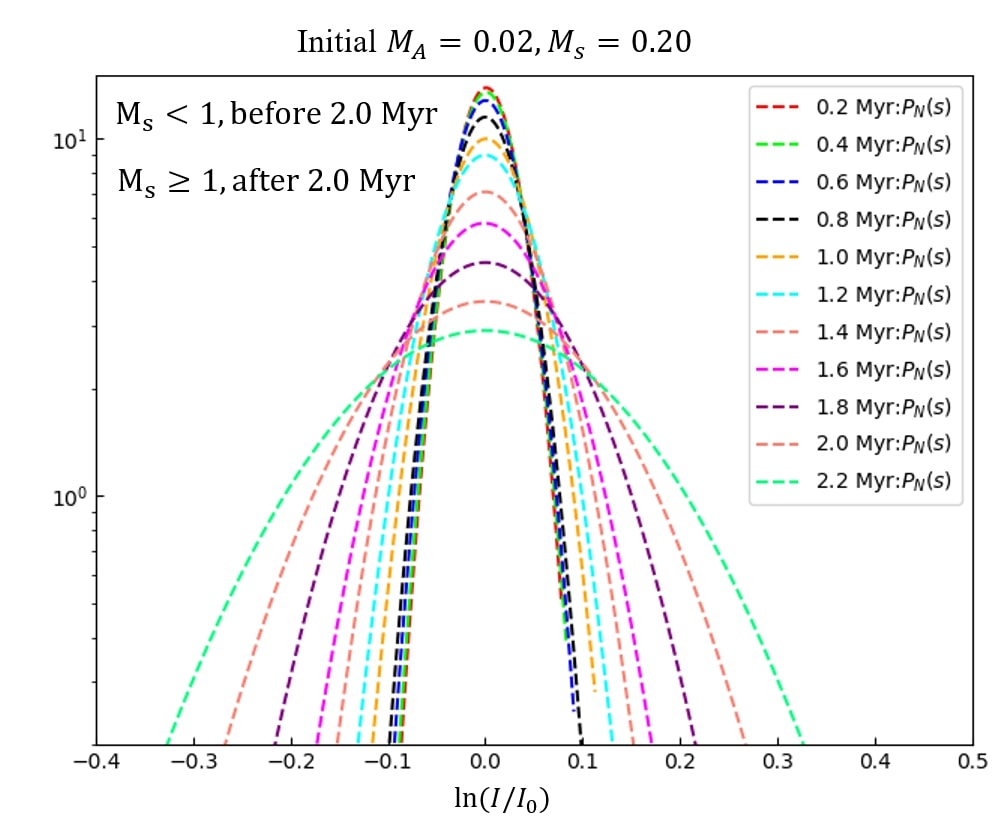}
		}
		\subfigure[]{
			\centering
			\includegraphics[width=.95\linewidth,height=0.77\linewidth]{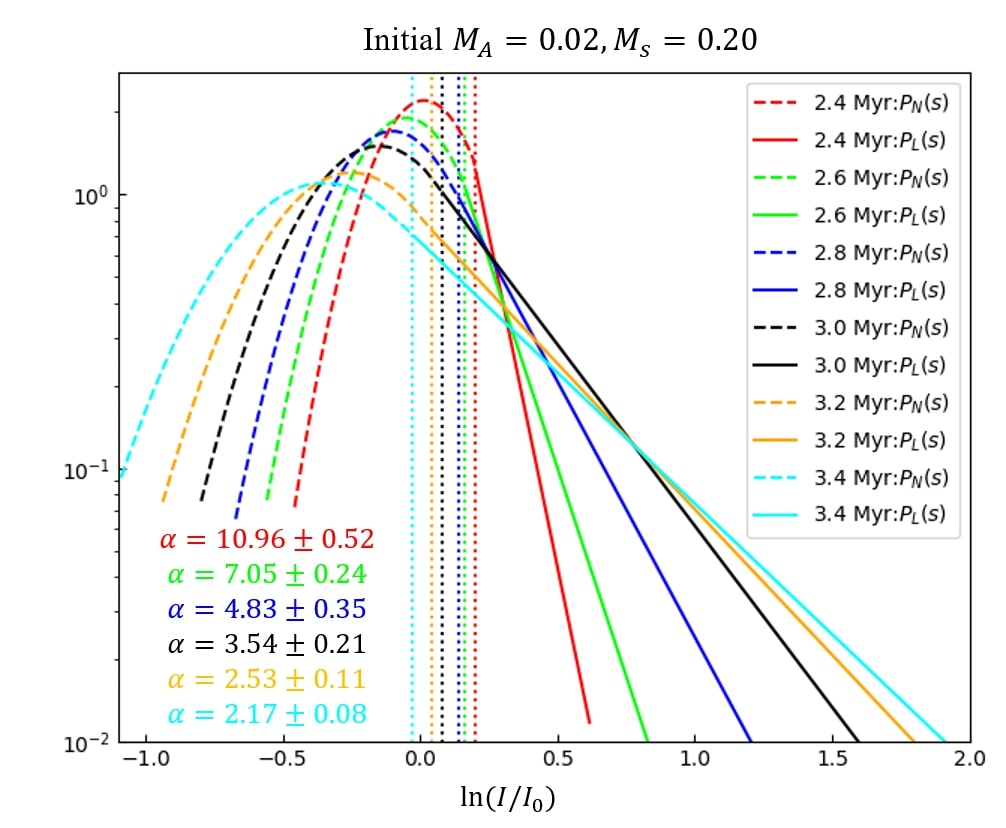}
		}
		\subfigure[]{
			\centering
			\includegraphics[width=.95\linewidth,height=0.77\linewidth]{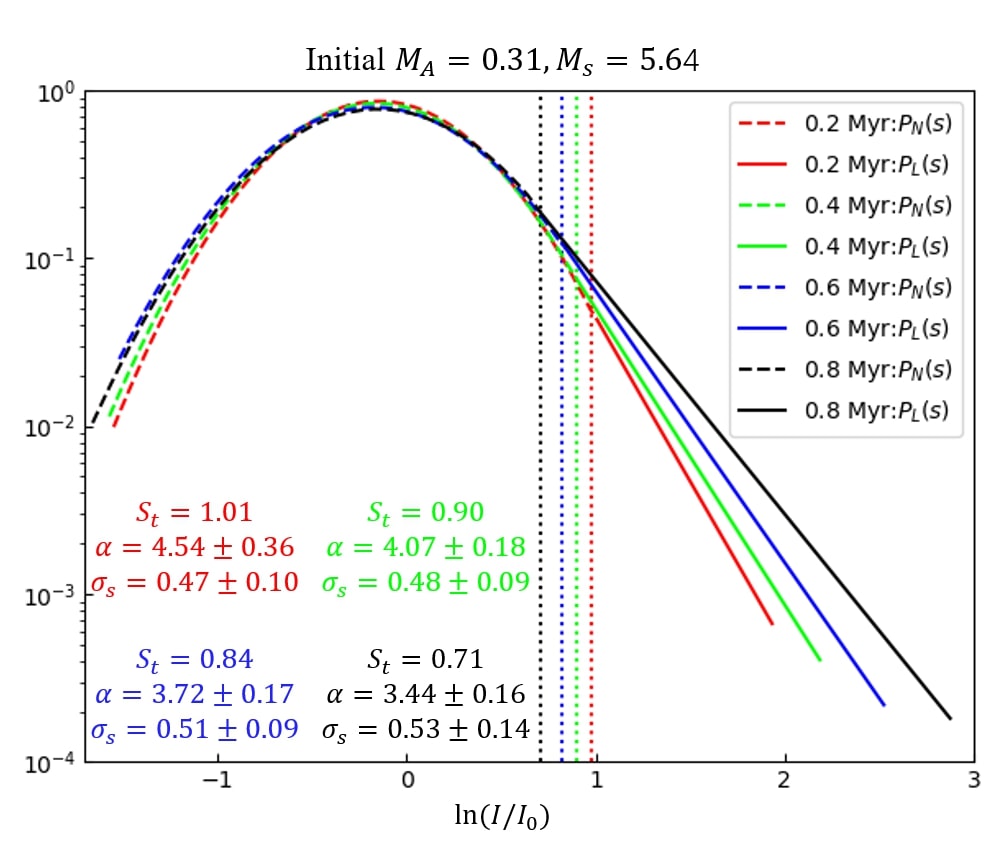}
		}
		\caption{\label{fig:pdf} The log-normal plus power-law models of normalized PDF with bin-size 100. The dotted line outlines all the density past the transition density which is the dense self-gravitating gas. $\alpha$ is the slope of the power-law part, $I_0$ is the mean intensity value, S$_t$ denotes the transition density, and $\sigma_s$ represents the standard deviation of the log-normal part.}
	\end{figure}
	To exam the behavior of PDF at different collapsing stages, we apply it to eighteen snapshots of A2 for each interval of the evolution time $\Delta t_r=0.20 Myr$, after the gravity is turned on. In Fig.~\ref{fig:pdf}, we plot the density PDF for the simulation A2. The N-PDFs are fitted within a $95\%$ confidential level from initial data analysis. Fig.~\ref{fig:pdf} (a) shows that the N-PDFs are in log-normal format until $t_r=2.2Myr$, while the width of the N-PDFs is increasing. The prior studies \citep{2005ApJ...630..250K,2009ApJ...693..250B} explained that for isothermal turbulence the width of log-normal N-PDFs is given by M$_s$ and turbulence driving parameter b:
	\begin{equation}
	\sigma_s^2=ln(1+b^2M_s^2)
	\end{equation}
	
	Theoretically, $\sigma_s$ should include the contribution from the magnetic field. However, $\sigma_s$ is an auxiliary way for our paper which is dealing with gradients. We do not provide a dedicated study of this quantity in this paper. In any case, this equation is widely used in many MHD studies \citep{2011MNRAS.416.1436B,2011ApJ...727L..21P,1995ApJ...441..702V,2008ApJ...680.1083R,2012ApJ...750...13C,2018ApJ...863..118B}, so we keep it has the original form. In this situation, we expect A2 tends to be super-sonic with the increment of self-gravity, and we do see the M$_s\ge1$ after $t_r>2.0Myr$. Note that although the fitted N-PDFs in Fig.~\ref{fig:pdf} (a) is log-normal, fewer parts of un-fitted N-PDFs are similar to a power-law format. One explanation is that the small volume gravitational center becomes self-gravitating earlier. 
	
	Fig.~\ref{fig:pdf} (b) shows that the N-PDFs of A2 with $t_r\ge2.4Myr$. We see the N-PDFs become the combination of a log-normal component for low-density gas and a power-law component for high-density gas. The increasing $\sigma_s$ also indicates the M$_s$ of A2 simulation is continuously getting larger. The gas within the power-law part is expected to be self-gravitating. The shallower slope $\alpha$ and the smaller value of transitional density S$_t$ reveal that the volume of self-gravitating gas is increasing \citep{2012ApJ...750...13C,2018ApJ...863..118B}.
	
	However, the super-sonic simulation A1 behaves differently in the presence of self-gravity. The power-law part appears at $t_r\le0.2$ Myr. The transition from an overall log-normal N-PDF to a hybrid (i.e., the combination of a log-normal part and a power-law part) PDF disappears for A1. The slope $\alpha$ gets shallower, and the value of transitional density S$_t$ becomes smaller with the evolution. The width of the log-normal part is increasing insignificantly. We then expect the volume of gravitational collapsing gas is small so that it gives a little contribution to the overall supersonic gas.
	
	\subsection{Adaptive sub-block averaging}
	\label{subsec:asb}
	\begin{figure*}
		\centering
		\includegraphics[width=.99\linewidth,height=0.85\linewidth]{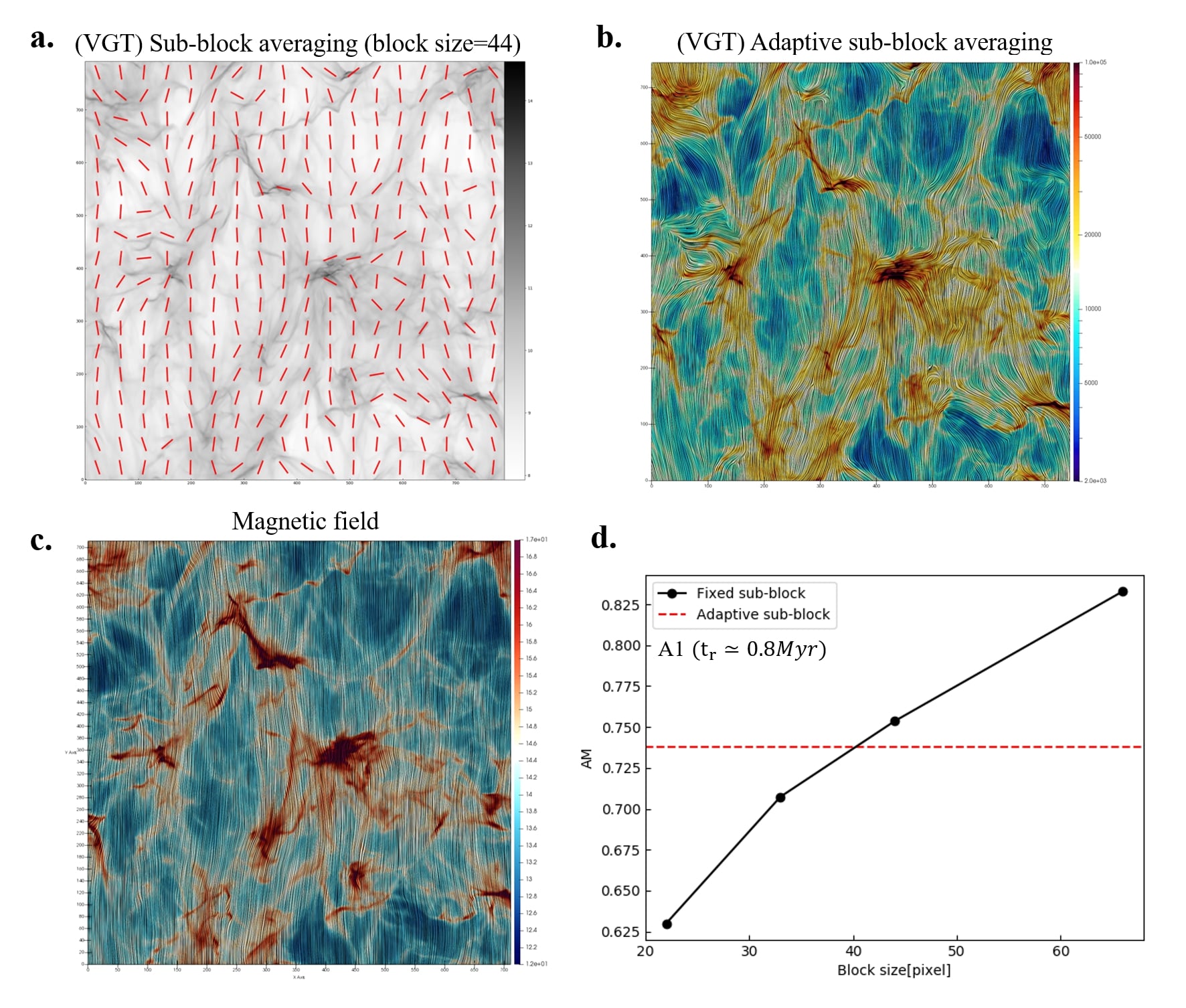}
		\caption{\label{fig:asb} The comparison of adaptive sub-block averaging and old sub-block averaging (fixed sub-block size) in terms of the alignment of rotated gradients and magnetic fields. We use the super-sonic simulation A1 at $t_{r}\simeq0.8Myr$. \textbf{Panel a:} the magnetic field (red segments) predicted by VGT with fixed sub-block size 44 pixels. \textbf{Panel b:} the magnetic field (the streamlines) predicted by VGT using the adaptive sub-block averaging. \textbf{Panel c:} the magnetic field morphology inferred from synthetic polarization. \textbf{Panel d:} the AM of VGT and actual magnetic field in in terms of various fixed block size. }
	\end{figure*}
	
	In order to identify the gravitational collapsing region, a high-resolution gradient map produced by VGT is indispensable. In \cite{YL17a}, the velocity gradient calculation requires the use of a {\it sub-block} that would eventually degrade the input observable map in order to have a statistically viable magnetic field prediction. The size of such block not only has to be large enough to obey the statistical requirement (which is the Gaussian fitting requirement) as listed in \cite{YL17a} \& \cite{LY18a}, but has to be not too large so that most of the information is retained. For our particular purpose in this paper, we are more concerned with the signature of gravitational collapse in the observables.  As we know from previous works \citep{YL17b,LY18a,IGs} that a large fixed block size would provide low spatial resolution and may miss the small gravitational collapsing region, at which the gradients flip their direction. It is inevitable that we need a block averaging algorithm that can provide a high-resolution magnetic field output yet not violating the statistical requirement as posted in \cite{YL17a}. 
	
	Therefore in the current work, we introduce an algorithm called {\bf adaptive sub-block averaging.} Pictorially for a selected pixel on the sky, a rectangular box is selected with the selected pixel to be at the center. By increasing the size of the sub-block, we can obtain a better fit in terms of the squared error of the Gaussian statistics requirement as listed in \cite{YL17a} under 95\% confidence, we shall pick that as the "optimal block size" for that pixel. This fitting procedure will ensure that (1) each pixel will have a block size for later statistical purposes (2) the output magnetic field has the comparable resolution as the input observable map. Although this procedure is similar to the convolution with a kernel, it takes the peak value of the fitted Gaussian curve rather than adds each element of the image to its local neighbors, weighted by the kernel. Without specification, we adopt the adaptive sub-block averaging method in what follows. As a remark, the Gaussian fitting requirement in \cite{YL17a} can be replaced by a more theoretically driven cosine fitting requirement as suggested by \citet{2019arXiv191002226L}.
	
	In Fig.~\ref{fig:asb}, we compare the adaptive sub-block averaging method with the old version, i.e., fixed sub-block size, using simulation A1 at $t_r\simeq0.8Myr$. We can clearly see the adaptive sub-block averaging method outputs a  magnetic field map with higher resolution than the fixed sub-block (size = 44 pixels). We calculate the AM between the rotated velocity gradients and magnetic fields. For the fixed sub-block averaging, we vary the block size from 22 to 33, 44, and 66. We can see the AM is increasing from 0.625 to 0.825 with the increment of block size. If we implement the adaptive sub-block averaging method, the AM is approximately 0.73, which is similar to the case that the fixed block size is 40 pixels. Note that the adaptive block method does not trace per-pixel magnetic field, but the mean magnetic field within approximately the effective block. The effective block size gives the global mean resolution of the magnetic field resolved by VGT. Importantly, the adaptive sub-block averaging method guarantees the gradient map sufficiently high resolution for the implementation of the double-peak algorithm, as we will discuss in \S~\ref{subsec:double-peak}.
	
	\subsection{The change of velocity gradients' orientation}
	\label{subsec:v-change}
	\begin{figure*}
		\centering
		\includegraphics[width=.99\linewidth,height=0.8\linewidth]{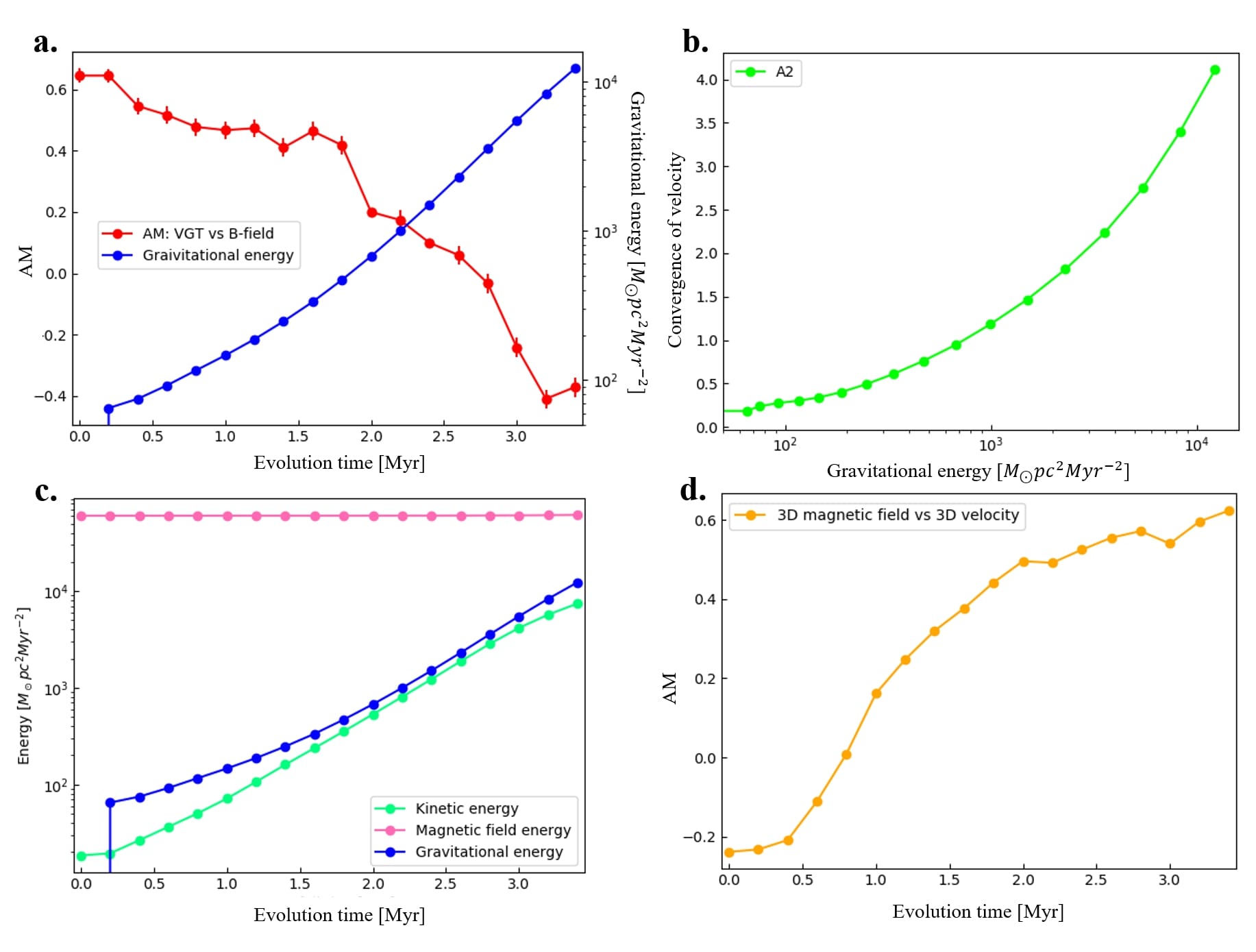}
		\caption{\label{fig:a1conv} \textbf{(a)}: the AM (red line) between magnetic fields and velocity gradients concerning the time since self-gravity is turned on (0 Myr), using the sub-sonic A2 simulation set. The blue line indicates the total gravitational energy in each cube. \textbf{(b)}: the correlation of total gravitational energy and the mean convergence of the 3D velocity field, using sub-sonic A2 simulations. \textbf{(c)}: the variation of total kinetic energy, magnetic field energy, and gravitational energy for A2 simulations. \textbf{(d)}: the AM of the 3D magnetic field angle and 3D velocity angle for A2 simulations.}
	\end{figure*}
	
	\begin{figure*}
		\centering
		\includegraphics[width=1.0\linewidth,height=0.53\linewidth]{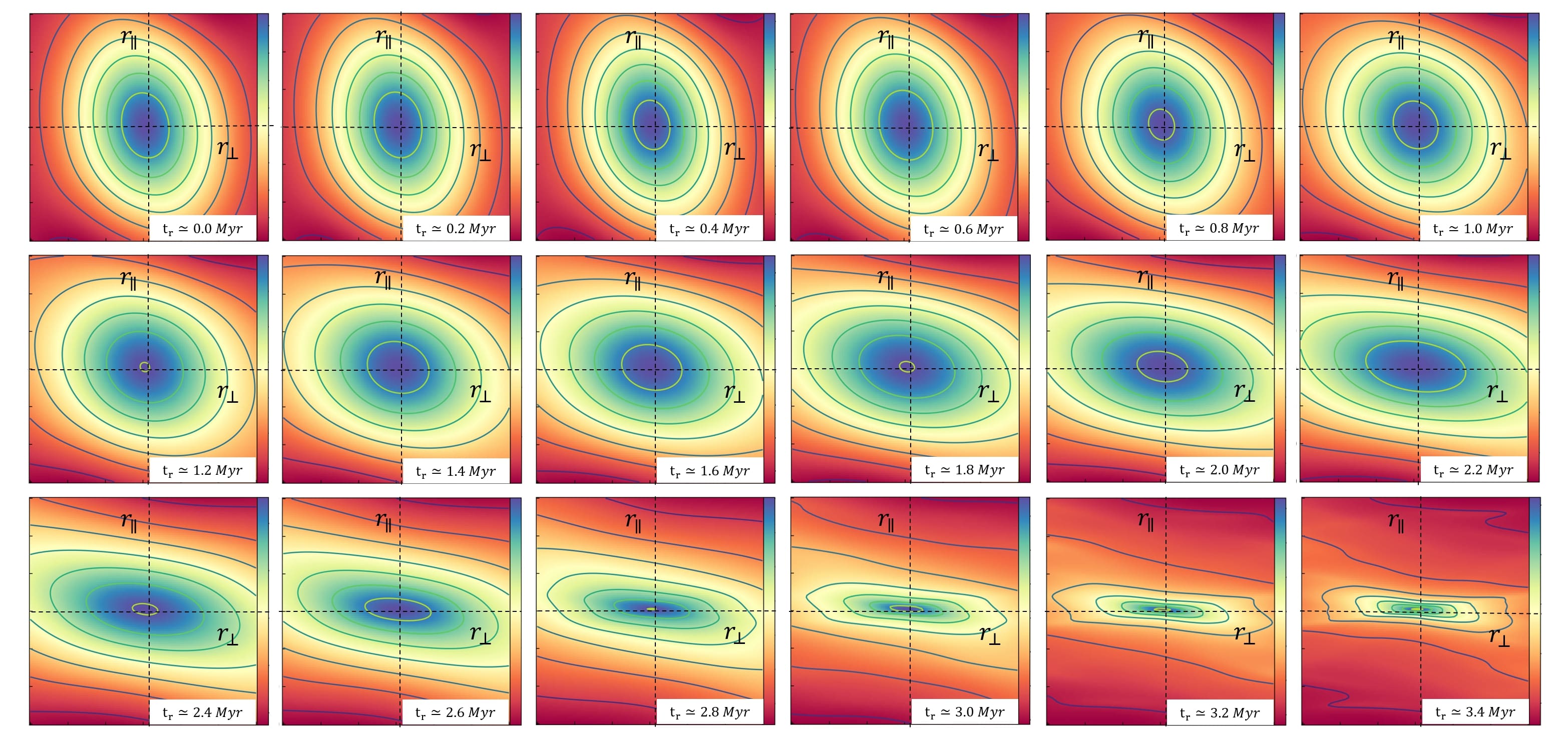}
		\caption{\label{fig:cf} The correlation function of 2D intensity maps at different snapshots, using simulation set A2. $r_\bot$ and $r_\parallel$ are the real space scales perpendicular and parallel to the magnetic field respectively. For all plots, $r_\bot$ and $r_\parallel$ are in scales less than 60 pixels.}
	\end{figure*}
	
	\begin{figure}
		\centering
		\includegraphics[width=.99\linewidth,height=0.80\linewidth]{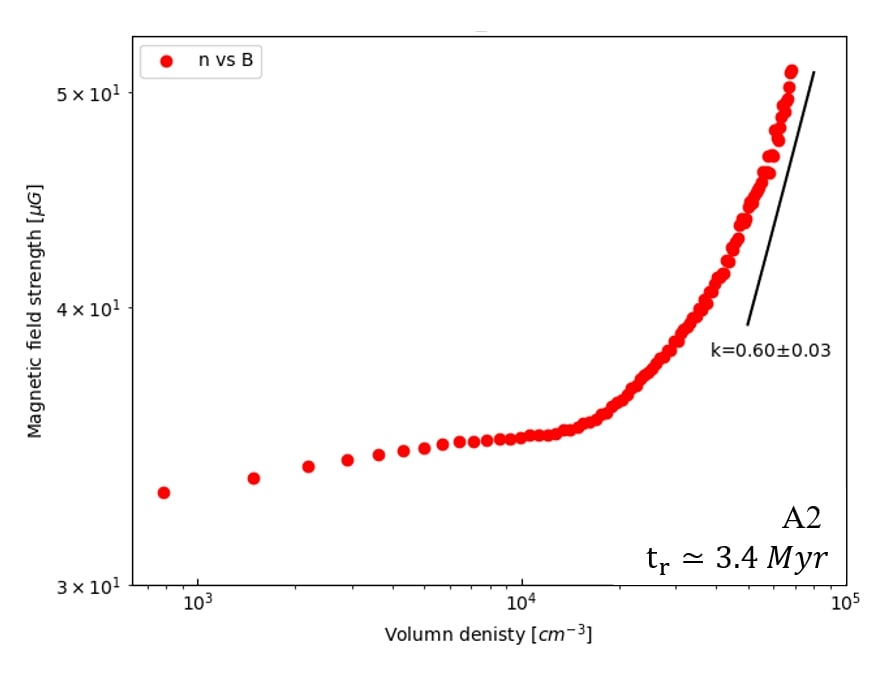}
		\caption{\label{fig:a1BN} The correlation of averaged volume density $n$ in uniformly spaced bins and 3D magnetic field strength $B$ using A2 simulation at $t_r\simeq 3.4 Myr$. $k$ is the slope of fitted line.}
	\end{figure}

	\begin{figure}
		\subfigure[]{
			\centering
			\includegraphics[width=.99\linewidth,height=0.78\linewidth]{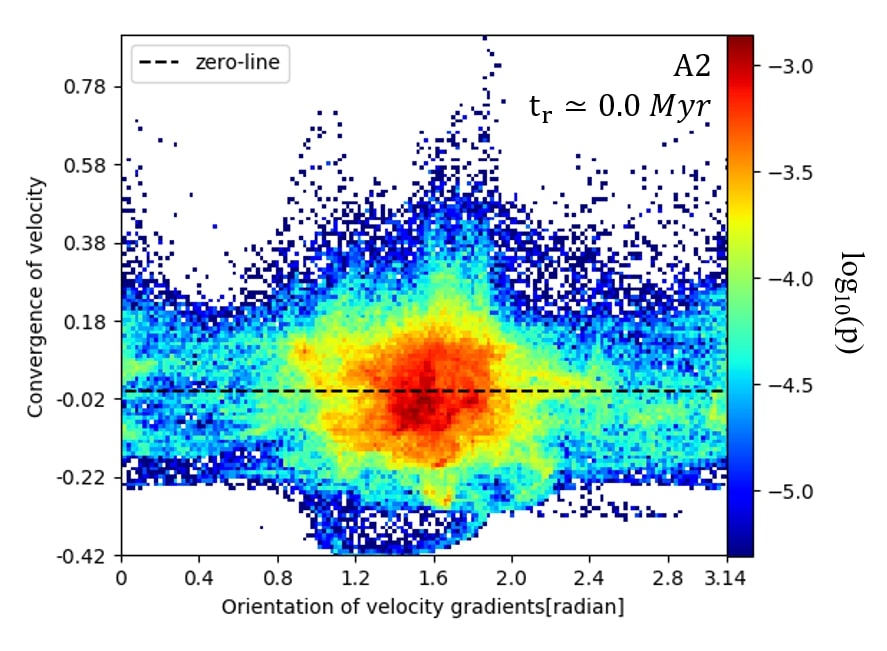}
		}
		\subfigure[]{
			\centering
			\includegraphics[width=.99\linewidth,height=0.78\linewidth]{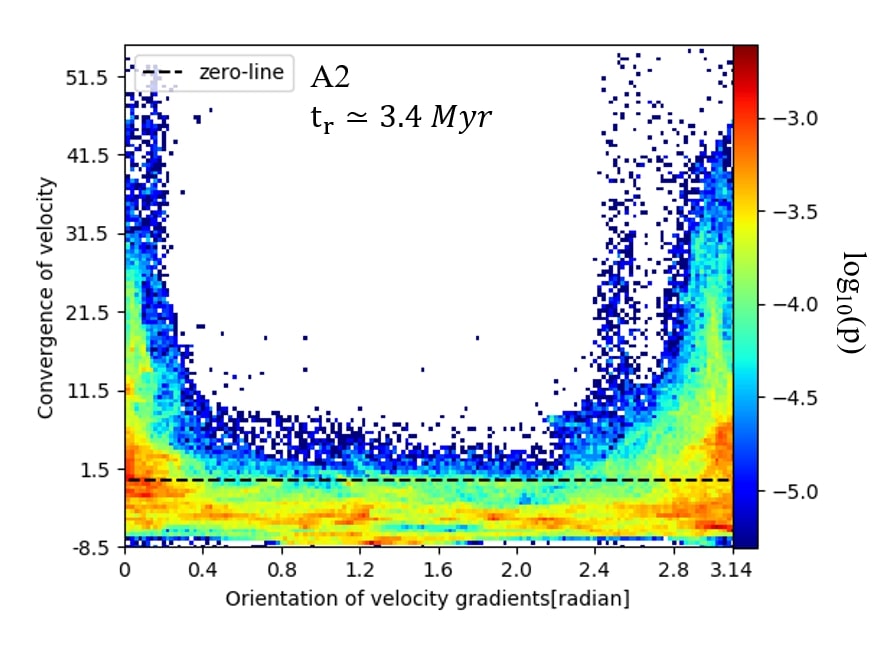}
		}
		\caption{\label{fig:a1amp} \textbf{(a)}: the 2D histogram of velocity gradients' orientation and velocity's convergence using A2 simulation at $t_r\simeq 0.0 Myr$. p gives the volume fraction of each data point. \textbf{(b)}: the 2D histogram of velocity gradients' orientation and projected velocity's convergence using A2 simulation at $t_r\simeq 3.4 Myr$. p gives the volume fraction of each data point and bin-size is 200 for the 2D histograms. Note that the orientation is measure in typical cartesian coordinate, i.e., with respect to the right horizontal direction.}
	\end{figure}

	\begin{figure}[t]
		\centering
		\includegraphics[width=.99\linewidth,height=1.0\linewidth]{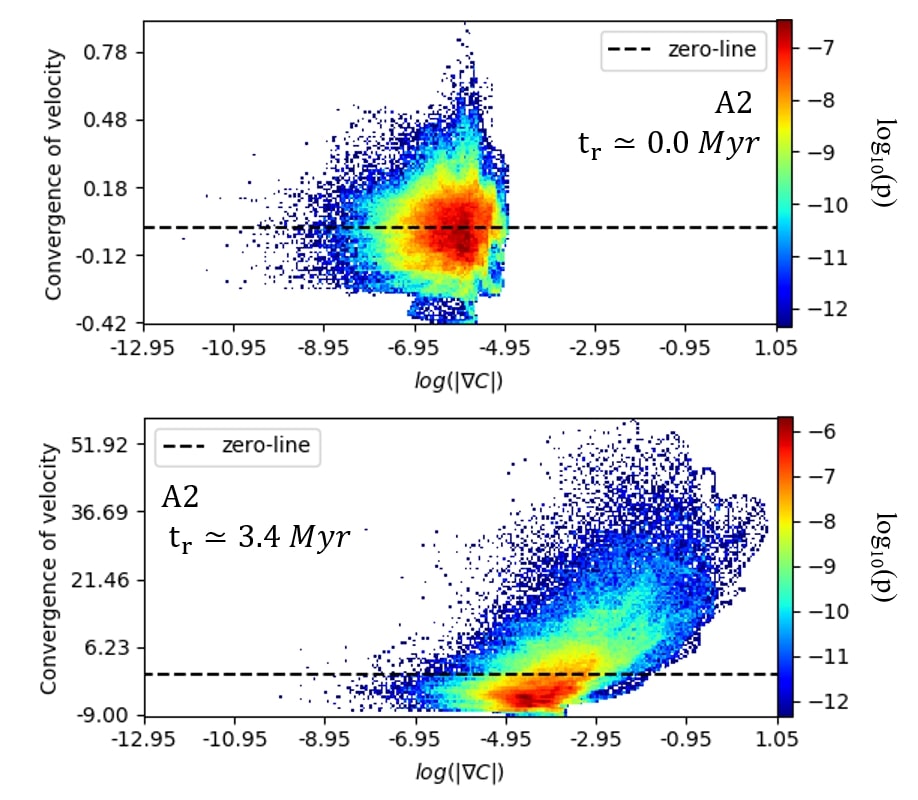}
		\caption{\label{fig:gradc_conv} The 2D histogram of velocity gradients' amplitude, i.e., $log(|\nabla C|)$, and velocity's convergence using A2 simulation at $t_r\simeq 0.0 Myr$ \textbf{(top)} and $t_r\simeq 3.4 Myr$ \textbf{(bottom)}. p gives the volume fraction of each data point. Bin-size is 200 for the 2D histograms.}
	\end{figure}
	
	Velocity gradients are expected to change their relative orientation with respect to magnetic fields in the presence of gravitational collapsing. To study the significance of gravity in the properties of gradients, we further explore the evolution of gradients concerning different stage gravitational collapsing by taking eighteen snapshots for simulation A2. We calculate the AM of rotated gradients and magnetic fields. The results are shown our in Fig.~\ref{fig:a1conv} (a). At the time $t_r = 0$, the gravity is introduced into the simulation that already stirred with magnetized turbulence and we take the snapshot until $t_r \simeq 3.4 Myr$. At $t_r \simeq 2.4 Myr$, the AM falls below zero. After this, the negative AM indicates that most of the rotated gradient vectors tend to be orthogonal to the magnetic field direction. To confirm this change is induced by gravitational collapsing, an estimation of the collapsing material is necessary. However, a cloud may not be collapsing because of the support of the magnetic field and turbulence, even the gravitation energy is large. A more reliable measurement of collapsing gas is the convergence of the 3D velocity field, since when there exists gravitational collapse, flows of matter are converging into the collapsing center. As a result, the convergence of velocity is expected to increase significantly. We then calculate the convergence $\mathfrak{C}$, which is defined as the negative divergence of velocity: 
	\begin{equation}
	\mathfrak{C}=-\nabla\cdot \Vec{v}=-(\frac{\partial v_x}{\partial x}+\frac{\partial v_y}{\partial y}+\frac{\partial v_z}{\partial z})
	\end{equation}
	where $v_x$, $v_y$, $v_z$ are the x, y, z components of the 3D velocity field respectively. We project it along line-of-sight into a 2D map. Since the projection sums up all divergence and convergence, a positive 2D convergence is expected to indicate the dominance of convergent flow in 3D. In Fig.~\ref{fig:a1conv} (b), we plot the correlation of the mean projected convergence and gravitational energy. We can see that convergence is positively proportional to the increment of gravitational energy. The supersonic convergent fluid is, therefore, induced by self-gravity. Also, In Fig.~\ref{fig:a1conv} (c), we show the variation of magnetic field energy, kinetic energy, and gravitational energy $|E_g|$. We find kinetic energy $E_k$ and gravitational energy are both increased since the self-gravity is accelerating the convergent flow. Approximately, the energy relation satisfies $|E_g|\simeq2E_k$. However, the magnetic energy is approximately keeping constant, which is the resultant of turbulence reconnection. The reconnection diffusion tends to make the magnetic distribution uniform and not correlated with the density enhancement \citep{2020arXiv200100868L}. 
	
	According to our theoretical consideration in \S~\ref{sec:theory}, the convergent flow shall follow the magnetic field line. In Fig.~\ref{fig:a1conv} (d), we calculate the AM of magnetic field angle and velocity angle in 3D Position-Position-Position (PPP) space. We see the AM is increasing from -0.2 to +0.6 with the increment of free fall time, i.e., gravitation energy. It implies that the velocity field is becoming parallel to the magnetic field. As the convergent fluid is following the same direction as velocity, the convergent flow is also parallel to the magnetic field in the case of gravitational collapsing. It agrees with our theoretical consideration. Also, at the full range of evolution time, the magnetic field energy is always more significant than others.
	
	In addition, we apply the second-order correlation function to the projected intensity map $I(x,y)$ at different snapshots, which is defined as:
	\begin{equation}
	CF(\Vec{R})=\langle I(\Vec{r})I(\Vec{r}+\Vec{R})\rangle
	\end{equation}
	here $\Vec{r}=(x,y)$ and $\Vec{R}$ is a lag vector. Fig.~\ref{fig:cf} shows how the correlation function should behave in terms of the contour plot. We can observe that when $t_r\le$ 1.0 Myr, the contours are elongated along the $r_\parallel$ direction, i.e., the mean magnetic field direction. As explained in \citet{2018ApJ...865...54Y}, the large contours are slightly misaligned from the $r_\parallel$ direction because of the limited inertial range in numerical simulations. In terms of correlation and structure functions, only the small scale contributions are considered to be meaningful. Importantly, with the increment of evolution time, the contours change its orientation becoming elongated along the $r_\bot$ direction, i.e., perpendicular to the mean magnetic field. We also observed the elongation along $r_\bot$ becomes more significant after $t_r\ge2.6$ Myr, i.e., the ratio of $\frac{r_\parallel}{r_\bot}$ get decreasing. According to our theoretical consideration, in the case of a strong magnetic field, the convergent flow can only have the motion along the magnetic field line. Therefore, initially, the turbulent eddies are anisotropic in the direction parallel the magnetic field, as described in \citet{LV99}. After the self-gravity starts controlling the cloud, the collapsing material is converging along the magnetic field direction. The material in the gravitational center, hence, is accumulated as filamentary structures perpendicular to the magnetic field,  i.e., elongating perpendicular to the magnetic field.
	
	\cite{2010ApJ...725..466C} showed that in molecular gas there is a power-law relation between the value of LOS component of magnetic field strength $B_{los}$ inferred from Zeenman splitting and the volume density $n$ as $B_{los}\propto n^{0.65}$. Later \cite{2015MNRAS.452.2500L} numerically demonstrated this power-law also holds for the total magnetic field strength $B_{tot}\propto n^{0.62 \pm 0.11}$. Similarly for simulation A2 at $t_r\simeq 3.4 Myr$, we bin the volume density in uniformly spaced bins with bin size 100 and take the averaged total magnetic field strength within each corresponding bin. Note different from Fig.~\ref{fig:a1conv}, the magnetic field used here is the local value instead of the global summation. Due to the finite rate of reconnection, the local magnetic field will be bent, and its strength can increase in the vicinity of 
	the gravitational collapse. In Fig.~\ref{fig:a1BN}, we plot the correlation between the volume density and magnetic field strength. We can see the density tends to increase with the increment of the magnetic field. At high density range $n\ge5\times10^4$, there is a power-law relation $B_{tot}\propto n^{0.60 \pm 0.03}$.  Summing up the results from Figs.~\ref{fig:a1conv},\ref{fig:cf},\ref{fig:a1BN}, we can then conclude that star formation can be efficient in strongly magnetized and fully ionized media.

	As above, the convergence of velocity is a proper probe of gravitational collapse.  We can then study the behavior of velocity gradients in collapsing regions using the convergence $\mathfrak{C}$. We plot the 2D histogram of velocity gradients' orientation and the projected velocity's convergence using A2 simulation in Fig.~\ref{fig:a1amp} (a) and (b). At the snapshot $t_r\simeq 0.0 Myr$, we see that the velocity gradients' orientation is concentrated at the surrounding of $\pi/2$. However, at $t_r = 3.4 Myr$, we see a extreme increment of the convergence. More importantly, the velocity gradients flip their direction by $\pi/2$, becoming 0 or $\pi$. The whole fraction of the gradients in the range of [0: 0.4] and [2.74: 3.14] is ~63.5\% comparing with the total volume (summation of all gradient). For some gradients, the change is less than 90 degrees. The reason is presented in Sec.~\ref{appendix:A}, there should be enough self-gravitating gases along LOS to see this change. It, therefore, confirms that velocity gradients change orientation by 90$^\circ$ in the presence of gravitational collapse. 
	
	The change of velocity gradients' orientation provides the possibility to identify the gravitational collapsing regions without the assistance of polarimetry. For example, as shown in Fig.~\ref{fig:illu}, the histogram of velocity gradients' orientation on the boundary of the collapsing region appears a double-peak feature. The corresponding angular difference between the two peaks is 90$^\circ$. This double-peak feature can, therefore, be used to distinguish diffuse and gravitational collapsing regions (see \S~\ref{subsec:double-peak}). The curvature of velocity gradients will also get its maximum value on the boundary of the collapsing region since the gradients rapidly flip their direction by 90$^\circ$ (see \S~\ref{subsec:curvature}). There are also several ways to detect this change. For instance, one can take calculate the gradient of the resultant velocity gradients' map again, i.e., $\nabla(\nabla v(x,y))$. The amplitude $|\nabla(\nabla v(x,y))|$ will arrive maximum on the boundary of the collapsing region. Also, we can perform the extrapolation along the direction of velocity gradients in diffuse regions. Since the rotated velocity gradients in the diffuse region are following magnetic field lines, the extrapolated gradients in the collapsing region do not flip their direction. When the difference between the extrapolated gradients and the actual gradient is 90$^\circ$,  we can locate the collapsing region.

	\begin{figure}[t]
		\centering
		\includegraphics[width=.99\linewidth,height=1.8\linewidth]{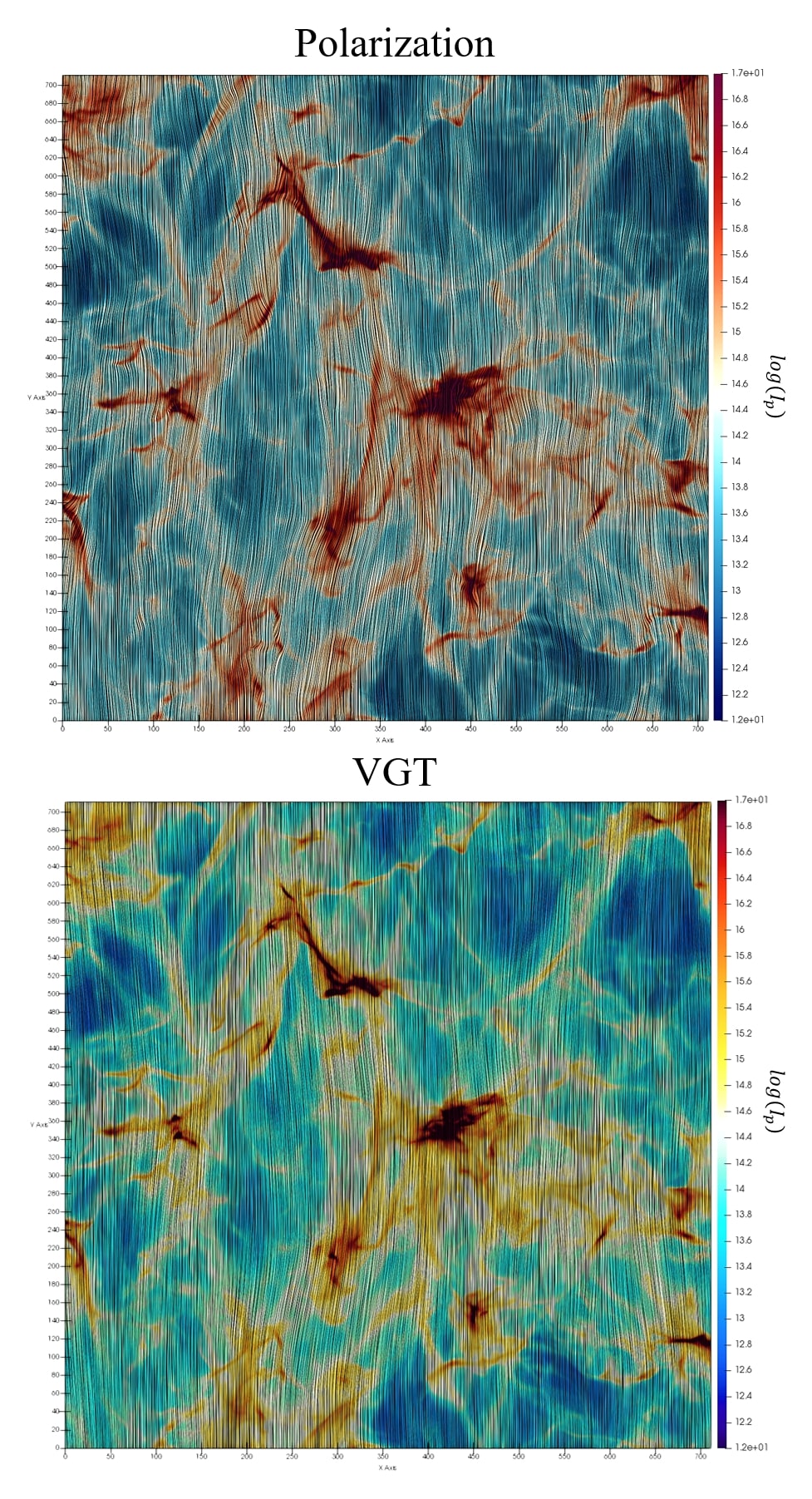}
		\caption{\label{fig:constant} The magnetic field derived from polarization (top) and the magnetic field inferred from VGT (bottom) using a synthetic PPV cube. The magnetic field is superimposed on the polarized intensity map $I_p$ and visualized using the LIC. The cube is produced from the MHD simulation A1 at t$_{r}=0.8 Myr$, using a unity density field while keeping the original velocity field unchanged M$_S=6.52$, i.e. all collapsing materials are removed. The corresponding AM = 0.98. }
	\end{figure}

	\subsection{The change of velocity gradients' amplitude}
	In addition to velocity gradients' orientation, the self-gravity also affects velocity gradients' amplitude. Self-gravitating gas induces an additional force and acceleration to turbulent plasma. As a result, the gradients' amplitude is expected to increase in the gravitational collapsing region. As shown in \citet{2018arXiv180200024Y}, the dispersion of gradients' amplitude is positively proportional to the evolution time. We, therefore, expect the analysis of velocity gradients' amplitude can also give a distinguishable feature in self-gravitating regions.
	
	In Fig.~\ref{fig:gradc_conv}, we describe the response of velocity gradients' amplitude in the presence of self-gravity. We plot two 2D histograms of logged velocity gradients' amplitude\footnote{Note we are using velocity centroid map \textbf{C(x,y)} for the calculation of the normalized velocity gradients amplitude here. The PCA technique employs the pseudo-Stokes parameters, which erases the information of gradients' amplitude.} and the projected velocity's convergence using A2 simulation at $t_r\simeq 0.0 Myr$ and $t_r\simeq 3.4 Myr$ respectively. When the self-gravity is absent, we see the 2D histogram give no preferential direction, i.e., the probability of obtaining large gradients' amplitude is similar for both high convergence and low convergence cases. However, the situation gets changed when gravitational collapsing starts. At $t_r\simeq 3.4 Myr$, we find that the 2D histogram becomes anisotropic. The high convergence corresponds to only large gradients' amplitude, and low convergence corresponds to small gradients' amplitude. Since the high velocity convergence is induced by self-gravity, large gradients' amplitude is, therefore, the reaction to self-gravity also. It then provides an alternative way to identify gravitational collapsing regions through gradients' amplitude.
	\begin{figure*}[t]
		\centering
		\includegraphics[width=.99\linewidth,height=0.8\linewidth]{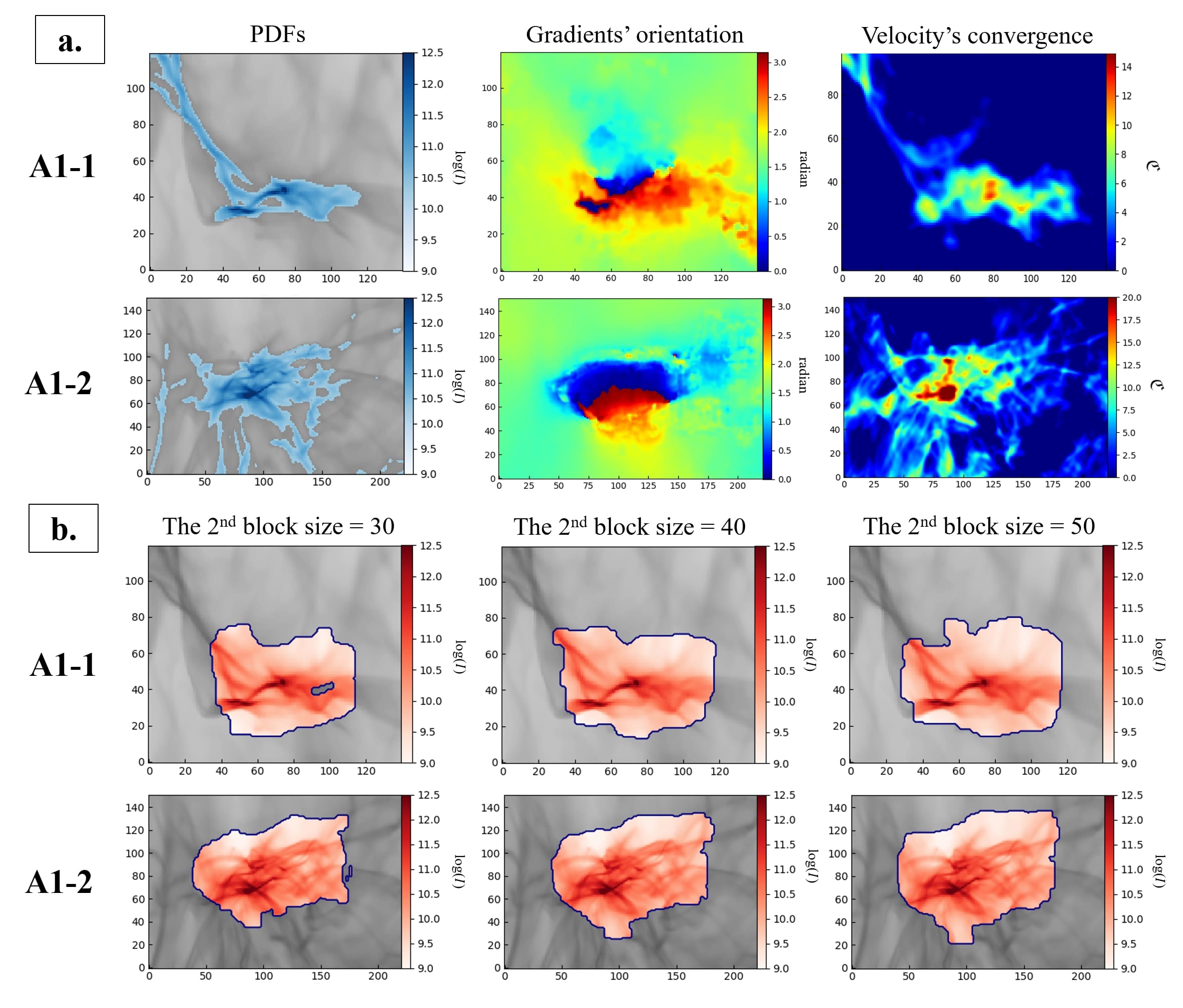}
		\caption{\label{fig:pdf_vgt} \textbf{Panel a:} the 1$^{st}$ and 2$^{nd}$ rows are two sub-regions extracted from simulation A1 at $t_r\simeq 0.8 Myr$, denoted as A1-1 (length scale $\simeq$ 1.7 pc) and A1-2 (length scale $\simeq$ 2.8 pc), respectively. The 1$^{st}$ column: the gravitational collapsing regions (blue regions) identified from the N-PDFs. The 2$^{nd}$ column: the orientation of velocity gradients in the range of [0, $\pi$] (i.e., red: $\simeq\pi$, blue: $\simeq0$, and green: $\simeq\pi/2$). The 3$^{rd}$ column: the projected velocity convergence of each corresponding region. The discontinuity comes from numerical effect.  \textbf{Panel b:} the gravitational collapsing regions (red regions) identified from the double-peak feature of velocity gradients morphology. The 2$^{nd}$ block size is implemented in the double-peak algorithm to plot the histogram of velocity gradients' orientation in a sub-region. We test three block sizes, i.e., 30, 40, and 50 pixels}.
	\end{figure*}
	\subsection{The fraction of collapsing gas}
	\label{appendix:A}
	The velocity gradient exhibits different reactions to the dominance of either turbulence or self-gravity. However, in the 2D map, the particular properties of gradients' change due to self-gravity can only be seen when the fraction of self-gravitating material is sufficiently large. In the case that the fraction of collapsing gas occupies a small volume of the clouds, the projection of both self-gravitating and non-self-gravitating gases can overwhelm the gravitational collapsing region in 2D. For example, in Fig.~\ref{fig:a1conv}, we see the change of AM, which quantifies the relative orientation between the velocity gradient and magnetic field, is gradually decreasing to negative values. At the same time, the fraction of collapsing gas is gradually increasing. The negative AM is therefore contributed by more collapsing regions resolved by VGT.
	
	To test the effect of a small fraction of collapsing gas, we produce a synthetic PPV cube from the MHD simulation A1 at t$_{r}=0.8$ Myr, using a unity density field while keeping the original velocity field unchanged M$_S=6.52$. The unity density field erases all self-gravitating gas. In Fig.~\ref{fig:constant}, we plot the actual magnetic field inferred from synthetic dust polarization and also the magnetic fields inferred from VGT (see \S~\ref{sec:method}).  We can see the magnetic field orientation is distributing around $\pi/2$ (with respect to the right horizontal direction) and VGT gives a good agreement with the actual magnetic field, i.e., AM = 0.98. The magnetic fields orientation inferred VGT using the actual density field is plotted in Fig.~\ref{fig:illu} showing AM = 0.73. It, therefore, confirms that when most of the matter along the line of sight is not collapsing, VGT can well trace the magnetic field comparing with dust polarization.
	
	\subsection{Identify gravitational collapsing regions through velocity gradient}
	\subsubsection{Double-peak histogram of velocity gradients orientation}
	\label{subsec:double-peak}
	
	\begin{figure*}[t]
		\centering
		\includegraphics[width=.99\linewidth,height=0.42\linewidth]{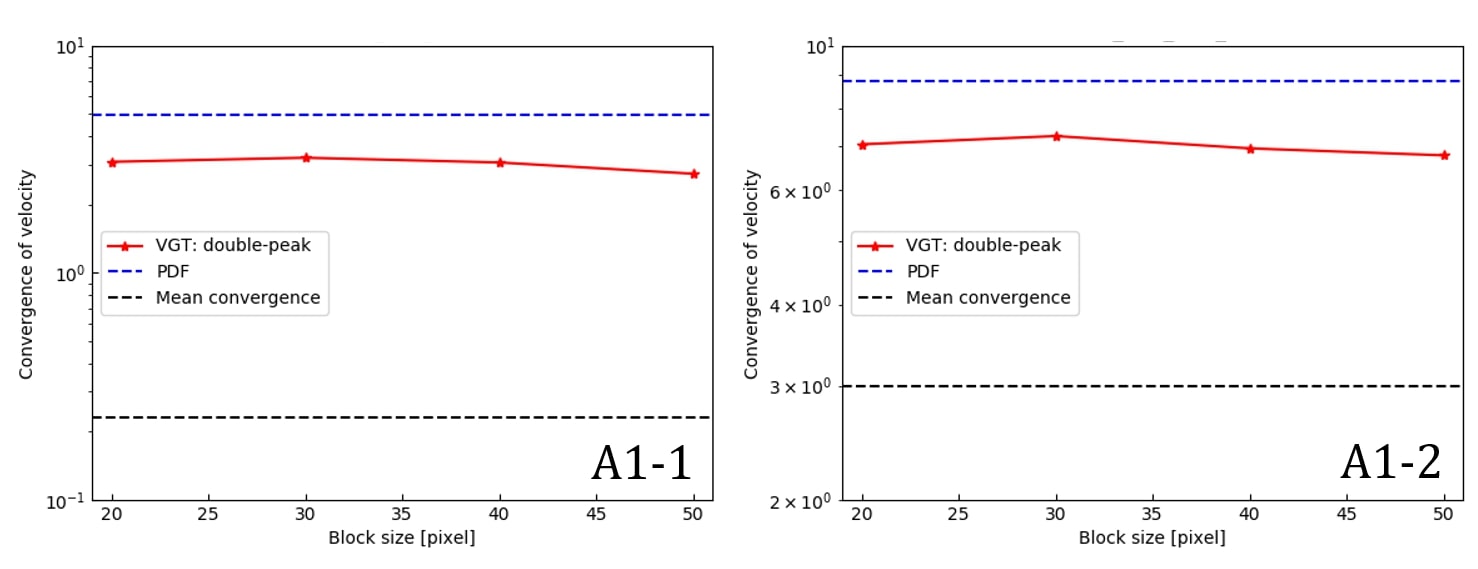}
		\caption{\label{fig:dp_conv} A comparison of velocity's mean convergence in the self-gravitating regions identified from N-PDFs (blue dashed line) and VGT's double-peak algorithm (red solid line). The two corresponding regions are shown in Fig.~\ref{fig:curvature}. The black dashed line indicates the global mean convergence in each region.}
	\end{figure*}
	
	As above, we confirm that velocity gradients change orientation by 90$^\circ$ in the presence of the gravitational collapsing (see \S~\ref{subsec:v-change}). We then expect when getting close to the boundary of self-gravitating gas the histogram of gradients' orientation appears a double-peak feature (see Fig.~\ref{fig:illu}). The algorithm of the double-peak feature is therefore used to locate gravitating collapsing gas (see \S~\ref{subsec:dp} for details).
	
	In Fig.~\ref{fig:pdf_vgt}, we zoom into two sub-regions A1-1 and A1-2 from simulation A1 at $t_r\simeq 0.8 Myr$. These sub-regions both contain well-defined convergent flows. We plot the gradients' orientation and the velocity convergence projected from 3D for each region. We see that amounts of gradients flip their direction by $\pi/2$. These resulting gradients are either 0 or $\pi$. One possible explanation is the different directions of infall gas. We compare the change of gradients' orientation with the N-PDF method (see Fig.~\ref{fig:pdf_vgt}). First of all, we observe that the majority of self-gravitating regions identified from the two methods are similar and reflect the convergent flows. However, for the A1-1 region, we find there exist a filamentary structure which is identified as self-gravitating by N-PDFs but not by the double-peak algorithm. For N-PDFs, it gives a density threshold to distinguish high-density gas showing self-similarity. However, the high-density gas contains not only self-gravitating materials but probably also non-self-gravitating density enhancement. As a result, the PDF method could mix self-gravitating structures and non-self-gravitating density enhancement. The second reason could come from velocity gradients. In Fig.~\ref{fig:a1conv}, velocity gradients hold their orientation little changed at the early stages of collapsing. We can expect the velocity field is significantly changed only when the gravitational energy dominates over the kinematic energy of turbulence. In this case, this filamentary structure could be at the beginning of collapsing. Also, if the collapsing fluid occupies only a small volume along LOS, the change of velocity gradients can be insignificant (see \S~\ref{appendix:A}).
	
	For the implementation of double-peak algorithm, we firstly produce a pixelized map of velocity gradients' orientation using the adaptive sub-block averaging method (see Fig.~\ref{fig:pdf_vgt}(a)). Based on the resultant velocity gradients' map after the adaptive sub-block, we draw the histogram of gradient orientation again in another sub-block (which it is denoted as the 2$^{nd}$ sub-block) and recognize the features of double peak based on the orientation distribution after the adaptive sub-block. Note the 2$^{nd}$ sub-block here is different from the one for the first step of magnetic fields tracing (see \S~\ref{subsec:dp}). Since the 2$^{nd}$ sub-block does not require a Gaussian fitting, we do not utilize an adaptive sub-block size, but keep it fixed (see \S~\ref{subsec:dp} for details).  In Fig.~\ref{fig:pdf_vgt}, we vary the size of the 2$^{nd}$ block and highlight the entire region enclosed by the boundary identified by the double-peak using red color. We can see that a large block size 50 produces a more significant boundary of the self-gravitating region, which is expected. We compare velocity's mean convergence in the self-gravitating regions identified from N-PDFs and VGT's double-peak algorithm, as well as the global mean convergence in the entire sub-region, see Fig.~\ref{fig:dp_conv}. For VGT with the double-peak recognition algorithm, the convergence is not constant. A large 2$^{nd}$ block, as well as the procedure of adaptive sub-block averaging, would include more non-collapsing gas so that the convergence becomes small. Both of the adaptive sub-block averaging method and the double-peak would include additional parts of non-convergent material due to the blocks on the boundary. This explains that the convergence obtained from the PDF method is always larger. To have the best performance, the block size should be selected until the histogram shows two apparent Gaussian distributions. In addition, in both of these two regions, we can see the convergences for VGT are positive and much significant than their corresponding global mean convergences. 
	
	\subsubsection{The curvature of velocity gradients}
	\label{subsec:curvature}
	
	\begin{figure*}[t]
		\centering
		\includegraphics[width=.99\linewidth,height=0.72\linewidth]{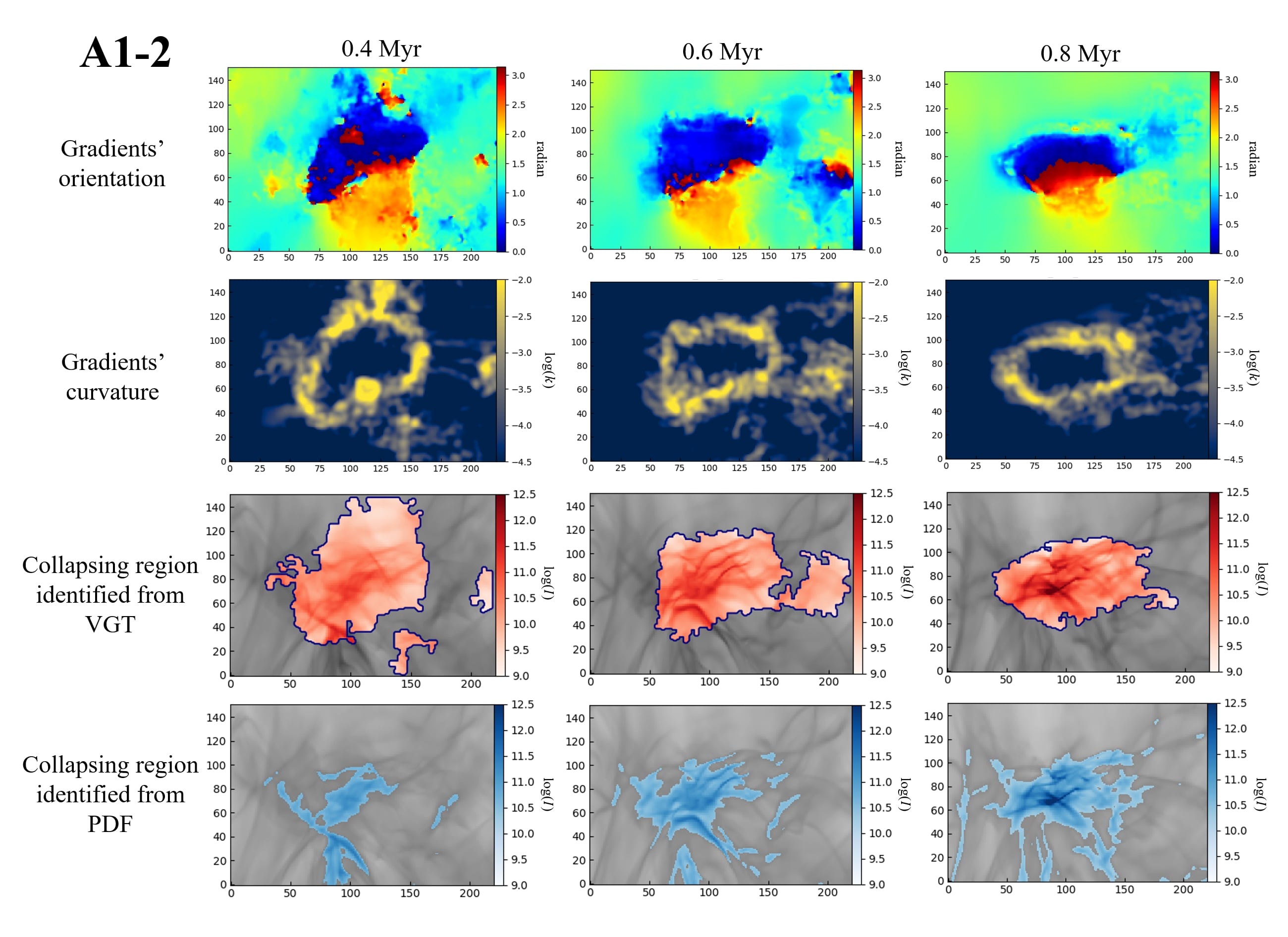}
		\caption{\label{fig:curvature} The example shows how to identify collapsing regions from the curvature of velocity gradients. We use three different collapsing stages of the sub-region A1-2, i.e., 0.4 Myr (the 1$^{st}$ column), 0.6 Myr (the 2$^{nd}$ column), and 0.8 Myr (the 3$^{rd}$ column). \textbf{The 1$^{st}$ row:} the orientation of velocity gradients in the range of [0, $\pi$) (i.e., red: $\simeq\pi$, blue: $\simeq0$, and green: $\simeq\pi/2$). \textbf{The 2$^{nd}$ row:} the curvature of velocity gradients calculated from the RK2 method. \textbf{The 3$^{rd}$ row:} the gravitational collapsing regions (red regions) identified from the curvature of velocity gradients. \textbf{The 4$^{th}$ row:} the gravitational collapsing regions (blue regions) identified from the N-PDFs.}
	\end{figure*}
	\begin{figure}[t]
		\centering
		\includegraphics[width=.99\linewidth,height=0.68\linewidth]{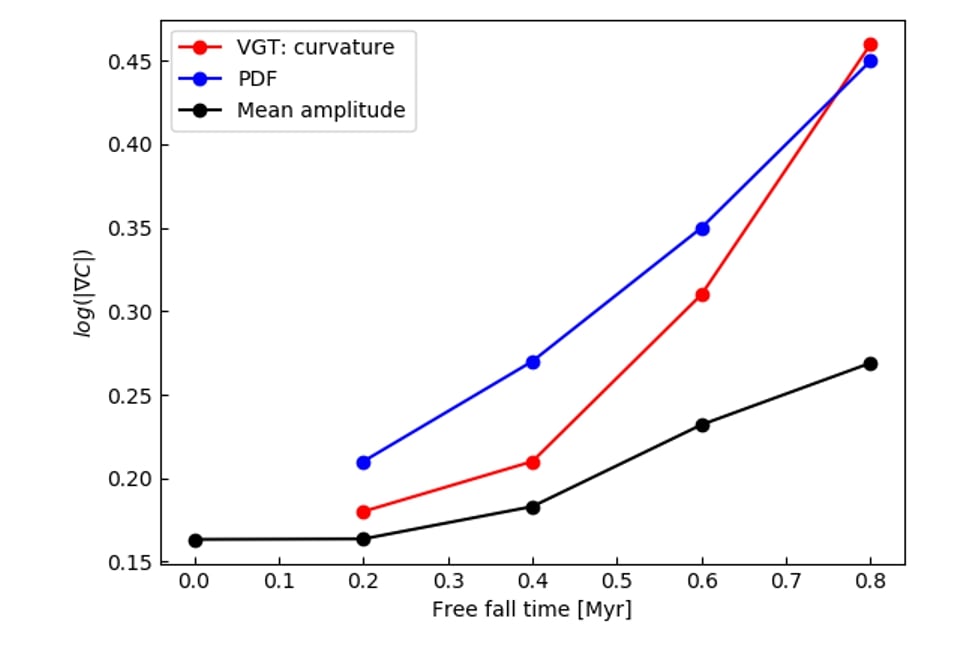}
		\caption{\label{fig:run1_cur} A comparison of the average velocity gradients' amplitude in the self-gravitating regions identified from N-PDFs (blue) and VGT's curvature algorithm (red). The corresponding regions are shown in Fig.~\ref{fig:curvature}. The black line indicates the global mean convergence in each region.}
	\end{figure}
	
	\begin{figure*}[t]
		\centering
		\includegraphics[width=.99\linewidth,height=1.2\linewidth]{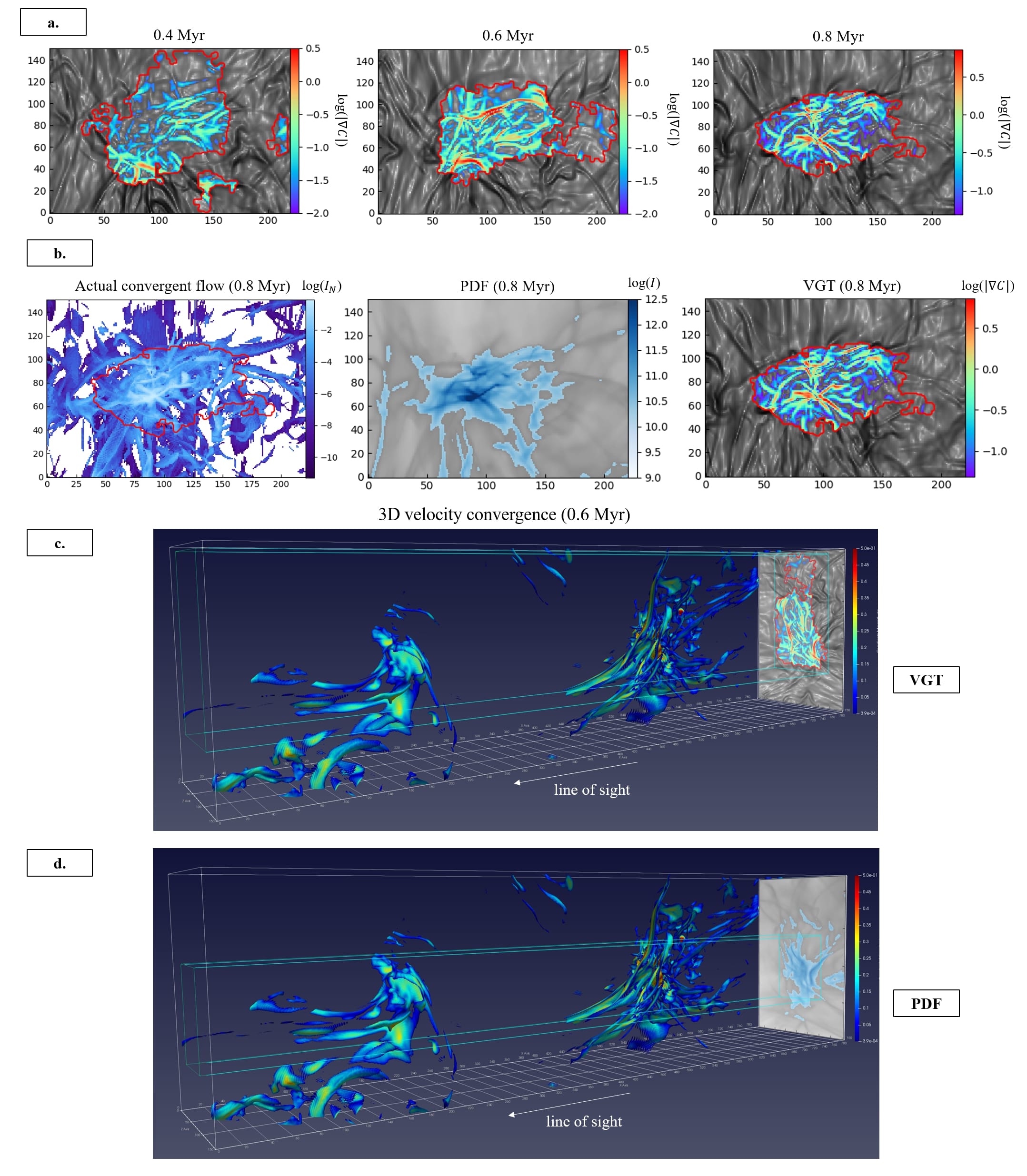}
		\caption{\label{fig:run1_amp} \textbf{Panel a:} The areas of the gradients amplitude corresponding to the collapsing regions identified by VGT (see Fig.~\ref{fig:curvature}) in A1-2 at $t_r\simeq$ 0.4 Myr (left), 0.6 Myr (middle), and 0.8 Myr (right). In the same regions, we mask the low amplitude pixel, i.e., its corresponding amplitude is less than the global mean value. The remaining colorful areas indicate a high gradients' amplitude and a high convergence, i.e., the convergent inflow. \textbf{Panel b:} A comparison of the convergent flow determined by the PDF (middle), VGT (right), and the actual convergent flow (left) at at $t_r\simeq 0.8 Myr$. $I_N$ is the projection of normalized volume density. \textbf{Panel c:} a visualization of 3D velocity convergence. The positive convergence outlines the convergent inflow. The inner blue box approximately outlines the boundary of the collapsing region identified from VGT. \textbf{Panel d:} a visualization of 3D velocity convergence. The inner blue box approximately outlines the boundary of the collapsing region identified from N-PDFs.}
	\end{figure*}
	\begin{figure}[t]
		\centering
		\includegraphics[width=.99\linewidth,height=0.7\linewidth]{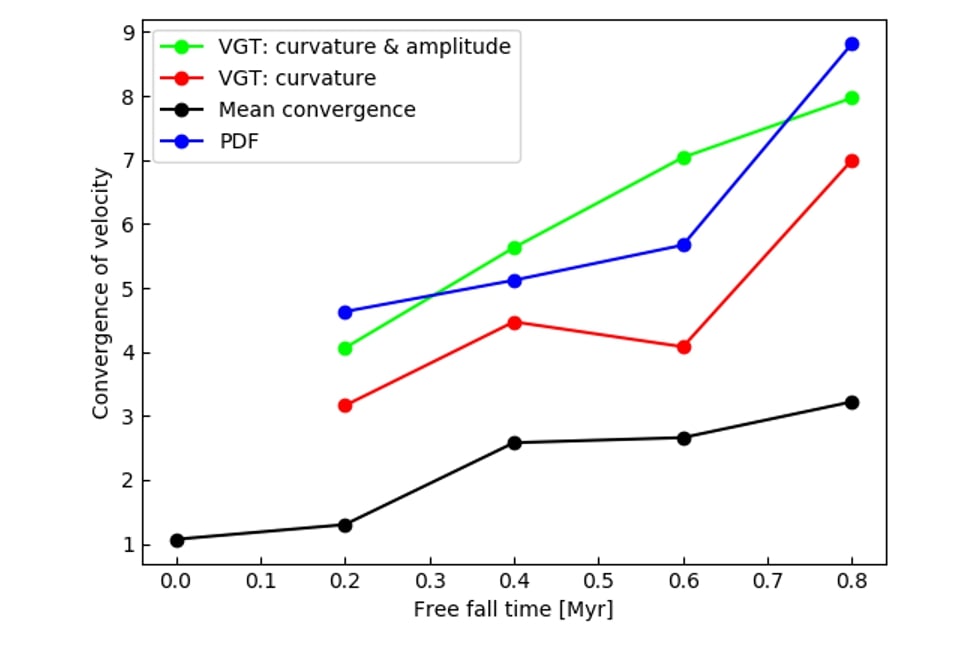}
		\caption{\label{fig:run1_cur_amp} A comparison of velocity's mean convergence for the regions corresponding to (i) the inflow outlined by the curvature and amplitude (lime), (ii) the collapsing region identified by the curvature algorithm (red), (iii) the collapsing region identified by N-PDFs (blue), and (iv) the entire A1-2 region (black).}
	\end{figure}
	The curvature of velocity gradients gets its maximum value when the gradients flip their direction by 90$^\circ$, but the ambient region usually has small curvature. The curvature therefore provides the second way to identify the self-gravitating regions by sorting out the curvatures. In Fig.~\ref{fig:curvature}, we give an example of how the curvature algorithm works. We use three different collapsing stages of the sub-region A1-2, i.e., 0.4 Myr, 0.6 Myr, and 0.8 Myr. At the initial collapsing stage, we expect that there exists a large amount of convergent gas, and the dynamics become slim at the later collapsing stage. This can be measured from the changes in gradients' orientation. Fig.~\ref{fig:curvature} shows that the area in which gradients change directions is decreasing its size when the evolution time goes up. It is clear that the red tail in this area is vanishing. In the second row of Fig.~\ref{fig:curvature}, one can see that the gradients' curvature becomes significantly large at the boundary of collapsing regions. However, the curvature does not reveal the region where the change of velocity gradients' direction is not sufficient for 90$^\circ$. For example, when the collapsing material only occupies a small fraction of the cloud along LOS, the resulting gradient can have a change of direction less than 90$^\circ$. We expect this is the reason we see three disconnected self-gravitating regions in the 3$^{rd}$ row, 0.4 Myr of Fig.~\ref{fig:curvature}. According to the enclosed boundary outlined by the curvature, we figure out the full collapsing regions using the "contourf" function of Julia \citep{2012arXiv1209.5145B}. One can see the formation of the core at t$_{r}=0.8$ Myr, and the collapsing region is shrinking its size. Comparing with the collapsing regions identified from N-PDFs (see Fig.~\ref{fig:curvature}, the 4$^{th}$ row), we get similar results for both VGT and N-PDFs methods at t$_{r}=0.8$ Myr. However, N-PDFs indicate a smaller collapsing area than VGT at t$_{r}=0.4$ Myr and  t$_{r}=0.6$ Myr. One reason is that N-PDFs do not measure the existence of the continuous inflow onto the star-forming cores at the early stages of star formation before enough gas is accumulated in the core, but VGT samples velocities and therefore it is sensitive to the inflow. 
	
	We show in Fig.~\ref{fig:a1amp} that the large convergence in collapsing regions guarantees a large amplitude of velocity gradients. We, therefore, study the gradient amplitude in the collapsing region identified from VGT and N-PDFs. In Fig.~\ref{fig:run1_cur}, we calculate the average value of gradients' amplitude in the corresponding collapsing regions and also the global mean value of the amplitude. We can see that both VGT and N-PDFs always give more significant gradients' amplitude than the global mean value. However, in some case, N-PDFs can also show a higher amplitude than VGT. The adaptive sub-block averaging method, as well as the 2$^{nd}$ sub-block, still contributes to the difference, because the utilization of both sub-blocks will take into account extra non-convergent inflow. However, the sub-block averaging is not implemented in the calculation of gradients' amplitude and high amplitude is only induced by self-gravity. With the assistance of gradients' amplitude, it is possible to get rid of the extra non-self-gravitating regions introduced by the both sub-blocks.
	
	In Fig.~\ref{fig:run1_amp}(a), we outline the areas of gradients amplitude corresponding to the collapsing regions identified by VGT (see Fig.~\ref{fig:curvature}), but at different collapsing stages $t_r\simeq$ 0.4 Myr, 0.6 Myr , and 0.8 Myr. In the same regions, we also mask the low amplitude pixel, i.e., its corresponding amplitude is less than the global mean value. The remaining areas, therefore, have high gradients' amplitude and high convergence, i.e., the convergent inflow. We can see, in Fig.~\ref{fig:run1_amp}, the inflow is slight but cover a larger area at t$_{r}=0.4$ Myr. At t$_{r}=0.6$ Myr, the inflow gets stronger and finally goes into the core of star-formation at t$_{r}=0.8$ Myr. In Fig.~\ref{fig:run1_amp}(b), we plot the actual 2D convergent flow map (i.e., the density structures correspond to positive convergence) normalized by the total column density along the particular line of sight. Comparing with the actual convergent flow, we can see both N-PDF and VGT give similar answers and can determine the projected convergent flow by themselves. Note that at t$_{r}=0.4$ Myr and t$_{r}=0.6$ Myr, the convergent flow obtained from VGT occupies a larger area than the one defined by the N-PDFs (see Fig.~\ref{fig:curvature}). We expect the reason is that  N-PDFs are sensitive to only the already formed accreting cores, but gradients trace all the convergent flow. The outcome of VGT is also affected by the fraction of collapsing gas in the cube. In \S~\ref{appendix:A}, we show that when most of the matter along the line of sight is not collapsing, VGT may not reveal the collapsing region. As for the N-PDFs, in addition to the issue of the low fraction of collapsing gas, the low-density part of collapsing gas may also be overwhelmed because of the projection effect. To justify this point, we show the 3D visualization of actual convergent flow at t$_{r}=0.6$ Myr in Fig.~\ref{fig:run1_amp}(c) and (d). We use a blue box to outline the corresponding 3D convergence identified by VGT and N-PDFs, respectively. We can see the convergent area defined by VGT included the majority of 3D convergent flows, while the N-PDFs give only the strongly convergent fluid, which does not cover the convergence on the top volume. We find the convergent fluid on the top corresponds to the low-intensity region of the 2D projected intensity map, see Fig.~\ref{fig:run1_amp} (d). The N-PDFs, therefore, do not resolve the low-intensity part since the transition threshold S$_t$ of N-PDFs only highlights the high-intensity part (see \S~\ref{sec:theory}). As for the bottom part of the 3D convergence, the fraction of collapsing gas is minimum, see Fig.~\ref{fig:run1_amp} (d). Its 2D projection is therefore dominated by non-collapsing gas.
	
	Besides, we calculate the mean convergence for the regions corresponding to (i) the inflow outlined by the VGT (curvature \& amplitude algorithm), (ii) the collapsing region identified by the curvature algorithm, (iii) the collapsing region identified by N-PDFs, and (iv) the entire A1-2 region. The result is presented in Fig.~\ref{fig:run1_cur_amp}. Firstly, we see the global mean convergence is increasing with the evolution of gravitational collapsing. Both VGT and N-PDFs always give larger convergence than the global mean value, while N-PDFs can also show a larger convergence than VGT employing only the curvature algorithm. In particular, at t$_{r}=0.4$ Myr and t$_{r}=0.6$ Myr, VGT (curvature \& amplitude) appears a larger range of inflow and the largest convergence. As for the very initial and final moments, i.e., t$_{r}=0.2$ Myr and t$_{r}=0.8$ Myr, the dynamics of collapsing are less significant, and the VGT (curvature \& amplitude) includes some non-convergent fluid. The performance of VGT in identifying gravitational collapsing region is therefore comparable or better to the N-PDFs.
	
	\section{Identify gravitational collapsing regions with intensity gradients}
	\label{sec:IGT}
	\begin{figure*}[t]
		\centering
		\includegraphics[width=.96\linewidth,height=.7\linewidth]{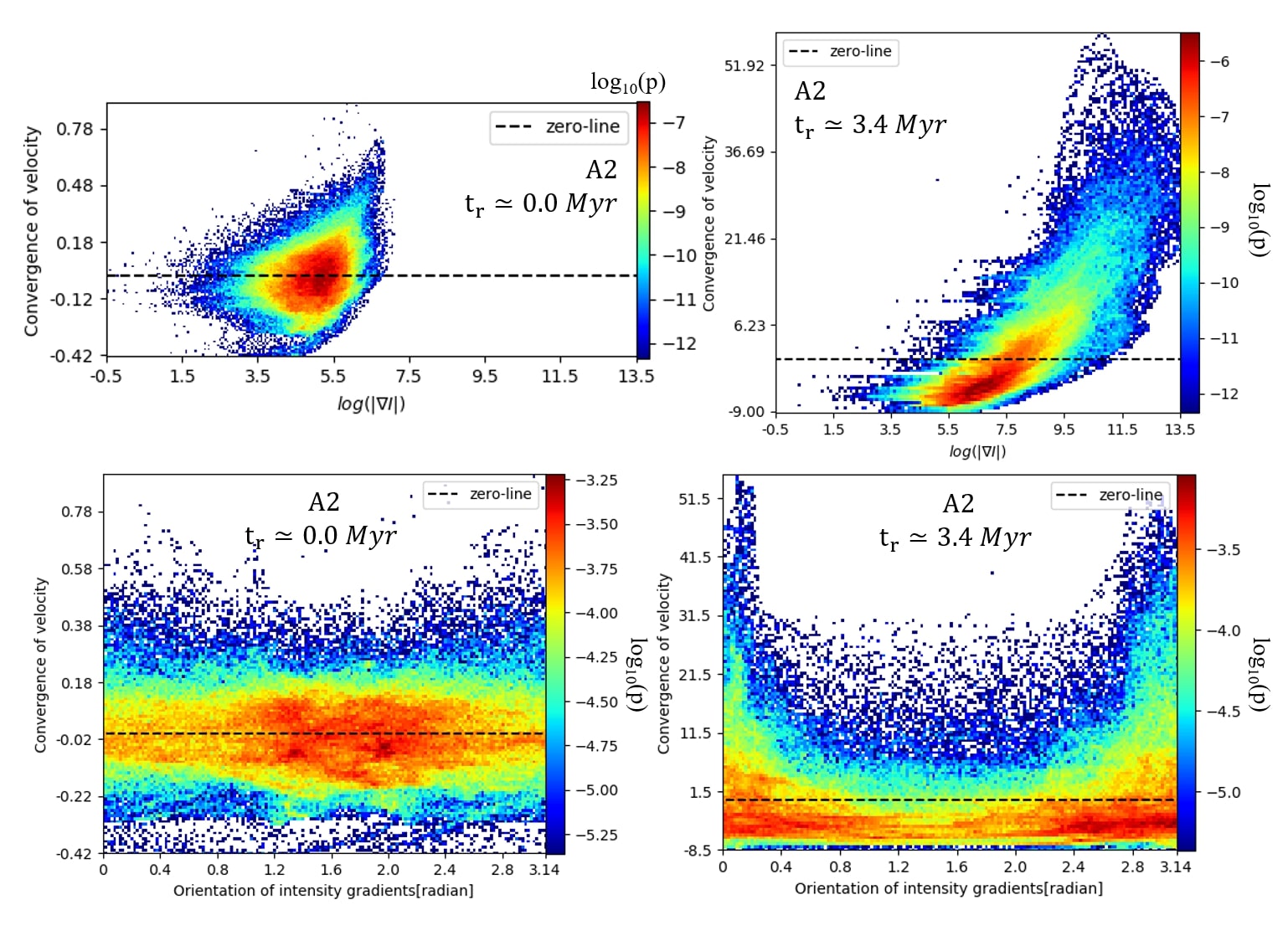}
		\caption{\label{fig:IGs} \textbf{Top row:} The 2D histogram of intensity gradients' amplitude, i.e., $log(|\nabla I|)$, and velocity's convergence using A2 simulation at $t_r\simeq 0.0 Myr$ (left) and $t_r\simeq 3.4 Myr$ (right). p gives the volume fraction of each data point. \textbf{Bottom row:} the 2D histogram of intensity gradients' orientation and velocity's convergence using A2 simulation. Note that the orientation is measure in typical cartesian coordinate, i.e., with respect to the right horizontal direction. Bin-size is 200 for the 2D histograms.}
	\end{figure*}
	It has been shown that intensity gradients also change their direction by 90$^\circ$ in the presence of gravitational collapsing \citep{YL17b, IGs}. Here we apply the tools developed in the paper, i.e., the adaptive sub-block averaging and curvature algorithm to identify the gravitational collapsing region using the Intensity Gradient Technique (IGT). The intensity map \textbf{I(x,y)} is produced through the integration of PPV cubes along the line of sight:
	\begin{equation}
	I(x,y)=\int\rho(x,y,v)dv
	\label{eq3}
	\end{equation}
	where $\rho$ is gas density, $v$ is the velocity component along the line of sight. By doing convolution with Sobel kernels $G_x$ and $G_y$, the intensity gradient orientation at individual pixel $(x,y)$ is obtained:
	\begin{equation}
	\begin{aligned}
	\bigtriangledown_x I(x,y)=G_x * I(x,y)  \\  
	\bigtriangledown_y I(x,y)=G_y * I(x,y)  \\
	\psi(x,y)=ASB[tan^{-1}(\frac{\bigtriangledown_y I(x,y)}{\bigtriangledown_x I(x,y)})]
	\end{aligned}
	\end{equation}
	where $\bigtriangledown_x I(x,y)$ and $\bigtriangledown_y I(x,y)$ are the x and y components of intensity gradients respectively. $ASB$ denotes the implementation of the adaptive sub-block averaging method (see \S~\ref{subsec:asb}) and $\psi$ is the resultant intensity gradient map.
	
	\begin{figure*}[t]
		\centering
		\includegraphics[width=.96\linewidth,height=.53\linewidth]{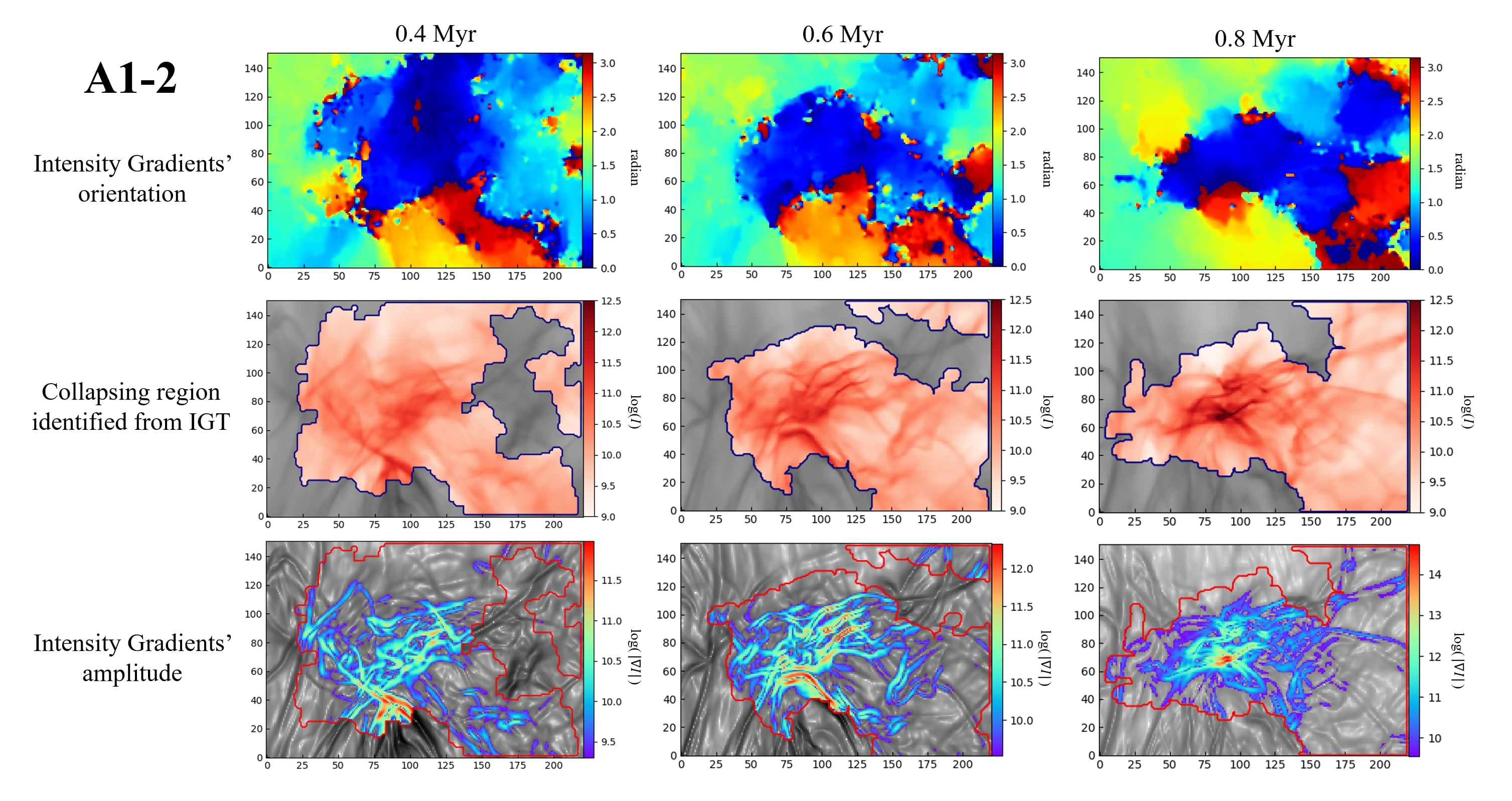}
		\caption{\label{fig:IGT} The example shows how to identify collapsing regions from the curvature of intensity gradients. We use three different collapsing stages of the sub-region A1-2, i.e., 0.4 Myr (the 1$^{st}$ column), 0.6 Myr (the 2$^{nd}$ column), and 0.8 Myr (the 3$^{rd}$ column). \textbf{The 1$^{st}$ row:} the orientation of intensity gradients in the range of [0, $\pi$) (i.e., red: $\simeq\pi$, blue: $\simeq0$, and green: $\simeq\pi/2$). \textbf{The 2$^{nd}$ row:} the gravitational collapsing regions (red regions) identified from the curvature of intensity gradients. \textbf{The 3$^{rd}$ row:} the areas of the gradients amplitude corresponding to the collapsing regions identified by intensity gradients. In the same regions, we mask the low amplitude pixel, i.e., its corresponding amplitude is less than the global mean value.}
	\end{figure*}
	
	\begin{figure*}[t]
		\centering
		\includegraphics[width=.88\linewidth,height=1.2\linewidth]{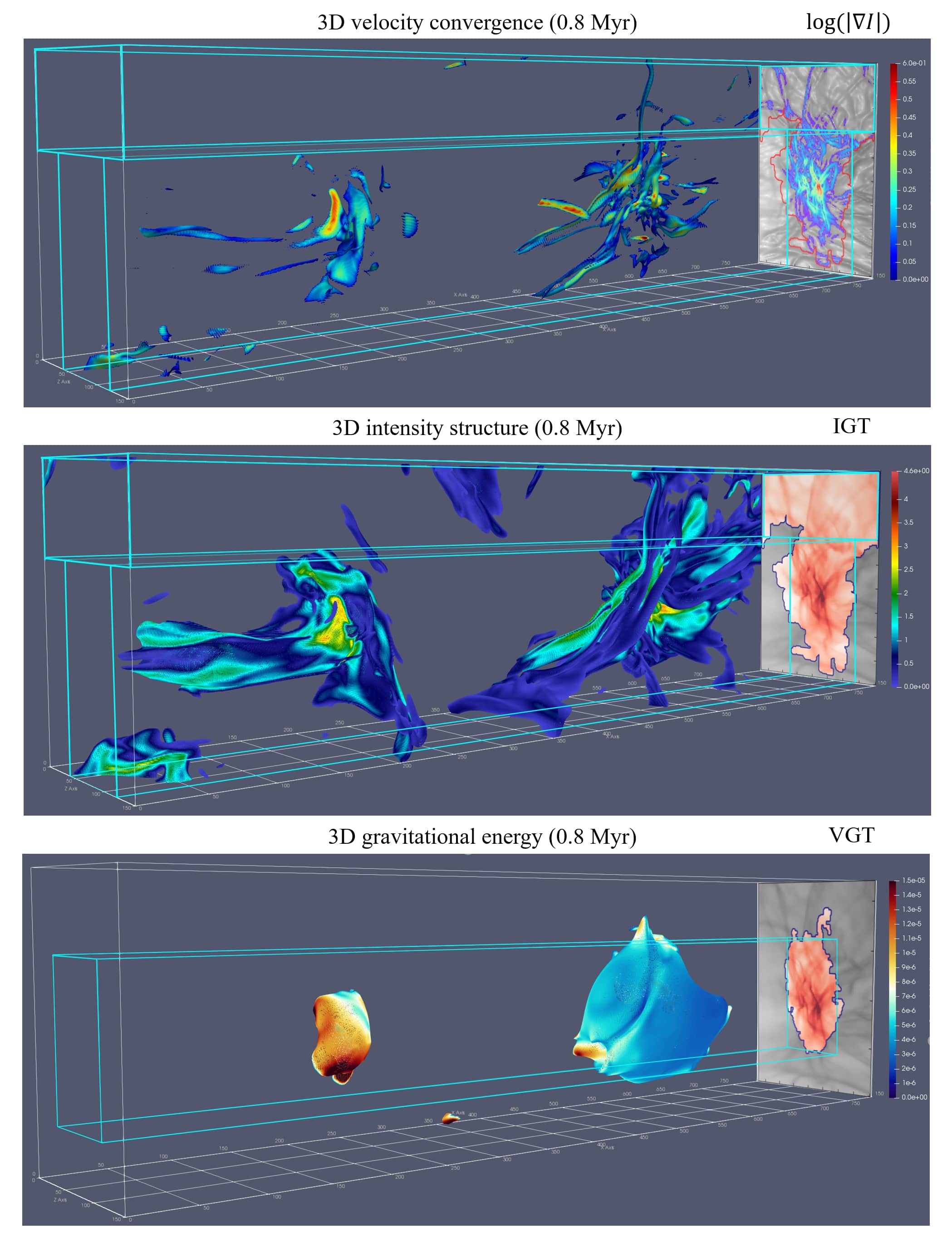}
		\caption{\label{fig:IG_conv} \textbf{Top:} a visualization of 3D velocity convergence using simulation A1 at $t_r\simeq 0.8 Myr$. The positive convergence outlines the convergent inflow. The inner blue box approximately outlines the boundary of the 3D collapsing region identified from IGT. The projected 2D plot on the right side shows the corresponding intensity gradients amplitude log$(|\nabla I|)$. We mask the low amplitude pixel, i.e., its corresponding amplitude is less than the global mean value. \textbf{Middle:} a visualization of 3D intensity structures. The inner blue box and approximately outlines the boundary of the 3D collapsing region identified from IGT (red region in the 2D projection). We highlight high-intensity structures, i.e., its corresponding intensity is five times larger than the global mean value. \textbf{Bottom:} a visualization of 3D gravitational energy. The inner blue box and approximately outlines the boundary of the 3D collapsing region identified from VGT (red region in the 2D projection). We keep only the pixels with high gravitational energy, i.e., its corresponding gravitational energy is five times larger than the global mean value. }
	\end{figure*}
	As an analogy to the velocity gradient, we repeat the same analysis for the intensity gradient to study its properties in collapsing regions. In Fig.~\ref{fig:IGs}, we plot the 2D histograms of intensity gradients' amplitude, i.e., $log(|\nabla I|)$, and velocity's convergence using A2 simulation at $t_r\simeq 0.0 Myr$ and $t_r\simeq 3.4 Myr$. In the absence of self-gravity, i.e., at $t_r\simeq 0.0 Myr$, the convergence of velocity is low, and the histogram is slight deviates towards the part with high-intensity gradients amplitude. In general, the probability of obtaining large intensity gradients amplitude is similar for both high convergence and low convergence cases. While Fig.~\ref{fig:a1amp} shows the similar histogram of velocity gradients amplitude is more isotropic. When the self-gravity grows strong, i.e., at $t_r\simeq 3.4 Myr$, the 2D histogram becomes very anisotropic. The high convergence corresponds to only large gradients' amplitude, and low convergence corresponds to small gradients' amplitude. Since the high-velocity convergence is induced by self-gravity, large gradients' amplitude is, therefore, the outcome of self-gravity. In the bottom row of Fig.~\ref{fig:IGs}, we give the 2D histograms of intensity gradients' orientation and velocity's convergence using still A2 simulation at $t_r\simeq 0.0 Myr$ and $t_r\simeq 3.4 Myr$. We see that the histogram is close to a uniform distribution. For either positive convergence part or negative convergence part, one can equally find the orientation of intensity gradient in the full range of $[0,\pi)$. However, the distribution of velocity gradients only concentrates on the $\pi/2$ (see Fig.~\ref{fig:a1conv}), which is the mean magnetic field direction in our simulation. Velocity gradients are, therefore, more accurate in terms of magnetic field probing than intensity gradients. As for $t_r\simeq 3.4 Myr$, the distribution of intensity gradients deviates towards either 0 or $\pi$, i.e., gradients become perpendicular to the magnetic field (note we rotate gradients by 90$^\circ$ in this work). Therefore, in the case of gravitational collapsing, both velocity gradient and intensity gradient flip their direction by 90$^\circ$, and large gradients' amplitudes are induced simultaneously. The intensity gradient then provides an alternative way to identify gravitational collapsing region.
	
	In Fig.~\ref{fig:IGT}, we apply the curvature algorithm to identify the collapsing part in the sub-region A1-2 at $t_r\simeq 0.4Myr$, $0.6Myr$, and $0.8 Myr$ respectively, which has been analyzed in Fig.\ref{fig:curvature}. Firstly, we see intensity gradients change their orientation at the surroundings of the core. However, the area showing the change is significantly larger than the one seen in velocity gradients' orientation, while its evolution is similar. The top part and bottom part are initial at $t_r\simeq 0.4Myr$ showing the change of gradients' direction, but at $t_r\simeq 0.8Myr$ the change disappears in these two areas. According to our theoretical consideration,  both intensity and velocity gradients are sensitive to the motion of infall fluid. Once the collapsing has been completed without surrounding infall fluid, the gradients do not distinguish the core. We distinguish the collapsing region and see some diffuse regions on the right side of the intensity map are also covered. Since the intensity gradients are sensitive to both self-gravity and shock \citep{IGs}, the change in the diffuse area might come from shocks. Recall that self-gravity will also induce a larger amplitude of intensity gradients, but this is not the case for shock, i.e., the upper limit of the amplitude depends on the compressibility $\beta$ \citep{2018arXiv180200024Y,2019ApJ...878..157X}. We then also highlight the high amplitude part (i.e., its corresponding amplitude is larger than the global mean value) in Fig.~\ref{fig:IGT}. We see that large gradients amplitude only appears in the surrounding of the collapsing core, while the diffuse region only shows small values of gradients amplitude. As for velocity gradients, we show in Fig.~\ref{fig:run1_cur_amp} that they disclose the self-gravitating region.
	
\begin{figure*}[t]
		\centering
		\includegraphics[width=.99\linewidth,height=0.9\linewidth]{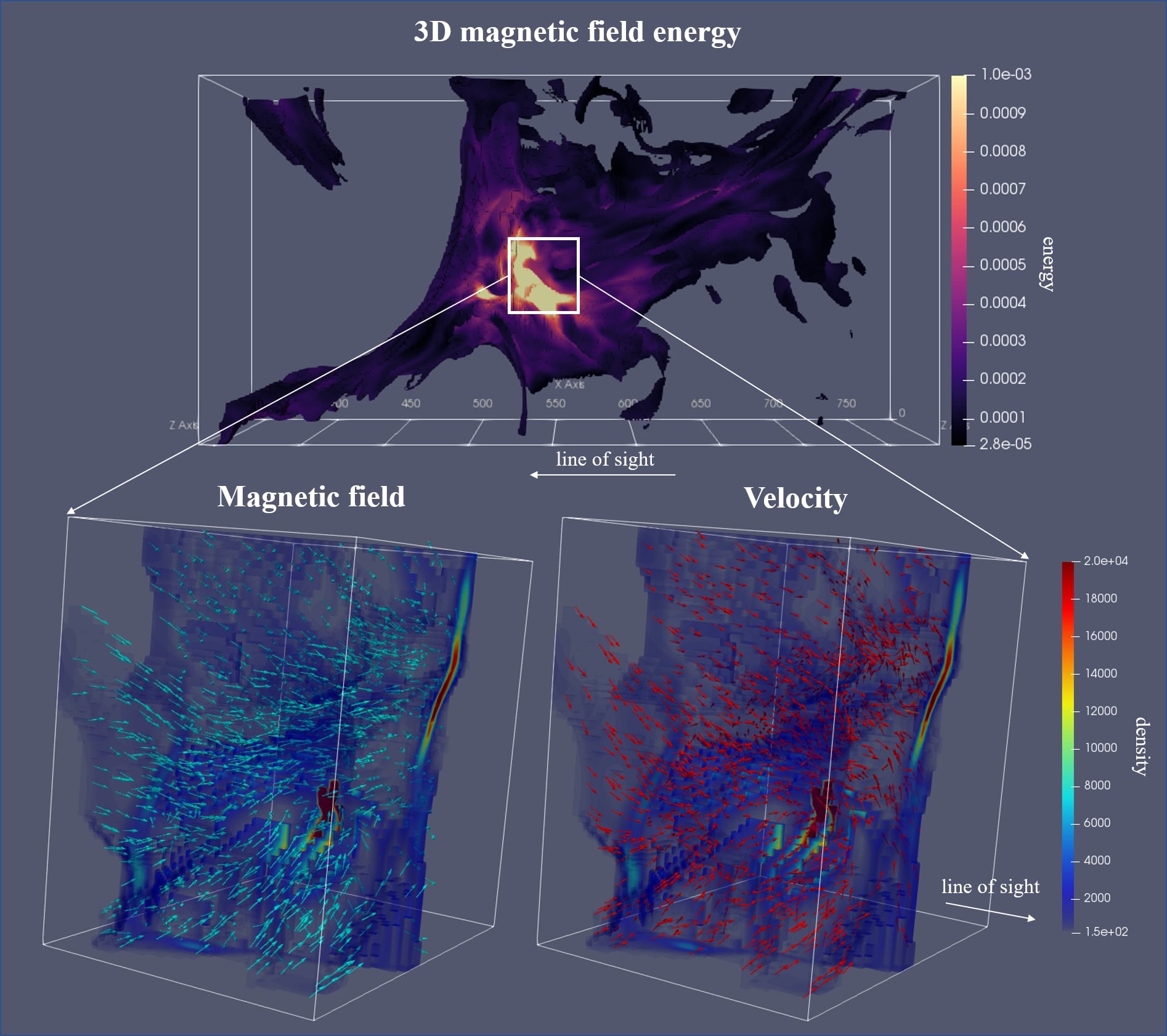}
		\caption{\label{fig:3dBV} \textbf{Top:} a visualization of 3D magnetic field energy using simulation A1 at $t_r\simeq 0.8 Myr$. We highlight high-intensity structures, i.e., its corresponding intensity is five times larger than the global mean value. \textbf{Bottom:} a visualization of magnetic field's orientation (left, blue vectors) and velocity's direction (right, red vectors). Note this sub-region exhibits high magnetic field energy and the velocity is indicating the direction of convergent flow. The unit of energy and volume density is $M_{\odot}\cdot pc^2 \cdot Myr^{-2}$ and $M_{\odot}\cdot pc^{-3}$ respectively.}
	\end{figure*}

	To justify our consideration about shocks, we give the 3D visualizations of velocity convergence in Fig.~\ref{fig:IG_conv}, using simulation A1 at $t_r\simeq 0.8 Myr$. Similar to Fig.~\ref{fig:run1_cur_amp}, we remove negative convergence to outline the convergent inflow and mask the low amplitude pixel, i.e., its corresponding amplitude of intensity gradients is less than the global mean value. The large amplitude region agrees with the high convergence region. As for the small-amplitude region on the top, the convergence is insignificant and is not expected to be resolved by gradients. In addition, in Fig.~\ref{fig:IG_conv}, we outline the high-intensity structures whose intensity is five times larger than the global mean value. Recall that the magnetic field is pointing in or out the paper in this visualization. We can see those high-intensity structures are perpendicular to magnetic fields, as explained in \citet{2019ApJ...878..157X}. In the bottom panel of Fig.~\ref{fig:IG_conv}, we show the  3D visualizations of high gravitational energy, i.e., the corresponding gravitational energy in each pixel is five times larger than the global mean value. As we expected, the upper right volume is not bounded by gravitational energy but full will high-intensity structures, which is believed to be shocks. After projection, this shock space matches the area in which the intensity gradients change their orientation. The change of intensity gradients' orientation in the diffuse region is, therefore, mostly coming from the contribution of shocks. We then can distinguish the collapsing regions and shock through the comparison of intensity gradients' orientation and amplitude, as well as velocity gradients. More importantly, the gravitational collapsing region identified by VGT agrees with the 2D projection of high gravitational energy space. Also, in Fig.~\ref{fig:3dBV}, we find in the zoom-in collapsing region, the local magnetic field energy gets maximum in the center of collapsing clump. Although the turbulent reconnection tries to minimize the increment of the magnetic field, the effect is constrained by its finite rate in high-density regions. We can, therefore, observe a stronger magnetic field in the collapsing center. 
	
	Fig.~\ref{fig:3dBV} also shows that our finding of the gravity-induced convergent flow is converged in terms of resolution. In Fig.~\ref{fig:a1conv}, we use the A2 simulation with resolution $480^3$ showing the magnetic field tends to follow the convergent flow. Fig.~\ref{fig:3dBV} visualizes this phenomenon using A1 simulation with resolution $792^3$. Indeed, as the gravitational collapse is not bounded, the convergent flow will spread to all surrounding pixels with the increment of evolution time. Otherwise, to see the convergent at the very beginning of the collapse, we need a very high resolution to resolve the collapsing core.
	
	The combination of VGT and IGT can also be used to determine the stages of gravitational collapsing. For instance, comparing Fig.~\ref{fig:IGT} and Fig.~\ref{fig:curvature}, we find the dense collapsing region identified by IGT is still larger than the one identified by VGT. Considering the physical evolution of intensity and velocity fields, the velocity field is dramatically changed only when the gravitational energy dominates over the kinematic energy of turbulence, but the intensity field's change is accumulating. Therefore, the change of intensity gradients would be observed in the initial collapsing stages. As a result, we can determine the collapsing stages through the comparison of VGT and IGT: (i) the onset of collapsing. The rotated velocity gradients and intensity gradients, together with the magnetic field, are aligned with each other; (ii) the middle stage of collapsing. The rotated velocity gradients are still parallel to magnetic fields, while the intensity gradients show no alignment to both velocity gradients and magnetic fields; (iii) the final stage of collapsing. The rotated velocity gradients and intensity gradients are anti-aligned (perpendicular) to the magnetic field, while velocity gradients and intensity gradients are aligned parallel to each other.
	\section{Discussion} \label{sec.dis}
	\subsection{Identify self-gravitating regions from velocity gradients and intensity gradients}
	
	\subsubsection{Implication to velocity gradients}
	The novel VGT technique utilizes the velocity gradient to trace the magnetic field based on the anisotropic properties of MHD turbulence. The properties of turbulence are modified in self-gravitating environments. As shown in Fig.~\ref{fig:3dBV}, in a gravitational collapsing region with sub-Alfv\'{e}nic condition, the magnetic field is strong enough to provide supporting force. A magnetic force counteracts any gravitational pull inducing the acceleration, which is perpendicular to the magnetic field. The gravitational force, therefore, produces the most significant convergent plasma accelerating in the direction parallel to the magnetic field. The maximum spatial difference of velocity is then following the same direction of the convergent flow, which means the velocity gradients are parallel to the magnetic field. Comparing with the non-self-gravitating turbulence, in which the velocity gradients are perpendicular to the magnetic field, the velocity gradient flips its direction by 90$^\circ$ when self-gravity dominates over turbulence. Such changes mark the transition from a magnetic field and turbulence dominated regime at low densities, to higher density regions that are collapsing under gravity. 
	
	Based on the change of gradients' orientation in the collapsing region, we propose two novel approaches to identify gravitational collapsing regions through velocity gradients, i.e., the double-peak algorithm (see \S~\ref{subsec:dp}) and the curvature algorithm (see \S~\ref{subsec:cur}). The double-peak algorithm quantifies the change of gradients' orientation in a histogram format. Supposing we zoom into a sub-region and draw the corresponding histogram of velocity gradients' orientation. When the sub-region is diffuse, the velocity gradient is perpendicular to the magnetic field, and the histogram is a single Gaussian distribution. However, when the sub-region contains enough collapsing materials, i.e., on the boundary of the collapsing region, amounts of velocity gradients flips their direction by 90$^\circ$, becoming parallel to the magnetic field. The histogram, therefore, appears two Gaussian peak values at angle $\theta$ and $\theta+\pi/2$, which we denote as the double-peak feature. This feature reveals the boundary of the gravitational collapsing region without the usage of polarimetry measurements. As for the curvature algorithm, it utilizes the geometry of the velocity gradient. When the velocity gradient goes from diffuse region to collapsing region, i.e., on the boundary of the collapsing region, its direction is rapidly changed by 90$^\circ$. This significant change induces a maximum curvature of velocity gradients, but the ambient region usually has small curvature. The calculation of curvature, therefore, allows us to highlight the boundary of the gravitational collapsing region.
	
	\subsubsection{Implication to intensity gradients}
	
	Intensity gradients also change their direction in the presence of self-gravity \citep{YL17b,IGs}. In particular, \citet{IGs} elaborated on the difference of intensity gradients and velocity gradients in the presence of self-gravity with more numerical detail. The study clearly showed that in self-gravitating regions, the change of intensity gradients' orientation is more dramatic and fast than velocity gradients. The expected reason is that the change of the density field is an accumulating process, while the velocity field is significantly changed only when the gravitational energy dominates over the kinematic energy of turbulence. This thus provides a way of, first of all, locating regions dominated by self-gravity and second, identifying the stage of gravitational collapsing. 
	
	By utilizing the properties of intensity gradients under the influencing of gravity, in \S~\ref{sec:IGT} we apply the tools developed in this paper,i.e., the double-peak algorithm and the curvature algorithm, to intensity gradients. We see that in the case of gravitational collapsing, intensity gradients flip their direction by 90$^\circ$, and large gradients' amplitudes are induced simultaneously. It then provides a self-consistent way to identify gravitational collapsing regions through intensity gradients. However, the change of the direction was shown to happen at different stages of the collapse \citep{LY18a}. By comparing the velocity and intensity gradients, one could distinguish the stage of the gravitational collapse.
	
	\subsection{Comparison with other techniques}
	In decades, several techniques have been proposed to probe magnetic field morphology based on the anisotropic properties of turbulence (see \S~\ref{sec:theory}). The Correlation Function Anisotropy (CFA) technique employs the correlation function to predict the magnetic field \citep{2005ApJ...631..320E,2014ApJ...790..130B}. At the same time, the CFA advantageously gives analytical predictions of different anisotropies associated with slow, fast, and Alfv\'{e}n modes \citep{KLP16, KLP17a,KLP17b}. However, the CFA requires a larger statistical area for averaging than the VGT, thus the CFA can only estimate a coarse structure of the magnetic field \citep{2018ApJ...865...54Y}.
	
	\citet{2014ApJ...789...82C} proposed the rolling Hough Transform (RHT) to study the magnetic field in the diffuse interstellar region probed by neutral hydrogen 21-cm emission. However, the RHT requires linear structures elongating along the magnetic field in the ISM \citet{2014ApJ...789...82C}. As for molecular clouds, there exist structures perpendicular to the magnetic field \citep{2016A&A...586A.138P,2019A&A...629A..96S}. Since RHT does not distinguish the perpendicular structures and the parallel structures with respect to the magnetic field, it is therefore incapable of revealing the magnetic field direction in the molecular cloud \citep{2016MNRAS.460.1934M}. Compared with RHT, VGT is parameter-free (since the block size is self-consistently found) while RHT requires three parameters as inputs: a smoothing kernel diameter (DK), window diameter (DW), and intensity threshold \citep{2014ApJ...789...82C}. Although CFA and RHT can be complementary to the study of magnetic fields in some phases of interstellar media, neither of them can be used to identify the gravitational collapse, but VGT can do. 
	
	The N-PDFs of column density is an alternative approach to study the gravitational collapse \citep{2011MNRAS.416.1436B,2011ApJ...727L..21P,1995ApJ...441..702V,2008ApJ...680.1083R,2012ApJ...750...13C,2018ApJ...863..118B}. The N-PDFs, however, are strongly limited by radiation transfer effects \citep{pdf}, the effects of line-of-sight contamination on the column density structure \citep{2015A&A...575A..79S,2015A&A...578A..29S,2019MNRAS.484.3604L}, and the stellar feedback via ionization \citep{2014A&A...564A.106T}.
	
	In addition to velocity gradients, intensity gradients are also widely used in the study of molecular clouds. The Histogram of Relative Orientation (HRO) developed in \citet{Soler2013} employs the intensity gradient to characterize the relative orientation of column density structures and magnetic fields. The HRO relies on the comparison of polarimetry measurement and does not distinguish between self-gravitating regions and shocks. Also, \citet{2012ApJ...747...79K} uses the intensity gradient of polarized dust emission to estimate the magnetic field strength in the molecular cloud. Their technique is different. It relies on polarimetry measurement and does not take sub-block averaging which is used in our IGT \citep{IGs}.
	
	\subsection{Shock identification}
	The change of the direction of intensity gradients can happen in the presence of shocks \citep{YL17b,IGs}, i.e., intensity gradients flip their direction by 90$^\circ$ in front of shocks, but this is not the case for velocity gradients. For instance, in \S~\ref{sec:IGT}, we show that the area in which intensity gradients flip their direction is significantly larger than the one seen in velocity gradients' orientation (see Fig.~\ref{fig:curvature} and Fig.~\ref{fig:IGT}). In Fig.~\ref{fig:IG_conv}, we confirm that the change that happened in diffuse regions is caused by shocks. At the same, the self-gravity will induce a large amplitude of intensity gradients than shocks. We can then separate shocks and gravitational collapsing through the comparison with either the intensity gradients' amplitude or velocity gradient's orientation (see \S~\ref{sec:IGT}). We summarize the difference of shocks and gravitational collapse in terms of the gradients in Tab.~\ref{tab:diss}.
	
	\begin{table}
		\centering
		\label{tab:diss}
		\begin{tabular}{| c | c | c | c | c |}
			\hline
			& $\nabla v_l$ v.s. $B$ & $\nabla \rho_l$ v.s. $B$  & $|\nabla v_l|$  & $|\nabla \rho_l|$ \\ \hline \hline
			Shock & $\perp$ & $\parallel$ & Low & Low \\  \hline
			Collapse & $\parallel$ & $\parallel$ & High & High \\
			\hline
		\end{tabular}
		\caption{Description of velocity gradient $\nabla v_l$ and density gradient $\nabla\rho_l$'s reaction with respect to shock and gravitational collapsing. The 2nd and 3rd columns are the relative orientation between gradients and the magnetic field $B$. The 4th and 5th columns indicate the gradients' amplitude.}
	\end{table}
	
	\subsection{Combining different approaches to trace magnetic fields in star-forming regions}
	Many polarimetry surveys showing the projected magnetic field morphology of molecular clouds, including the Planck survey of diffuse dust polarized emission \citep{2018arXiv180104945P}, which provided us with a comprehensive picture of magnetic field morphology across the full sky. The problem that grain alignment efficiency drops near the gravitational object center brings more concern on the accuracy of such surveys \citep{Aetal15}. It is well known that in high-density regions, grain alignment frequently fails if the radiation field is not strong enough, while in the vicinity of the radiation sources, the alignment can happen with respect to radiation rather than the ambient magnetic field \citep{2007MNRAS.378..910L,2018ApJ...852..129H}. Inside Giant Molecular Clouds (GMCs), this occurs most likely in the region under severe gravitational collapse. The gravitational collapse prevents observers from studying the contribution of the magnetic field to the star formation process. 
	{
		
		The VGT has been developed as a vast ecosystem which not only can trace magnetic field tracing method in both the diffuse atomic gas region and dense molecular region \citep{PGCA, 2019ApJ...873...16H,survey,velac}, but also measure the magnetization \citep{LY18a,2020arXiv200201926Y}, study the turbulent properties of the ISM, i.e., measuring M$_S$ \citep{2018ApJ...865...54Y}, distinguish shocks \citep{IGs}, and probing the magnetic field in galaxy clusters \citep{cluster}. The magnetization is correlated with the dispersion of the velocity gradient's orientation. When the media is highly magnetized, the compressed gas exhibits small dispersion, while the dispersion gets increment for a weakly magnetized environment. The magnetization, therefore, can be computed through the power-law correlation between M$_A$ and the dispersion given in \citet{2018ApJ...865...46L}. A similar argument has also been extended to the intensity gradient by \citet{IGs}. In addition to the dispersion, the geometry of the velocity gradient is still sensitive to the magnetic field strength. A strong magnetic field bends the stream-path formed by the velocity gradient showing small curvatures. The corresponding study of measuring M$_A$ from the curvature in done by  \citet{2020arXiv200201926Y}. The velocity gradient also contains fruitful information about the turbulence. Supersonic turbulence induces the compressible motion of the fluid, which accordingly shows a large amplitude of the gradients, including both velocity and intensity gradients. \citet{2018ApJ...865...54Y} gives the power-law relation of M$_s$ and the dispersion of the amplitude. Also, to study the magnetic field and turbulence, one particular property of velocity gradients is that they flip their directions by 90$^\circ$ in the presence of gravitational collapsing. This property gives two bonuses, i.e., identifying the collapsing regions and restoring the magnetic field morphology in a self-gravitating GMC. As shown in Fig.~\ref{fig:illu}, we firstly identify the collapsing region through the double-peak algorithm and then re-rotate the velocity gradients by 90$^\circ$. The re-rotated velocity gradients still align with the magnetic field inferred from synthetic polarization measurement, showing AM = 0.75. The comprehensive picture of the parameter-free VGT is clear. Using the molecular line observations, VGT can probe the magnetic field morphology, magnetization, sonic Mach number over scales ranging from tens of parsecs to $<<0.1pc$, being synergistic to dust polarimetry. Once the observation has a sufficiently high resolution which resolves the collapsing structures in the scale larger than the minimum sub-block size (empirically the minimum size is 20$\times$ 20 pixels), VGT reveals the self-gravitating region.
		
		To ensure the high accuracy of VGT, several approaches have been proposed to remove the noise, i.e., the Principal Component Analysis \citep{PCA} and extract the most crucial velocity gradients components, i.e., the sub-block averaging method \citep{YL17a} and moving window \citep{LY18a}. In this work, based on our old sub-block averaging method, we develop the adaptive sub-block averaging method which guarantees the high resolution of VGT. The uncertainty of VGT measurement can also be accurately quantified through the standard deviation of velocity gradients' orientation in each sub-block. Comparing with polarimetry, VGT shows advantages in probing the local magnetic fields getting rid of the contribution along LOS with high resolution. Besides the study of GMC, VGT can then assist in modeling the three-dimensional Galactic magnetic fields, which is indispensable for removing the Galactic foreground and detecting primordial B-mode polarization in the CMB. Our prior works have applied VGT to the full data of the GALFA-\ion{H}{1} survey and showing good correspondence, including both magnetic field tracing and E/B modes decomposition, with those reported by Planck 353 GHz \citep{PGCA}. 
		
		\subsection{Observational application of VGT}
		The development of the VGT was initially motivated by studying magnetic fields in turbulent diffuse gas. In molecular clouds, it was shown that the velocity gradients change their directions with respect to the magnetic field \citep{YL17b,LY18a,survey}. However, the identification of these regions with polarimetry devalued the VGT technique, which is the technique of tracing the magnetic field on its own. This paper provides a detailed study of how the VGT can work on its magnetic fields tracing in self-gravitating molecular clouds. In addition to magnetic field tracing, the VGT is shown to be able to identify the collapsing self-gravitating regions. The problem of identifying such regions is significant for the understanding of star formation.
		
		The abundant spectroscopic data sets of different molecular tracers, for instance, the CO data obtained from JCMT \citep{2019ApJ...877...43L}, GAS \citep{2017A&A...603A..89K}, COMPLETE \citep{2006AJ....131.2921R}, FCRAO \citep{1995ApJS...98..219Y}, ThrUMMS \citep{2015ApJ...812....6B}, CHaMP \citep{2005ApJ...634..476Y}, and MALT90 \citep{2011ApJS..197...25F} surveys, extends the applicability of VGT. \citet{velac} showed VGT could construct the 3D magnetic fields model in molecular clouds utilizing different emission lines of Vela C molecular cloud. For example, using $^{12}CO$, $^{13}CO$, and $C^{18}O$ data, VGT tells us about the POS component of the magnetic field over three different volume density ranges, from $10^2cm^{-3}$ to $10^4cm^{-3}$. VGT is also applicable to higher density tracers, such as CS, HCN, HCO$^{+}$ \citep{velac}. We can, therefore, expect that VGT can reveal the volume density range in which the collapsing occurs using multiple emission lines. A similar argument also holds for intensity gradients.
		
		\section{Summary}
		\label{sec:con}
		The paper continues advancing the Velocity Gradient Technique and extends it to the regions where the effect of self-gravity is important. We attempt to solve two inter-related problems. First of all, we explore the ability of gradients to trace of the magnetic field in the presence of gravitational collapse. In parallel, we study how to identify different stages of the gravitational collapse observationally. In our study, we present star formation simulations for different magnetization of the media and different levels of turbulence. We observe that star formation can successfully happen in strongly magnetized media and the non-ideal effects like ambipolar diffusion are not necessary for this. In our simulations, after the gravitational collapsing starts, the dispersion of intensity N-PDFs, as well as the $M_S$, are increasing with the super-sonic turbulence is driven by self-gravity.
		Our main advances are:
		\begin{enumerate}
			
			\item We confirm that intensity gradients and velocity gradients flip their orientation by 90$^\circ$ in the presence of gravitational collapsing, becoming parallel to magnetic fields.
			
			\item Further developing the VGT we introduce a new adaptive sub-block averaging method, which increases the resolution of the magnetic field map that we obtain with our technique. We test it using the synthetic data from MHD simulations.
			
			\item We demonstrate that without polarization measurements VGT is capable to identify regions of gravitational collapse through:
			\begin{enumerate}
				\item The double-peak feature in the histogram of gradients' orientation, i.e., in the boundary of collapsing regions, the histogram shows two Gaussian distribution with peak values $\theta$ and $\theta\pm\pi/2$.
				\item The curvature of velocity gradients, which reaches its maximum value in a gravitational collapsing region since intensity gradients and velocity gradients rapidly change their direction by 90$^\circ$.
				\item The synergetic utilization of velocity gradients' morphology and amplitude that can reveal the inflow motion in collapsing regions.
			\end{enumerate}
			
			\item We demonstrate the VGT's ability in tracing magnetic fields in both diffuse and gravitational collapsing regions. 
			
			\item We demonstrate the ability of VGT to reveal the boundaries of the transitional regions and determining the density range for which the gravitational collapse takes place. 
			
			\item Comparing the VGT to the analysis of N-PDFs, we demonstrate that N-PDFs are sensitive to the strongly convergent flows induced by the already formed accreting cores. The density threshold given by N-PDFs does not distinguish self-gravitating gas and non-self-gravitating density enhancement, while VGT traces all gravity-induced converging flow.  
			
			\item We explore the synergy of the VGT and its offshoot, the Intensity Gradient Technique (IGT) to identify different stages of the gravitational collapse as well as identify shocks. 
		\end{enumerate}
		\noindent
		{\bf Acknowledgment}  {We acknowledge Siyao Xu and Ka Wai
			Ho for the fruitful discussions. We thank the reviewer for the numerous suggestions for improving the manuscript. A.L. acknowledges the support of the NSF grant AST 1715754, and 1816234. K.H.Y. acknowledges the support of the NSF grant AST 1816234. Y.H. acknowledges the support of the NASA TCAN 144AAG1967. We acknowledge the allocation of computer time by NERSC TCAN 144AAG1967 and the Center for High Throughput Computing (CHTC) at the University of Wisconsin. }
		\software{Julia \citep{2012arXiv1209.5145B}, ZEUS-MP/3D \citep{2006ApJS..165..188H}, Paraview \citep{ayachit2015paraview}}
		
		\newpage
		\appendix
		\section{The relative angle between magnetic field and line-of-sight}
		As above, in both the strong magnetic field and weak magnetic field environments, the convergent flow induced by the gravitational collapse is following the magnetic field direction. As a result, the convergent flow significantly modified the properties of velocity gradients so that their direction becomes parallel to the magnetic field and exhibit a larger amplitude (see \S~\ref{sec:theory}). 
		
		Nevertheless, in observation, one can only get access to the  the velocity component parallel to LOS. Ideally, if the flow induced by the gravitational collapse is exactly perpendicular to LOS and the flow does not allow the transfer of energy between different components, the change of velocity gradient may not be observed. However, as discussed in \S~\ref{subsec:shockvschannel}, in reality, the gravitational collapse accelerates the turbulent flow in all directions changing all the components of velocity. For example, in our sub-Alfv\'{e}nic numerical simulation, the magnetic field is set to be perpendicular to LOS, which indicates ideally the convergent flow should lie on the POS. However, as shown in Fig.~\ref{fig:3dBV}, near the collapse region, gravity bends the magnetic field so that all velocity components get converged. The emission lines therefore always contain the velocity information about the gravitational collapse.
		\begin{figure*}[p]
			\centering
			\includegraphics[width=.99\linewidth,height=.94\linewidth]{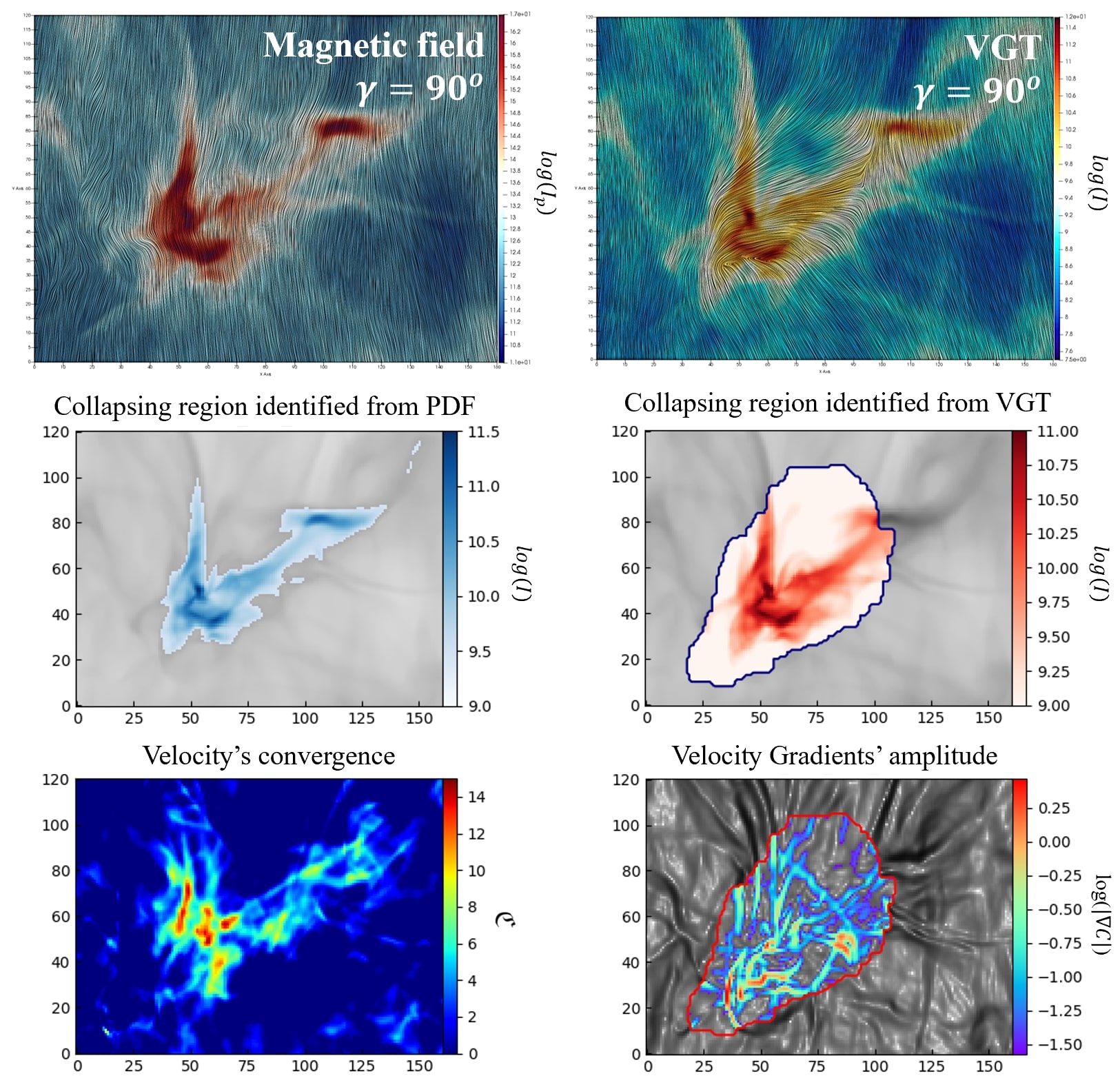}
			\caption{\label{fig:run2} An example of how the VGT identifies collapsing regions from the double-peak algorithm, in the condition that the global magnetic field is perpendicular to the line-of-sight. We use simulation A3 with $t_r\simeq1.0$ Myr and length scale $\simeq 1.96$ pc. \textbf{Top right:}: the magnetic field morphology inferred from VGT. The magnetic field is superimposed on the projected intensity map and visualized using the LIC. $\gamma$ is the relative angle between the mean magnetic field and LOS. \textbf{Top left:} the magnetic field morphology inferred from synthetic polarization, which is superimposed on the polarized intensity map. \textbf{Middle left:} the gravitational collapsing regions (blue regions) identified from the N-PDFs. \textbf{Middle right:} the collapsing regions identified by the double-peak algorithm velocity gradients with the 2$^{nd}$ block size 30. \textbf{Bottom left:} the projected velocity convergence. \textbf{Bottom right:} the areas (red contour) of the velocity gradient's amplitude corresponding to the collapsing regions identified by the double-peak algorithm. In the same regions, we mask the low amplitude pixel, i.e., its corresponding amplitude is less than the global mean value.}
		\end{figure*}
		\begin{figure*}[p]
			\centering
			\includegraphics[width=.99\linewidth,height=.94\linewidth]{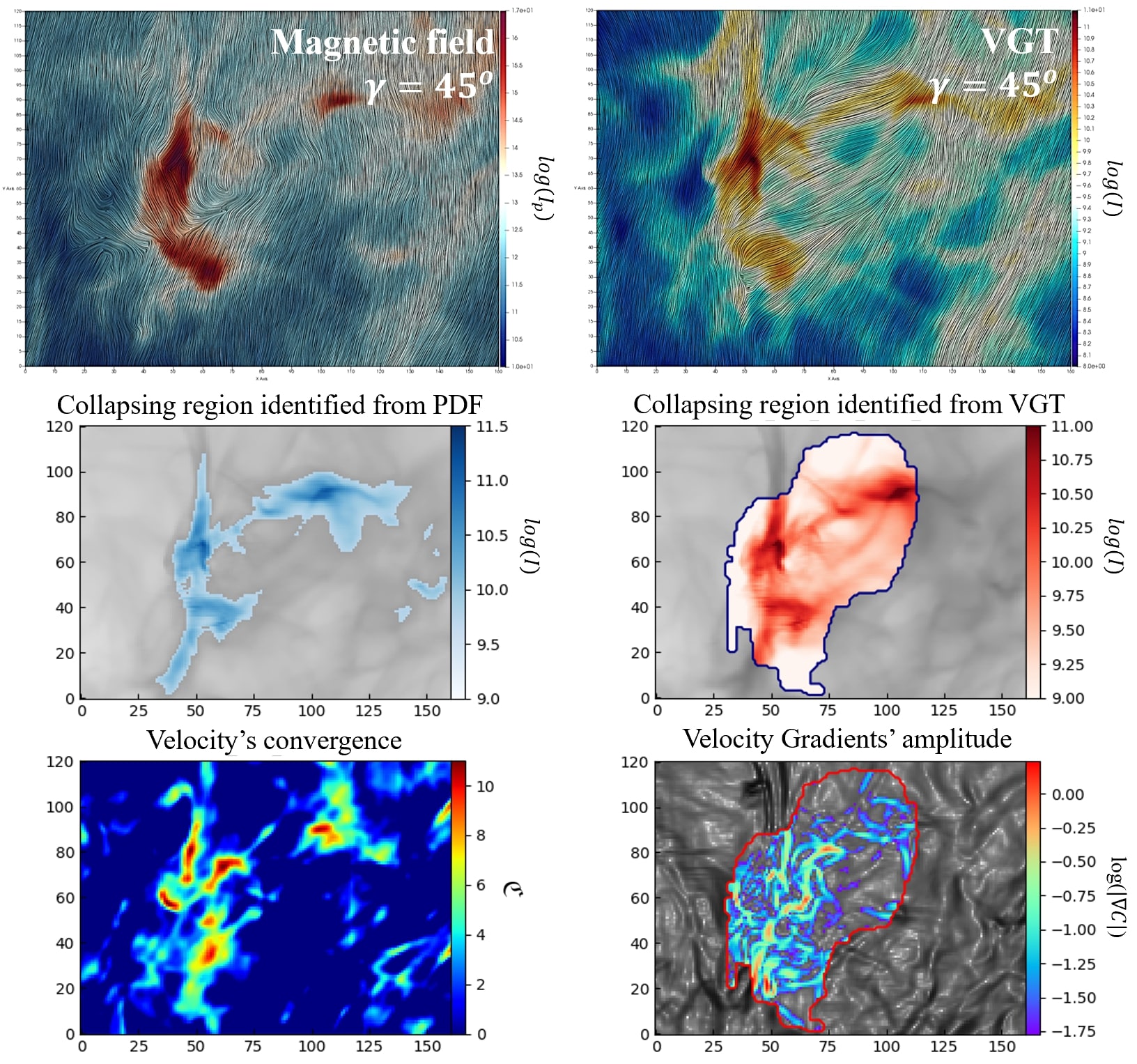}
			\caption{\label{fig:run2_rotation} An example of how the VGT identifies collapsing regions from the double-peak algorithm, in the condition that the mean magnetic field is 45$^\circ$ inclined to the line-of-sight. We use simulation A3 with $t_r\simeq1.0$ Myr and length scale $\simeq 1.96$ pc. \textbf{Top right:}: the magnetic field morphology inferred from VGT. The magnetic field is superimposed on the projected intensity map and visualized using the LIC. $\gamma$ is the relative angle between the mean magnetic field and LOS. \textbf{Top left:} the magnetic field morphology inferred from synthetic polarization, which is superimposed on the polarized intensity map. \textbf{Middle left:} the gravitational collapsing regions (blue regions) identified from the N-PDFs. \textbf{Middle right:} the collapsing regions identified by the double-peak algorithm velocity gradients with the 2$^{nd}$ block size 30. \textbf{Bottom left:} the projected velocity convergence. \textbf{Bottom right:} the areas (red contour) of the velocity gradient's amplitude corresponding to the collapsing regions identified by the double-peak algorithm. In the same regions, we mask the low amplitude pixel, i.e., its corresponding amplitude is less than the global mean value.}
		\end{figure*}
		To test how the relative angle between the magnetic field and LOS affects VGT, we use a new MHD simulation A3 with M$_{s}\simeq6.47$, M$_{A}\simeq0.61$, $\beta\simeq0.018$, and resolution $792^3$. We consider the simulation at $t_r\simeq1.0$ Myr snapshot to have a best match VGT with the N-PDFs, as Fig.~\ref{fig:curvature} shows the N-PDFs is not sensitive to the initial stages of gravitation collapse. A3 is still isothermal with temperature T = 10.0 K, sound speed $c_s$ = 187 m/s and cloud size L = 10 pc. The total mass $M_{tot}$ in the simulated cubes A3 is $M_{tot} \sim 8191.45$ M$_\odot$, as well as the magnetic Jean mass $M_{JB}\sim 143.57$ M$_\odot$, average magnetic field strength $B\sim14.18\mu G$, mass-to-flux ratio $\Phi\sim1.06$, volume density $\rho\sim141.47$ cm$^{-3}$, and free fall time $t_{ff}\sim$ 2.82 Myr. Similar to the simulation described in \S~\ref{Sec.data}, we keep driving both turbulence and self-gravity until the simulation violates the Truelove criterion \citep{1997ApJ...489L.179T}. The jeans length $\sim$1.77 pc, therefore, occupies 140 pixels. The density can be enhanced by a factor of $\sim$ 1225.0, i.e., we can have maximum volume density $\rho_{max}\sim$ 173300.75 cm$^{-3}$, which informs us when to stop the simulation before having numerical artifacts due to self-gravitating collapse.
		
		In Fig.~\ref{fig:run2}, we set the global magnetic field perpendicular to LOS and zoom-in one sub-region A3-1 from the simulation A3. This sub-region contains well-defined convergent flows and formed cores. We plot the magnetic field morphology predicted by synthetic dust polarization and VGT. We observed in the high-density regions, the rotated velocity gradients are perpendicular to the magnetic field rather than align with it. According to our study presented in \S~\ref{sec:results}, this is potentially a gravitational collapsing region. Through the density-threshold determined from the N-PDFs, we highlight the region in which the N-PDFs is a power-law distribution, indicating this region is under gravitational collapse. We find the collapsing region identified from N-PDFs agrees with the one showing 90$^\circ$ change of velocity gradient. There is one exception on the top right region, in which the N-PDFs are indicating gravitational collapse which there is no change of velocity gradient observed. The possible explanation includes (i) the density threshold given by N-PDFs does not distinguish self-gravitating gas and non-self-gravitating density enhancement; (ii) the fraction of collapsing material is small in this region so that the velocity gradient does not reveal its existence, see the discussion in \S~\ref{subsec:dp}. Nevertheless, we apply the double-peak algorithm with the 2$^{nd}$ block size 30 to identify the collapsing regions from the velocity gradient (see Fig.~\ref{fig:run2}). First of all, we observe that the majority of self-gravitating regions identified from the two methods, i.e, N-PDFs, and VGT,  are similar and reflect the convergent flows. However, VGT shows larger identified areas than N-PDFs, since the double-peak is required to cover both diffuse and the self-gravitating regions to identify the boundary. To reduce the boundary effect, we highlight the region showing a high gradient's amplitude, i.e., its corresponding amplitude is larger than the global mean value. Since the gravitational collapse also induces a high gradient's amplitude (see \S~\ref{sec:results}), this procure helps exclude the additional diffuse regions introduce by the double-peak algorithm. 
		
		The above result illustrated the case that the global magnetic field is perpendicular to LOS. We then rotate the MHD simulation A3 so that the magnetic field is 45$^\circ$ inclined to LOS. The corresponding convergent flow is changed together with the magnetic field. We repeat the calculation of VGT as above. The results are presented in Fig.~\ref{fig:run2_rotation}. We can see that for the low-density regions, the magnetic field morphology inferred from VGT agrees well with the one derived from synthetic polarization measurement. While for the intermediate regions, the velocity gradient starts appearing misalignment with the magnetic field. This misalignment comes from the spread convergent flow in the space, since the inclination angle $\gamma=45^\circ$. An insufficient collapsing material along LOS will also induce the change of velocity gradient, but this change does not arrive at $90^\circ$. Nevertheless, in high-density regions, the VGT measurement gets perpendicular to the magnetic field and the magnetic field is also partially bent by the self-gravity, as we see in Fig.~\ref{fig:3dBV}. We can see this high-density region corresponds to the collapsing region determined by N-PDFs.
		By repeating the double-peak \& gradient amplitude algorithm, we get similar results as the N-PDFs, and also determined the convergent flow. We, thus, conclude the relative angle between the magnetic field and LOS does not degrade the VGT.


		
	}
\end{document}